\documentclass[twocolumn,showpacs,floatfix,superscriptaddress, showkeys,prc]{revtex4-2}%

\usepackage{mathrsfs}
\usepackage{graphicx}
\usepackage{dcolumn}
\usepackage{bm}
\usepackage[colorlinks,linkcolor=blue,citecolor=blue]{hyperref}
\usepackage{amsmath, amssymb}
\usepackage{CJK}
\usepackage{multirow}
\usepackage[section]{placeins}

\newcommand{\ivec}[1]{\vec{#1}}
\newcommand{\svec}[1]{\boldsymbol{#1}}
\newcommand{\scr}[1]{{\mathscr #1}}

\newcommand{\cals}[1]{{\mathcal #1}}
\newcommand{\ff}[1]{\frac{1}{#1}}

\newcommand{\lrlc}[1]{\left|#1\right>}
\newcommand{\lrcl}[1]{\left<#1\right|}
\newcommand{\lrs}[1]{\left[#1\right]}

\newcommand{\Lrb}[1]{\left\{#1\right\}}

\newcommand{\sigs}{{\sigma\text{-S} }}
\newcommand{\omev}{{\omega\text{-V} }}
\newcommand{\rhov}{{\rho  \text{-V} }}

\newcommand{\couv}{{ A    \text{-V} }}

\usepackage{ulem}

\newcommand{\delete}{\bgroup\markoverwith{\textcolor{red}{\rule[0.5ex]{2pt}{1pt}}}\ULon}

\begin{document}


\title{Relativistic Hartree-Fock model for axial-symmetric nuclei with quadruple and octupole deformations}
\author{Yong Peng}
\affiliation{Frontier Science Center for Rare isotope, Lanzhou University, Lanzhou 730000, China}
\affiliation{School of Nuclear Science and Technology, Lanzhou University, Lanzhou 730000, China}
\author{Jing Geng}
\affiliation{Frontier Science Center for Rare isotope, Lanzhou University, Lanzhou 730000, China}
\affiliation{School of Nuclear Science and Technology, Lanzhou University, Lanzhou 730000, China}
\author{Wen Hui Long}\email{longwh@lzu.edu.cn}
\affiliation{Frontier Science Center for Rare isotope, Lanzhou University, Lanzhou 730000, China}
\affiliation{School of Nuclear Science and Technology, Lanzhou University, Lanzhou 730000, China}
\affiliation{Joint Department for Nuclear Physics, Lanzhou University and Institute of Modern Physics, CAS, Lanzhou 730000, China}

\begin{abstract}

\noindent\textbf{Background:} The initial observation of a negative-parity state in proximity to the ground state in the 1950s marked the advent of extensive research into octupole deformed nuclei. Since then, the physics of octupole deformed nuclei has consistently held a special interest within the field of nuclear physics. In the present era, with the advent of sophisticated radioactive ion beam (RIB) facilities and advanced detectors, coupled with the remarkable capabilities of high-performance computing, extensive and intensive explorations are being conducted from both experimental and theoretical perspectives to elucidate the physics of octupole deformed nuclei.

\noindent\textbf{Purpose:} The objective of this study is to develop an axially Octupole-quadruple Deformed Relativistic Hartree-Fock (OD-RHF) model for the reliable description of octupole deformed nuclei. This will be achieved by utilizing the spherical Dirac Woods-Saxon (DWS) basis to expand the single-particle wave functions.

\noindent\textbf{Method:} The Lagrangian of nuclear systems, the starting point of the OD-RHF model, is achieved by considering the degrees of freedom associated with the isoscalar ($\sigma$ and $\omega$) and isovector ($\rho$ and $\pi$) meson fields. Following the standard procedure, the RHF energy functional is further obtained by considering the expectation of the Hamiltonian with respect to the Hartree-Fock ground state. Due to the non-local Fock terms, the variation of the RHF energy functional yields the integro-differential Dirac equation. After incorporating both quadruple and octupole deformations, such an integro-differential Dirac equation is solved by expanding the Dirac spinors on a spherical DWS basis. For open-shell nuclei, the pairing correlations are treated within the BCS scheme, in which the finite-range Gogny force D1S is employed as the pairing force.

\noindent\textbf{Results:} The reliability of the newly developed OD-RHF model is illustrated by taking the octupole nucleus $^{144}$Ba as an example. Furthermore, the octupole deformation effects in $^{144}$Ba is verified by using the RHF Lagrangians PKO$i$ ($i=1,2,3$) and the RMF one DD-ME2. 

\noindent\textbf{Conclusions:} This work establishes the OD-RHF model, which provides a reliable tool for studying octupole nuclei over a fairly wide range. The intrusion of the neutron $1i_{13/2}$ and proton $1h_{11/2}$ components is demonstrated to play an essential role in determining the notable octupole deformation of $^{144}$Ba using PKO$i$ ($i=1,2,3$) and DD-ME2. It is indicated that the Fock terms play an important role in stabilizing the octupole deformation. More specifically, due to the repulsive tensor coupling between the intrude components and the core of $^{144}$Ba, the tensor force component carried by the $\pi$-PV coupling, that contributes only via the Fock terms, plays an opposing role in the formation of the octupole deformation of $^{144}$Ba.

\end{abstract}

\pacs{21.60.-n, 21.30.Fe, 21.10.-k}
\maketitle

\section{Introductions}

The development of radioactive ion beam (RIB) facilities and advanced detectors \cite{Sturm2010NPA834.682c, Thoennessen2010NPA834.688c, Zhan2010NPA834.694c, Motobayashi2010NPA834.707c, Gales2010NPA834.717c} has led to significant advances in nuclear physics. These advances have enriched the field of nuclear physics by enabling the observation of the nuclei lying far from the stability line in nuclear landscape, namely the exotic nuclei \cite{Jonson2004PhyRep389.1, Tanihata1995PPNP35.505, Jensen2004RMP76.215, Casten2000PPNP45.S171, Ershov2010JPG:NP37.064026}. These nuclei exhibit plenty of novel nuclear phenomena, including the disappearance of traditional magic shells and the emergence of the new shells \cite{Hoffman2008PRL100.152502, Simon1999PRL83.496, Motobayashi1995PLB346.9, TshooPRL109.022501, Ozawa2000PRL84.5493, Kanungo2009PRL102.152501, Gade2006PRC74.021302, Steppenbeck2015PRL114.252501, Steppenbeck2013Nature502.207}, as well as the halo phenomena \cite{Minamisono1992PRL69.2058, Tanihata1985PRL55.2676, Schwab1995ZPA350.283}. Such intriguing novel phenomena not only challenge our understanding of traditional nuclear physics, but also present rich opportunities for nuclear theoretical and experimental investigations.

On the other hand, it is widely acknowledged that the majority of nuclides in nuclear chart are deformed, except for a few close to the magic numbers. For most deformed nuclei, a description in terms of axial and reflection symmetry is sufficient, as the shape is symmetric under space inversion. However, with the first observation of a negative-parity state in proximity to the ground state in the 1950s \cite{Asaro1953PR92.1495, Stephens1954PR96.1568, Stephens1955PR100.1543}, it became evident that some nuclei may exhibit an asymmetric shape under reflection, such as a pear-like shape \cite{LeePR1957108.774}. Microscopically, the appearance of those nuclei with a pear-like shape is attributed to the coupling between two single-particle orbits near the Fermi surface which differ by $\Delta l$ = 3 and $\Delta j$ = 3. Such nuclei are located close to proton and neutron numbers of 34, 56, 88 and neutron number of 134 \cite{Butler1996RMP68}. It is notable that strong octupole deformation has been confirmed in the nuclei of $^{224}$Ra \cite{Gaffney2013NATURE477}, $^{144}$Ba \cite{Bucher2016PRL116}, $^{146}$Ba \cite{Bucher2016PRL118}, $^{228}$Th \cite{Chishti2020NARURE16} and $^{96}$Zr \cite{Zhang2022PRL128} in recent years. Connecting with the octupole deformation, a multitude of novel nuclear phenomena have been discovered \cite{Butler1996RMP68}, as evidenced by the alternating-parity rotational bands, low-lying $1^-$ and $3^-$ states and the $E3$ transition. In parallel with the experimental progress, considerable theoretical efforts have been dovoted to understanding these colorful phenomena, which are potentially related to the octupole deformation \cite{Butler1996RMP68, Butler2016JPG43.073002}.

Under the meson-propagated picture of nuclear force, the relativistic Hartree-Fock (RHF) theory \cite{Bouyssy1987PRC36.380}, which includes both the Hartree and Fock diagrams, provides us a powerful tool for exploring the structural properties of nuclei. It is clear that only working through the Fock terms allows us to take into account the important degrees of freedom associated with the $\pi$-pseudo-vector ($\pi$-PV) and $\rho$-tensor ($\rho$-T) couplings. Nowadays, the RHF theory and the extension, which incorporate the density-dependent meson-nucleon coupling strengths \cite{Long2006PLB640.150, Long2007PRC76.034314, Long2010PRC81.024308, Geng2020PRC101.064302, Geng2022PRC105.034329}, have shown comparable accuracy to popular mean-field models in describing nuclear structure properties, with the proposed RHF Lagrangians PKO$i$($i$=1,2,3) \cite{Long2006PLB640.150, Long2008EPL82.12001} and PKA1 \cite{Long2007PRC76.034314}. Over the past twenty years, with the presence of the Fock terms, significant improvements have been achieved in the self-consistent description of nuclear structure properties, such as shell evolutions \cite{Long2008EPL82.12001, Long2009PLB680.428, Li2016PLB753.97, Wang2013PRC87.047301}, symmetry energy \cite{Sun2008PRC78.065805, Long2012PRC85.025806, Zhao2015JPG42.095101}, new magicity \cite{Li2016PLB753.97, Li2014PLB732.169, Li2019PLB788.192}, novel phenomena \cite{Long2010PRC81.031302, Lu2013PRC87.034311, Li2019PLB788.192}, pseudo-spin symmetry \cite{Long2006PLB639.242, Long2007PRC76.034314, Liang2010EPJA44.119, Geng2019PRC100.051301R}, and spin and isospin excitations \cite{Liang2008PRL101.122502, Liang2009PRC79.064316, Liang2012PRC85.064302, Niu2013PLB723.172, Niu2017PRC95.044301}. In particular, the important ingredient of nuclear force --- the tensor force can be taken into account naturally via the Fock terms \cite{Jiang2015PRC91.025802, Jiang2015PRC91.034326, Zong2018CPC42.024101}, e.g., by the $\pi$-PV coupling. It should be noted that the tensor force was found to play an important role in nuclear shell evolution \cite{Long2008EPL82.12001, Wang2013PRC87.047301}, symmetry energy \cite{Jiang2015PRC91.025802}, and nuclear excitations \cite{Bai2009Phys.Lett.B675.28, Bai2009Phys.Rev.C79.041301, Bai2010Phys.Rev.Lett.105.072501, Bai2011Phys.Rev.C83.054316, Bai2011Phys.Rev.C84.044329}.

In order to extend the RHF theory for deformed nuclei, significant efforts were devoted to solving the integro-differential Dirac equation, which contains the non-local mean fields contributed by the Fock terms. In solving the differential Dirac equation, Dirac spinors were typically expanded on the harmonic oscillator (HO) basis \cite{Pannert1987PRL59.2420, Price1987PRC36.354, Gambhir1990AP198.132} or the Dirac Woods-Saxon (DWS) basis \cite{Zhou2003PRC68.034323, Li2012PRC85.024312, Chen2012PRC85.067301}. It is noteworthy that analogous attempts have been made to expand the Dirac spinor and the RHF mean field on the deformed fermionic and bosonic HO bases, respectively, thereby providing the first RHF description of deformed nuclei \cite{Ebran2011PRC83.064323}. However, the calculation of the non-local mean fields is prohibitively time-consuming, particularly in dealing with the rearrangement terms originating from the density dependence of the coupling strengths, which limits the further extension of the model. In contrast to the HO basis, the DWS basis \cite{Zhou2003PRC68.034323} offers the advantage of providing an appropriate asymptotic behavior of the wave functions, which is essential for a reliable description of loosely bound exotic nuclei. By expanding the Dirac spinors upon the spherical DWS basis, both the RHF and relativistic Hartree-Fock-Bogoliubov (RHFB) theories were extended for axially quadruple deformed nuclei \cite{Geng2020PRC101.064302, Geng2022PRC105.034329}, respectively the D-RHF and D-RHFB models. As a primary application, the D-RHFB model with PKA1 is capable of accurately reproducing well both the even-parity ground state and halo structure of $^{11}$Be \cite{Geng2023CPC47.044102}.

In this work, a new axially deformed RHF model, abbreviated as the OD-RHF model, will be established by incorporating both quadruple and octupole deformations. Similar to the D-RHF and D-RHFB models, which consider only the quadruple deformation, Dirac spinors are expanded on a spherical DWS basis, and the pairing correlations are treated by the BCS method with the finite-range Gogny force D1S \cite{Berger1984NPA428.23} as the pairing force. Furthermore, the octupole nucleus $^{144}$Ba is taken as an example to understand the octuple deformation, by focusing on the effects of the Fock terms. The paper is organized as follows. In Sec. \ref{sec:general formalism}, we present the general formalism of the OD-RHF model based on the spherical DWS basis. Sec. \ref{sec:result and discussions} presents the results and discussions, including the space truncation, the convergence check, and the description of the nucleus $^{144}$Ba. Finally, a summary is given in Sec. \ref{sec:summary}.

\section{General formalism}\label{sec:general formalism}
This section will provide a brief introduction of the general formalism of the relativistic Hartree-Fock (RHF) theory and the RHF energy functional for the nuclei with the reflection-asymmetric deformation. In order to provide readers with a comprehensive understanding, some details of the Dirac equations with the spherical DWS basis and the pairing correlations will also be introduced. For further details on the Dirac Woods-Saxon potentials and deformation parameters, please refer to Appendix \ref{app:DWS-DP}.

\subsection{RHF Energy Functional}
Under the meson-exchange diagram of the nuclear force, the Lagrangian of nuclear systems, the starting point of the RHF theory, can be obtained by considering the degrees of freedom associated with the nucleon ($\psi$) and the meson/photon fields, including the isoscalar $\sigma$- and $\omega$-mesons, the isovector $\rho$- and $\pi$-mesons, and the photon field ($A^\mu$). Among the select degrees of freedom, the strong attractions and repulsions between nucleons are carried by the isoscalar $\sigma$- and $\omega$-fields, respectively, the isovector ones are introduced to describe the isospin-related properties of nuclear force, and the photon field for the electromagnetic interactions between protons. Consequently, the Lagrangian of nuclear systems can be expressed as,
\begin{align}
\scr L=&\scr L_{M}+\scr L_{\sigma}+\scr L_{\omega }+{\mathscr L}_{\rho}+\scr L_{\pi}+\scr L_{A}+\scr L_{I}, \label{Lagrangian}
\end{align}
where the Lagrangians of the free fields $\scr L_\phi$ ($\phi = \psi,\sigma, \omega^\mu, \ivec\rho^\mu, \ivec\pi$ and $A^\mu$) read as,
\begin{align}
\scr L_{M}=&\bar{\psi}\left(  i\gamma^{\mu}\partial_{\mu}-M\right)\psi, \\
\scr L_{\sigma}=  &  +\frac{1}{2}\partial^{\mu}\sigma\partial_{\mu}\sigma-\frac{1}{2}m_{\sigma}^{2}\sigma^{2}, \\
\scr L_{\omega}=& -\frac{1}{4}\Omega^{\mu\nu}\Omega_{\mu\nu}+\frac{1}{2}m_{\omega}^{2}\omega_{\mu}\omega^{\mu}, \\
\scr L_{\rho}=& -\frac{1}{4}\vec{R}_{\mu\nu}\cdot\vec{R}^{\mu
\nu}+\frac{1}{2}m_{\rho}^{2}\vec{\rho}^{\mu}\cdot\vec{\rho}_{\mu},\\
\scr L_{\pi}= & +\frac{1}{2}\partial_{\mu}\vec{\pi}\cdot \partial^{\mu}\vec{\pi}-\frac{1}{2}m_{\pi}^{2}\vec{\pi}\cdot\vec{\pi},\\
\scr L_{A}= &  -\frac{1}{4}F^{\mu\nu}F_{\mu\nu},
 \end{align}
with the field tensors $\Omega^{\mu\nu} \equiv \partial^{\mu}\omega^{\nu} - \partial^{\nu} \omega^{\mu}$, $\vec{R}^{\mu\nu} \equiv \partial^{\mu}\ivec{\rho}^{\nu} - \partial^{\nu} \ivec{\rho}^{\mu}$, and $F^{\mu\nu} \equiv \partial^{\mu}A^{\nu} - \partial^{\nu}A^{\mu}$. Considering the Lorentz scalar ($\sigma$-S), vector ($\omega$-V, $\rho$-V and $A$-V), tensor ($\rho$-T) and pseudo-vector ($\pi$-PV) couplings, the Lagrangian $\scr L_I$ that describes the interactions between the nucleon and the mesons (photons) can be written as,
 \begin{align}
\scr L_I = \bar\psi &\Big( -g_\sigma\sigma - g_\omega\gamma^\mu\omega_\mu - g_\rho\gamma^\mu\ivec\tau\cdot\ivec\rho_\mu - e\gamma^\mu\frac{1-\tau_3}{2} A_\mu \nonumber \\ &\hspace{1em} + \frac{f_\rho}{2M}\sigma_{\mu\nu}\partial^\nu \ivec\rho^\mu\cdot\ivec\tau -\frac{f_\pi}{m_\pi}\gamma_5\gamma^\mu\partial_\mu \ivec\pi\cdot\ivec\tau\Big) \psi.
 \end{align}
In the Lagrangian densities, $M$ and $m_\phi$ are the masses of the nucleon and mesons, and $g_\phi$ ($\phi = \sigma, \omega^\mu, \ivec\rho^\mu$) and $f_{\phi'}$ ($\phi' = \ivec\rho^\mu, \ivec\pi$) represent the meson-nucleon coupling strengths. In this paper, we employ the use of arrows to signify isovector quantities and the use of bold font to indicate space vectors.

From the Lagrangian (\ref{Lagrangian}) the Hamiltonian can be derived following the Legendre transformation. After neglecting the time component of the four-momentum carried by the mesons (photon), namely ignoring the retardation effects, and substituting the meson/photon field equations, the Hamiltonian can be written as,
\begin{align}
  H =& T + \sum_\phi V_\phi,
\end{align}
where the kinetic energy term $T$ and the potential energy one $V_\phi$ read as,
\begin{align}\label{eq:Hamiltonian}
T=& \int d\svec x \bar\psi(\svec x)\big(-i\svec\gamma\cdot\svec\nabla + M\big)\psi(\svec x), \\
V_\phi=& \frac{1}{2} \int d\svec x d\svec x' \bar\psi(\svec x)\bar\psi(\svec x') \Gamma_\phi D_\phi  \psi(\svec x') \psi(\svec x) \label{eq:Pot}.
 \end{align}
In the potential energy terms $V_\phi$, $\phi$ represents the $\sigma$-S, $\omega$-V, $\rho$-V, $\rho$-T, $\rho$-vector-tensor ($\rho$-VT), $\pi$-PV and photon vector ($A$-V) couplings, and $\Gamma_{\phi}$ and $D_{\phi}$ correspond to the interaction vertex and the propagators, respectively. For further details, please refer to Ref. \cite{Geng2020PRC101.064302}.

Restricted on the level of the mean field approach, the no-sea approximation is introduced as usual, which is amount to neglect the contributions from the negative energy states. Thus, the nucleon field operator $\psi$ in the Hamiltonian can be quantized as,
\begin{align}\label{eq:quantize}
  \psi(x) = & \sum_i \psi_i(\svec x) e^{-i\varepsilon_i t} c_i,
\end{align}
where $c_i$ and $c_i^\dag$ are the annihilation and creation operators defined by the positive energy solutions of Dirac equation, $\varepsilon_i$ is the single-particle energy ($\varepsilon_i>0$), and $\psi_i(\svec x)$ is the wave function of state $i$.

Subsequently, the energy functional $E$ can be obtained from the expectation of the Hamiltonian with respect to the Hartree-Fock ground state $\lrlc{\Phi_0}$, namely $E = \lrcl{\Phi_0} H\lrlc{\Phi_0}$ \cite{Bouyssy1987PRC36.380, Geng2022PRC105.034329}. For the two-body interaction $V_\phi$, the expectation leads to two types of contributions, the direct Hartree and the exchange Fock terms. If only the Hartree terms are taken into account, this leads to the so-called relativistic mean field (RMF) theory. If both Hartree and Fock terms are considered, it gives the RHF theory.

Following the variational principle, one may derive the integro-differential Dirac equation from the energy functional $E$,
 \begin{align}\label{eq:Dirac-x}
\int d\svec x' h(\svec x, \svec x') \psi_i(\svec x') =& \varepsilon_i \psi_i(\svec x),
 \end{align}
where the single-particle Hamiltonian $h$ consists of three parts, the kinetic term $h^{\text{kin}}$, the local potential term $h^D$, and the non-local terms $h^E$ contributed by the Fock terms,
 \begin{align}
h^{\text{kin}} =& \lrs{\svec\alpha\cdot\svec p + \gamma^0 M }\delta(\svec x-\svec x'),\\[0.25em]
h^D =& \lrs{\Sigma_T(\svec x)\gamma_5 + \Sigma_0(\svec x) + \gamma^0\Sigma_S(\svec x)}\delta(\svec x-\svec x'),\label{eq:hD}\\[0.25em]
h^E =& \begin{pmatrix}Y_G(\svec x, \svec x') &Y_F(\svec x, \svec x')\\[0.5em]
X_G(\svec x, \svec x')&X_F(\svec x, \svec x')\end{pmatrix}.\label{eq:hE}
 \end{align}
In the above expressions, $h^D$ contains the local scalar self-energy $\Sigma_S$, the time component $\Sigma_0$ of the vector self-energy, and the tensor self-energy $\Sigma_T$, and $X_G$, $X_F$, $Y_G$ and $Y_F$ in $h^E$ are the non-local mean fields contributed by the Fock terms.

In the popular RHF Lagrangians PKO$i$ ($i=1,2,3)$ and PKA1, the density dependencies are introduced for the meson-nucleon coupling strengths $g_\phi$ ($\phi = \sigma, \omega^\mu, \ivec\rho^\mu$) and $f_{\phi'}$ ($\phi' = \ivec\rho^\mu, \ivec\pi$), which are taken as functions of nucleon density $\rho_b=\bar\psi\gamma^0\psi$ \cite{Long2006PLB640.150, Long2007PRC76.034314},
\begin{subequations}\label{eq:density-dependence}
\begin{align}
  g_\phi(\rho_b) = & g_\phi (\rho_0) F_\phi(\xi), & \text{for } \phi =& \sigma, \omega,\\
  g_{\phi'}(\rho_b) = & g_{\phi'}(0) e^{-a_{\phi'}\xi}, & \text{for } g_{\phi'} = & g_\rho, f_\rho, f_\pi,
\end{align}
where $\xi = \rho_b/\rho_0$, with $\rho_0$ being the saturation density, and the function $F_\phi$ reads as,
\begin{align}
  F_\phi(\xi) = & a_\phi \frac{1+b_\phi(\xi + d_\phi)^2}{1+c_\phi(\xi + d_\phi)^2}.
\end{align}
\end{subequations}
The density-dependent parameters $a$, $b$, $c$, $d$ in the above expressions, together with the coupling strengths and the masses of mesons, define an effective RHF Lagrangian for applications \cite{Long2006PLB640.150, Long2007PRC76.034314, Long2008EPL82.12001}. It is noteworthy that the density dependencies of the coupling strengths lead to an additional contribution to the self-energy $\Sigma_0$, i.e., the rearrangement terms $\Sigma_R$ \cite{Long2006PLB640.150, Long2007PRC76.034314, Long2010PRC81.024308}, which should be taken into account for promising the energy-momentum conservation \cite{Typel1999NPA656.331}.

\subsection{RHF energy functional for octupole deformed nuclei with the DWS base}

In this work, we restrict ourselves under the axial symmetry for the deformed nuclei with quadruple and octupole deformations. It shall be noted that due to the reflection asymmetry the parity does not remain as a good quantum number any more. Thus for the axially deformed nuclei, the complete set of good quantum numbers of the single-particle states are denoted as $(\nu m)$, where $m$ is the projection of total angular momentum and $\nu$ is the index of the orbits in a given $m$-block. In the following, we solve the integro-differential equation (\ref{eq:Dirac-x}) and the RHF field, by expanding the single-particle wave functions on the spherical DWS basis.

In expanding the wave functions, both positive and negative energy states in the spherical DWS basis shall be considered. Such treatment does not conflict with the no-sea approximation, which is considered in quantizing the nucleon field operator $\psi$ [see Eq. (\ref{eq:quantize})]. It is noteworthy that the no-sea approximation is amount to neglecting the contributions from the Dirac sea, e.g., to nucleon densities. Conversely, considering the negative energy states of the spherical DWS basis is demanded by the mathematical completeness. In terms of the DWS basis states, the expansion of the wave function reads as,
\begin{align}\label{eq:expansionS}
  \psi_{\nu m}(\svec x) = & \sum_a C_{a,i} \psi_{am}(\svec x) = \sum_{\kappa} \psi_{\nu \kappa m}(\svec x),
\end{align}
where the index $a$ represents the set of quantum numbers $a =(n\kappa)$, and the expansion coefficient $C_{a,i}$ is restricted as a real number. For giving compact expressions, $\psi_{ \nu \kappa m}$ is introduced as,
\begin{align}
  \psi_{\nu \kappa m} = & \sum_n C_{n\kappa, i} \psi_{n\kappa m} = \frac{1}{r} \begin{pmatrix} \cals G_{i\kappa} \Omega_{\kappa m} \\[0.5em] i\cals F_{i\kappa} \Omega_{-\kappa m} \end{pmatrix},\label{eq:expansionS1}
\end{align}
with $\cals G_{i\kappa} = \sum_n C_{n\kappa,i} G_{n\kappa}$ and $\cals F_{i\kappa} = \sum_n C_{n\kappa,i} F_{n\kappa}$, and $\Omega_{\kappa m}$ (also referred as $\Omega_{j m}^l$) is the spherical spinor. It shall be noted that in the expansion (\ref{eq:expansionS}) the expanded state $\lrlc{\nu m}$ and the DWS state $\lrlc{n\kappa m}$ will not share the same parity, due to the reflection asymmetry.

For the propagator in Equation (\ref{eq:Pot}), it can be expressed in the spherical coordinate space ($r,\vartheta,\varphi$) as \cite{Bouyssy1987PRC36.380, Varshalovich1988.165},
\begin{align}\label{eq:expansionD1}
  D_\phi  = & \sum_{\lambda_y\mu_y}(-1)^{\mu_y} R_{\lambda_y\lambda_y}^\phi(r,r') Y_{\lambda_y\mu_y}(\svec\Omega) Y_{\lambda_y-\mu_y} (\svec\Omega'),
\end{align}
where $\svec\Omega=(\vartheta,\varphi)$, the index $\lambda_y$ represents the expansion terms of the Yukawa propagator, and $R_{\lambda_y\lambda_y}$ contains the modified Bessel functions $I$ and $K$ as,
\begin{align}
  R_{\lambda_y\lambda_y}^\phi = & \frac{1}{2\pi}\frac{1}{\sqrt{rr'}} I_{\lambda_y+1/2}(m_\phi r_<) K_{\lambda_y+1/2}(m_\phi r_>),\\
  R_{\lambda_y\lambda_y}^A    = & \frac{1}{2\pi}\frac{1}{2\lambda_y+1} \frac{r_<^{\lambda_y}}{r_>^{\lambda_y+1}}.
\end{align}

Due to the density dependence, the coupling strengths become functions of the coordinate variable, which can be expanded as,
\begin{align}\label{eq:ExpansionP}
  g_\phi(\rho_b) = & \sqrt{2\pi}\sum_{\lambda_p} g_{\phi}^{\lambda_p}(r) Y_{\lambda_p0}(\vartheta,\varphi),
\end{align}
where $g_\phi$ represents the coupling strengths $g_\sigma$, $g_\omega$, $g_\rho$ and $f_\pi$. In contrast to the D-RHF model with only quadruple deformation,  $\lambda_p$ does not remain as even integer any more due to the reflection asymmetry.

In this work, we concentrate on the RHF Lagrangian PKO$i$ ($i=1,2,3$) \cite{Long2006PLB640.150, Long2008EPL82.12001}. For PKO2, that share the same degrees of freedom as the popular RMF models, it contains the $\sigma$-S, $\omega$-V, $\rho$-V and $A$-V couplings. In addition to that, PKO1 and PKO3 take the $\pi$-PV coupling into account. Thus, only the local self-energies $\Sigma_S$ and $\Sigma_0$ remain in $h^D$ (\ref{eq:hD}). Under the RHF scheme, the energy function can be expressed as,
\begin{align}
  E = & E^{\text{kin}} + \sum_\phi \big( E_\phi^D + E_\phi^E\big),
\end{align}
where $E^{\text{kin}} = \lrcl{\Phi_0} T\lrlc{\Phi_0}$, and $E_\phi^D$ and $E_\phi^E$ are the Hartree and Fock terms of the two-body potential energies $E_\phi = \lrcl{\Phi_0}V_\phi\lrlc{\Phi_0}$ with $\phi = \sigma$-S, $\omega$-V, $\rho$-V, $\pi$-PV and $A$-V.

Specifically, applying the expansion (\ref{eq:expansionS}), the kinetic energy functional $E^{\text{kin}}$ can be derived as,
\begin{align}
  E^{\text{kin}} = \sum_iv_i^2 \sum_{\kappa} &\int dr \Big\{\cals F_{i\kappa} \Big[\frac{d\cals G_{i\kappa}}{dr} + \frac{\kappa}{r} \cals G_{i\kappa} - M\cals F_{i\kappa}\Big] \nonumber\\
  & - \cals G_{i\kappa} \Big[\frac{d\cals F_{i\kappa}}{dr} - \frac{\kappa}{r} \cals F_{i\kappa}- M \cals G_{i\kappa}\Big]\Big\},
\end{align}
with $v_i^2$ ($ \in[0,2]$) being the occupations of the orbit $i$.

For giving a compact expression for the potential energies $E_\phi^D$, the local scalar and baryon densities $\rho_s$ and $\rho_b$ are expressed as,
\begin{subequations}\label{eq:densities}
\begin{align}
  \rho_s = & \sum_i \bar\psi_i(\svec x) \psi_i(\svec x) = \sum_{\lambda_d} \rho_s^{\lambda_d} (r) P_{\lambda_d}(\cos\vartheta),\\
  \rho_b = & \sum_i \bar\psi_i(\svec x)\gamma^0 \psi_i(\svec x) = \sum_{\lambda_d} \rho_b^{\lambda_d} (r) P_{\lambda_d}(\cos\vartheta),
\end{align}
\end{subequations}
where $P_{\lambda_d}$ is the Lengdre polynomial, and the cutoff of $\lambda_d$ is determined naturally by the truncation of the DWS basis in the expansion (\ref{eq:expansionS}). The radial part of the densities $\rho^{\lambda_d}(r)$ can be obtained as,
\begin{align}
  \rho^{\lambda_d}(r) = & \sum_i v_i^2 \sum_{\kappa\kappa'} (-1)^{m+\ff2} \scr D_{\kappa m,\kappa' m}^{\lambda_d 0} \nonumber\\ &\hspace{2em}\times\Big[\frac{\cals G_{i\kappa}(r) \cals G_{i\kappa'}(r)}{2\pi r^2} \pm \frac{\cals F_{i\kappa}(r) \cals F_{i\kappa'}(r)}{2\pi r^2}\Big],\label{eq:rhob-l}
\end{align}
where $\pm$ in squared brackets corresponds to the baryonic and scalar densities, respectively, and the symbol $\scr D$ is given in Eq. (\ref{eq:Dsymbol}). Thus, the energy functional $E_\phi^D$ can be written in a compact form as,
\begin{subequations}\label{eq:ED}
\begin{align}
  E_\sigs^D = & \frac{2\pi}{2}\sum_{\lambda_d} \int r^2 dr \Sigma_{S,\sigs}^{\lambda_d}(r)\rho_s^{\lambda_d}(r),\\
  E_\omev^D = & \frac{2\pi}{2}\sum_{\lambda_d} \int r^2 dr \Sigma_{0,\omev}^{\lambda_d}(r)\rho_b^{\lambda_d}(r),
\end{align}
\end{subequations}
where the expansion term of self-energies $\Sigma_{S,\sigs}^{\lambda_d}$ and $\Sigma_{0,\omev}^{\lambda_d}$ are given in Eq. (\ref{eq:sv-self}). For the $\rhov$ and $\couv$ couplings, the expressions can be obtained similarly as the $\omev$ ones, and the details can be found in Appendix \ref{app:Hartree}. In the above expressions, it shall be noted that $\lambda_d$ can be both even and odd numbers, due to the reflection asymmetry.

For the potential energy $E_\phi^E$ from the Fock terms, despite the complexity itself, the contribution can be written in a unified form as,
\begin{align}
  E_{\phi}^E =& \ff2\int drdr' \sum_{i} v_i^2\sum_{\kappa_1\kappa_2}\begin{pmatrix} \cals G_{i\kappa_1} & \cals F_{i\kappa_1} \end{pmatrix}_r \nonumber \\ &\hspace{2em}\times\begin{pmatrix} Y_{G, m}^{\kappa_1,\kappa_2; \phi} & Y_{F, m}^{\kappa_1,\kappa_2; \phi}  \\[0.5em] X_{G, m}^{\kappa_1,\kappa_2; \phi} & X_{F, m}^{\kappa_1,\kappa_2; \phi} \end{pmatrix}_{r,r'}\begin{pmatrix} \cals G_{i\kappa_2} \\[0.5em] \cals F_{i\kappa_2} \end{pmatrix}_{r'},\label{eq:ENE-F}
\end{align}
where $\phi$ represents the $\sigma$-S coupling, the time and space components of the vector ($\omega$-V, $\rho$-V and $A$-V) ones, and the $\pi$-PV one. The non-local densities, appearing in the non-local self energies $Y_G, Y_F, X_G$ and $X_F$ as the source terms, are denoted by symbol $\cals R$,
\begin{align}
  \cals R_{\kappa\kappa',  m}^{++}(r, r') = &  \sum_{\nu}v_i^2  \cals G_{\nu m,\kappa}(r) \cals G_{\nu m,\kappa'}(r'),\\
  \cals R_{\kappa\kappa',  m}^{+-}(r, r') = &  \sum_{\nu}v_i^2  \cals G_{\nu m,\kappa}(r) \cals F_{\nu m,\kappa'}(r'),\\
  \cals R_{\kappa\kappa',  m}^{-+}(r, r') = &  \sum_{\nu}v_i^2  \cals F_{\nu m,\kappa}(r) \cals G_{\nu m,\kappa'}(r'),\\
  \cals R_{\kappa\kappa',  m}^{--}(r, r') = &  \sum_{\nu}v_i^2  \cals F_{\nu m,\kappa}(r) \cals F_{\nu m,\kappa'}(r'),
\end{align}
where $\kappa$ and $\kappa'$ denote the $\kappa$-blocks of the spherical DWS basis in expanding the wave functions of a given $m$-block. The details of the non-local self-energies $Y_G, Y_F, X_G$ and $X_F$ are provided in Appendix \ref{app:Fock}.

\subsection{Dirac equations with the spherical DWS basis}
Since the Dirac spinors are expanded on the spherical DWS basis, the variation of the energy functional $E$ with respect to the expansion coefficient $C_{a,i}$ leads to a series of eigenvalue equations as,
\begin{align}\label{eq:eigen}
  H_{m} \widehat C_{i} = & \varepsilon_i \widehat C_{i}.
\end{align}
By diagonalizing the Hamiltonian $H_{m}$, the eigenvalue, i.e., the single-particle energy $\varepsilon_i$ can be obtained, as well as the eigenvector $\widehat C_i$ that represents a set of the expansion coefficients of orbit $i=(\nu m)$.

Similar as the single-particle Hamiltonian in the integro-differential Dirac equation (\ref{eq:Dirac-x}), the matrix $H_{m}$ for a given $m$-block in Eq. (\ref{eq:eigen}) consists of three parts, i.e., the kinetic $H^{\text{kin}}$, local $H^D$ and non-local $H^E$ terms,
\begin{align}
  H = & H^{\text{kin}} + H^{D} + H^E,
\end{align}
where the subscript $(m)$ is omitted. In terms of the spherical DWS basis, $H^{\text{kin}}$ can be derived as,
\begin{align}
  H^{\text{kin}}_{aa'} = \delta_{kk'}\int& dr  \Big\{-G_{a}\Big[\frac{dF_{a'}}{dr} - \frac{\kappa}{r}F_{a'} -M G_{a'}\Big] \nonumber \\
  & + F_{a} \Big[\frac{dG_{a'}}{dr} + \frac{\kappa}{r}G_{a'} -M F_{a'}\Big] \Big\}.
\end{align}
For the local term $H_{aa'}^D$, which contains the Hartree mean fields and rearrangement terms, the matrix elements $H_{aa'}^D$ read as,
\begin{align}
  H^D_{aa'} = & \int dr\sum_{\lambda_d} (-1)^{m+1/2}\scr D_{\kappa  m,\kappa' m}^{\lambda_d0}\begin{pmatrix} G_{a }  & F_{a } \end{pmatrix}\nonumber\\ &\hspace{1em} \times\Big[\gamma_0\Sigma_{S,\sigs}^{\lambda_d} + \sum_{\phi'}\Sigma_{0,\phi'}^{\lambda_d} + \Sigma_R^{\lambda_d}\Big]\begin{pmatrix} G_{a'}  \\[0.5em] F_{a'} \end{pmatrix},\label{eq:HDaa'}
\end{align}
where $\phi'$ represents the $\omev$, $\rhov$ and $\couv$ couplings, and the details of the rearrangement terms $\Sigma_R^{\lambda_d}$ are given in Appendix \ref{app:DDMNC}. For the non-local term $H^E$, the matrix elements can be uniformly expressed as,
\begin{align}
H^E_{aa'} = & \sum_\phi\int d r dr' \begin{pmatrix}  G_{a } & F_{a } \end{pmatrix}_r \nonumber\\ & \hspace{2em}\times \begin{pmatrix} Y_{G,\pi m}^{\kappa \kappa',\phi} & Y_{F,\pi m}^{\kappa \kappa',\phi}  \\[0.5em] X_{G,\pi m}^{\kappa \kappa ,\phi}  & X_{F,\pi m}^{\kappa \kappa',\phi}  \end{pmatrix}_{(r, r')} \begin{pmatrix} G_{a'} \\[0.5em] F_{a'} \end{pmatrix}_{r'},
\end{align}
where $\phi$ represents the coupling channels $\sigma$-S, $\omega$-V, $\rho$-V, $A$-V and $\pi$-PV.

\subsection{Pairing correlations: BCS method with Gogny type pairing force}

For the open-shell nuclei, the pairing correlations are considered within the BCS scheme, and the finite-range Gogny force D1S \cite{Berger1984NPA428.23} is adopted as pairing force, regarding the advantage of the natural convergence with respect to the configuration space. The Gogny-type pairing force is of the following general form as,
\begin{align}
  V^{pp}(\svec r, \svec r') = & \sum_{\chi=1,2} e^{{(\svec r-\svec r')^2}/{\mu_\chi^2}} \big(W_\chi + B_\chi P^\sigma\nonumber\\ & \hspace{6em}  - H_\chi P^\tau - M_\chi P^\sigma P^\tau\big),
\end{align}
where $\mu_\chi, W_\chi, B_\chi, H_\chi$ and $M_\chi$ ($\chi=1,2$) are parameters of the Gogny force, and $P^\sigma$ and $P^\tau$ are the spin and isospin exchange operators, respectively. Similar as the expansion of the propagator (\ref{eq:expansionD1}), the coordinate part of $V^{pp}$ expanded in spherical coordinate ($r,\vartheta,\varphi$) can be expressed as,
\begin{align}
  V(\svec r, \svec r') = & 2\pi\sum_{\chi=1,2} \big(A_\chi + D_\chi P^\sigma\big) \sum_{\lambda=0}^\infty V_{\chi,\lambda}(r,r') \nonumber\\
   & \hspace{5em} \times\sum_{\mu} Y_{\lambda\mu}(\vartheta,\varphi) Y_{\lambda\mu}^*(\vartheta',\varphi'),\label{eq:Gogny-EP}
\end{align}
where $A_\chi = W_\chi - H_\chi P^\tau$ and $D_\chi = B_\chi - M_\chi P^\tau$, and the radial part $ V_{\chi,\lambda}$ reads as,
\begin{align}
  V_{\chi,\lambda}(r,r') = e^{-(r^2+r'^2)/\mu_\chi^2} \sqrt{2\pi \frac{\mu_\chi^2}{2rr'}} I_{\lambda+1/2}\Big(\frac{2rr'}{\mu_\chi^2}\Big).
\end{align}

For given orbits $i$ and $i'$ of an axially deformed nucleus with quadruple and octupole deformation, the pairing interaction matrix element $V_{ii'}^{pp}$ can be derived as,
\begin{align}
  V_{ii'}^{pp} = & \int d r d r'\sum_{\kappa\kappa'}\begin{pmatrix}K_{i\kappa, i'\kappa'}^G & K_{i\kappa, i'\kappa'}^F \end{pmatrix}_r \nonumber \\ & \hspace{2em}\times\begin{pmatrix}\bar Y^G_{\kappa\kappa'} & \bar Y^F_{\kappa\kappa'} \\[0.5em] \bar X^G_{\kappa\kappa'} & \bar X^F_{\kappa\kappa'}\end{pmatrix}_{(r,r')} \begin{pmatrix} K_{i\kappa, i'\kappa'}^G \\[0.5em] K_{i\kappa, i'\kappa'}^F  \end{pmatrix}_{r'},\label{eq:vpp}
\end{align}
where $K_{i\kappa, i'\kappa'}^G$ and $K_{i\kappa, i'\kappa'}^F$ read as,
\begin{align}
  K_{i\kappa, i'\kappa'}^G(r) = & \cals G_{i\kappa} \cals G_{i'\kappa'},   &  K_{i\kappa, i'\kappa'}^F(r) = & \cals F_{i\kappa} \cals F_{i'\kappa'} .
\end{align}
It is noteworthy that the pairing potentials $\bar X^G$, $\bar X^F$, $\bar Y^G$ and $\bar Y^F$ are non-local due to the finite-range nature of the Gogny force. For further details, please refer to Appendix \ref{app:pair}.

\section{Results and Discussions}\label{sec:result and discussions}

Aiming at future applications of the newly developed OD-RHF model, we firstly test the convergence with respect to the space truncation of the spherical DWS basis, using $^{144}$Ba as an example, which is confirmed to have octupole deformation \cite{Bucher2016PRL116}. In determining the spherical DWS basis, the spherical Dirac equations are solved by setting the spherical box size to $20$ fm with a radial mesh step of $0.1$ fm. Using the RHF Lagrangians PKO$i$ ($i=1,2,3$) \cite{Long2006PLB640.150, Long2008EPL82.12001} and RMF one DD-ME2 \cite{Lalazissis2005PRC71.024312}, we then analyze the structural properties of $^{144}$Ba, by focusing on the quadruple and octupole deformation effects, and the role of the tensor force component carried by the $\pi$-PV coupling.

\subsection{Space truncations and Convergence check}

In the OD-RHF calculations, two independent truncations shall be tested carefully, i.e., the cutoff on the quantities $(n\kappa)$ of the spherical DWS basis [Eq. (\ref{eq:expansionS})] and the expansion term $\lambda_p$ of the density-dependent coupling strengths [Eq. (\ref{eq:ExpansionP})]. For the propagators (\ref{eq:expansionD1}) and the Gogny force (\ref{eq:Gogny-EP}), the expansions are truncated naturally by selected spherical DWS states and $\lambda_p$, as detailed in Eqs. (\ref{eq:Theta}) and (\ref{eq:scrXS}). For the sake of clarify, we briefly introduce the space truncations, which are similar to the D-RHFB calculations \cite{Geng2022PRC105.034329}, except for some special considerations related to the parity. Due to the reflection asymmetry considered in this work, the terms with both odd and even $\lambda_p$ should be taken into account in expanding the coupling strengths (\ref{eq:ExpansionP}). This differs from the D-RHF and D-RHFB models \cite{Geng2020PRC101.064302, Geng2022PRC105.034329}, in which only even $\lambda_p$ terms are considered because of the reflection symmetry. In general, it is sufficient to set $\lambda_p = 0, 1, 2, \cdots, 8$ for the majority of nuclei.

Practically, the maximum value of $m$, named as $m_{\max}$, depends on the specific nucleus under consideration. The number of $\kappa$-blocks included in the expansion of the Dirac spinors with $m_{\max}$ is given by $K_{m}$. In order to maintain consistency, the $m_{\max}$ and $K_m$ together determine the maximum value of $|\kappa|$, i.e., $\kappa_{\max} = m_{\max} + K_{m} - 1/2$. Specifically, for an arbitrary Dirac spinor $\psi_{\nu m}$, the $\kappa$-quantities in the expansion (\ref{eq:expansionS}) read as $\kappa = \pm (m + 1/2), \pm(m + 3/2), \cdots, \pm\kappa_{\max}$, including both even- and odd-parity states due to the reflection asymmetry. In general, the $m_{\text{max}}$ value is determined with reference to the conventional shell-model picture of the single-particle spectrum. For the nucleus $^{144}_{\ 88}$Ba$_{56}$, the $m_{\max}$ value is determined as $15/2$ for both neutron and proton orbits, in accordance with the traditional magic shell $N/Z = 184$. In addition, the value of $K_m$ is chosen to be $K_{m} = 4$, and both of these values are tested to be accurate enough. For the cases of large deformation, some large $j$-orbits may penetrate the well-known major shells. Therefore, it is necessary to perform careful test calculations, for instance by enlarging the $m_{\max}$ value.

In a manner analogous to the D-RHFB model \cite{Geng2022PRC105.034329}, the maximum values of the principal number $n$ for each $\kappa$-block in the expansion (\ref{eq:expansionS}) are determined by the energy cutoff $E_{\pm}^C$. This cutoff is defined by the single-particle energy $\varepsilon$ of the spherical DWS state, with the sign $+$ or $-$ indicating the positive or negative energy states, respectively. Specifically, the states with positive (negative) energies, that $\varepsilon-M < E_{+}^C$ $(\varepsilon + M > E_{-}^C)$, are considered in the expansion (\ref{eq:expansionS}). It is important to exercise caution when testing the energy cutoff $E_\pm^C$.

\begin{figure}[ht]
\begin{center}
\includegraphics[width=0.96\linewidth]{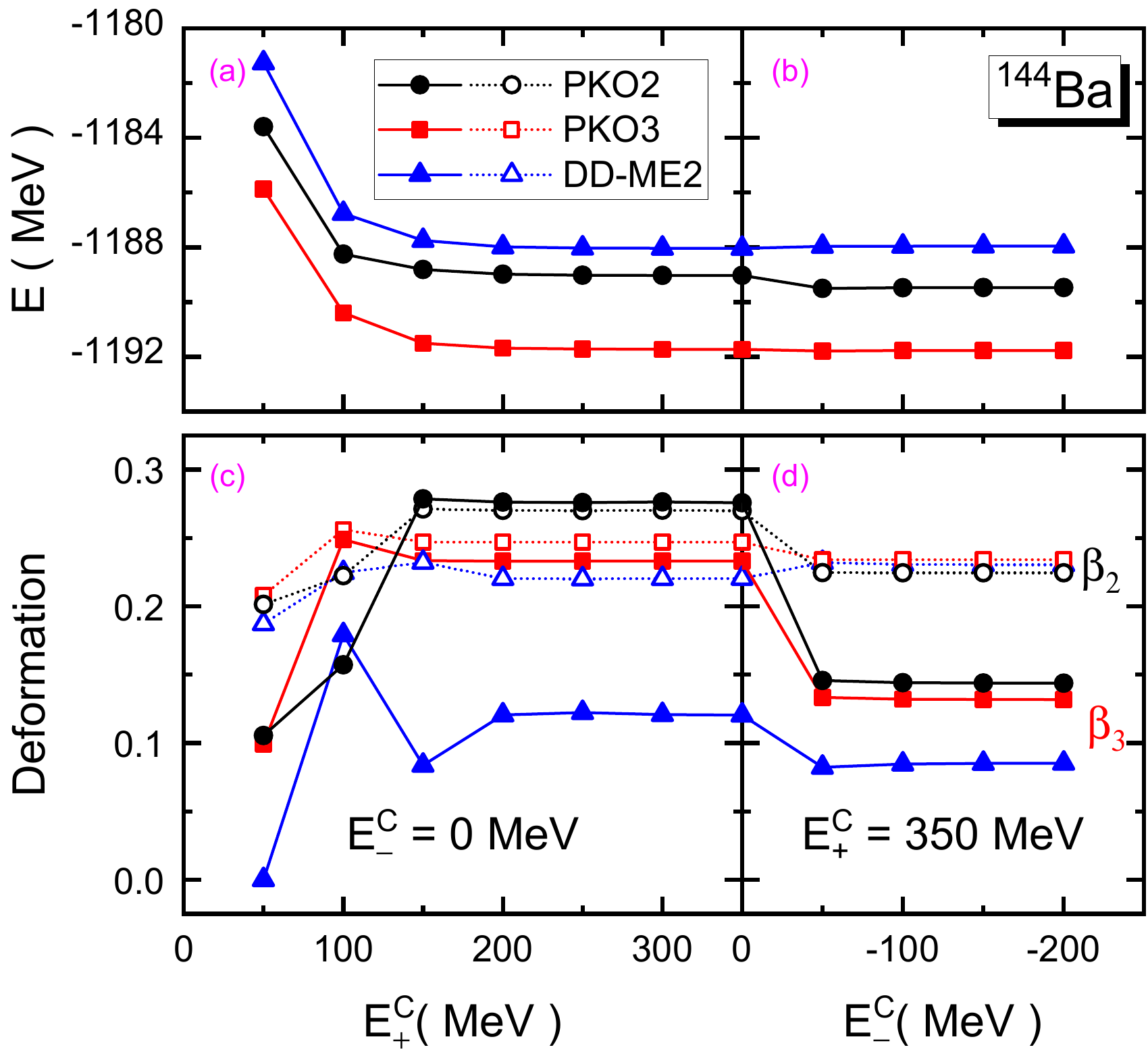}
\caption{(Color Online) Total energy $E$ (MeV) [plots (a) and (b)] and deformations $(\beta_2, \beta_3)$ [plots (c) and (d)] for $^{144}$Ba with respect to the positive ($+$) and negative ($-$) energy cutoff $E_{\pm}^C$ (MeV) in expanding the Dirac spinors $\psi_{\nu m}$. The results are calculated by DD-ME2, PKO2 and PKO3.} \label{Fig:Convergence}
\end{center}
\end{figure}

Figures \ref{Fig:Convergence} (a-b) and (c-d) present the tests of the energy cutoff $E_\pm^C$ for the total energy $E$ (MeV) and the deformation $(\beta_2,\beta_3)$ of $^{144}$Ba, respectively. The results were obtained by employing the RHF Lagrangians PKO2 and PKO3, and the RMF one DD-ME2. Plots (a) and (c) illustrate the convergence with respect to $E_{+}^C$, where $E_{-}^C$ is fixed as $0$ MeV, and plots (b) and (d) demonstrate the convergence with respect to $E_{-}^C$, in which $E_{+}^C$ is fixed as $350$ MeV. It is evident that both the total energy $E$ [plot (a)] and the deformation $(\beta_2, \beta_3)$ [plot (c)] exhibit a converged tendency when $E_{+}^C > 200$ MeV for $^{144}$Ba. When considering more negative energy states in the expansion (\ref{eq:expansionS}), the total energy $E$ remains almost unchanged for DD-ME2 and PKO3, while PKO2 shows slight but visual changes, as illustrated in Fig. \ref{Fig:Convergence} (b).

In contrast to the total energy $E$, the deformations $(\beta_2, \beta_3)$ appear to be more sensitive to the negative energy cutoff $E_-^C$. As illustrated in Figs. \ref{Fig:Convergence} (c, d), the changes in quadruple deformation $\beta_2$ are distinct in all the results given by selected Lagrangians. It is surprising to see that the octupole deformation $\beta_3$ exhibits notable alterations when considering more spherical DWS states with negative energy, particularly for PKO2 and PKO3. Despite the obvious changes, both deformations $\beta_2$ and $\beta_3$ demonstrate rapid convergence when the $E_{-}^C$-value is further augmented.

\subsection{Significance of the negative energy states of the DWS basis for $^{144}$Ba}

\begin{figure}[ht]
\begin{center}
\includegraphics[width=0.96\linewidth]{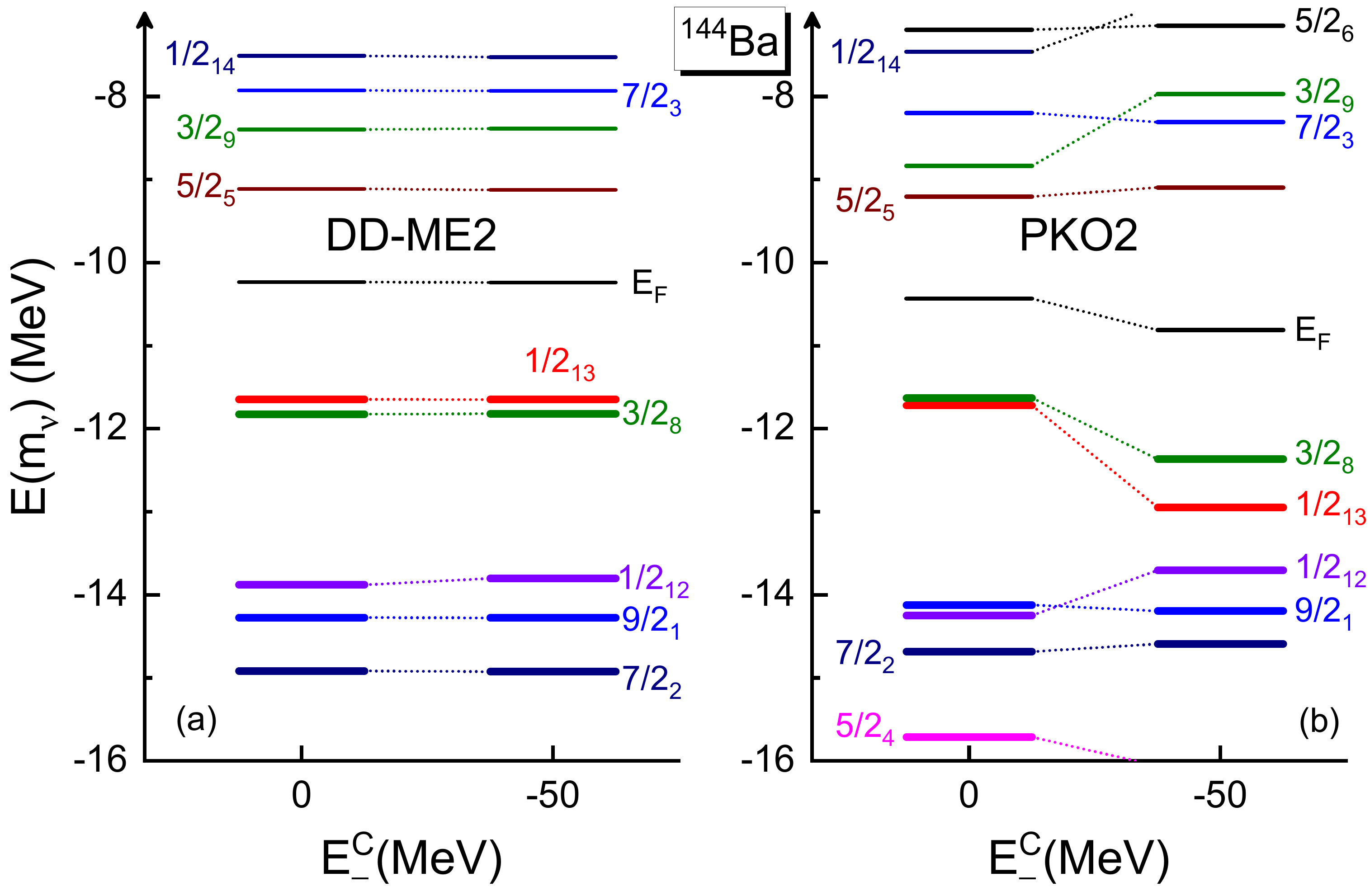}
\caption{ Proton single-particle spectra of $^{144}$Ba given by DD-ME2 [plot (a)] and PKO2 [plot (b)] with negative energy cutoff $E_{-}^C=0$ MeV and $-50$ MeV, in which the positive energy cutoff is fixed as $E_{+}^C = 350$ MeV. The ultra thick bars represent the occupation probabilities of the orbits $m_\nu$ and $E_F$ denotes the Fermi levels.}\label{Fig:Lev-CUT}
\end{center}
\end{figure}

In order to understand the remarkable alterations in the deformations $(\beta_2, \beta_3)$, particular for the octupole one $\beta_3$, as illustrated in Figs. \ref{Fig:Convergence} (c, d), Fig. \ref{Fig:Lev-CUT} shows the proton single-particle spectra of $^{144}$Ba calculated by DD-ME2 and PKO2 with $E_-^C=0$ and $-50$ MeV. In Fig. \ref{Fig:Lev-CUT}, the Fermi levels are denoted by $E_{\text{F}}$, and $m_\nu$ for the single-particle orbits. For the RHF Lagrangians PKO3, the detailed results are not shown since they provide a similar description to those of PKO2.

Consistent with the small alterations of the deformations from Figs. \ref{Fig:Convergence} (c) to \ref{Fig:Convergence} (d), the proton single-particle spectra yielded by DD-ME2 remain almost unaltered, when the negative energy cutoff $E_-^C$ is varied from zero to $-50$ MeV, see Fig. \ref{Fig:Lev-CUT} (a). Conversely, the results obtained by PKO2 exhibit remarkable alterations with respect to the $E_-^C$ value, as illustrated in Fig. \ref{Fig:Lev-CUT} (b). In fact, the neutron single-particle spectra, which are not shown here, exhibit similar systematics as the proton ones. That is, the results given by DD-ME2 remain almost unchanged, while the PKO2 results manifest rather distinct alterations, when the $E_-^C$ value changes from zero to $-50$ MeV.

In this work, the deformed single-particle orbits are expanded on the spherical DWS basis, which provides us insight into the microscopic properties of the octupole nucleus $^{144}$Ba. In order to comprehend the obvious alterations given by PKO2 in Fig. \ref{Fig:Lev-CUT} (b), Table \ref{tab:Ba144-ECUT} lists the main expansion proportions of the proton valence orbits $3/2_8$, $1/2_{13}$ of $^{144}$Ba. The results were extracted from the calculations of DD-ME2 and PKO2 with $E_-^C=0$ and $-50$ MeV. It is notable that a slight yet discernible change in the proportions of spherical components can be observed in the DD-ME results, when the $E_-^C$ value is altered from zero to $-50$ MeV. In contrast, the proportions of the main spherical components given by PKO2 undergo a significant shift, indicating a notable redistribution of the expansions of the selected orbits.

\begin{table}[htbp]
\caption{Proportions (in percentage) of the main expansion components of proton orbits $3/2_8$ and $1/2_{13}$, These results are calculated with DD-ME2 and PKO2 by selecting the negative energy cutoff $E_{-}^C$, respectively as 0 MeV and $-$50 MeV, with the positive energy cutoff $E_{+}^C = 350$ MeV. }   \label{tab:Ba144-ECUT}
\renewcommand{\arraystretch}{1.30}\setlength{\tabcolsep}{0.25em}
\begin{tabular}{c|c|c|ccc|cc}  \hline \hline
Proton               & orbits                      &$E_{-}^C$&$2d_{5/2}$          &$1g_{7/2}$&$1g_{9/2}$  & $1f_{5/2}$& $1h_{11/2}$        \\ \hline
\multirow{4}{*}{DD-ME2}& \multirow{2}{*}{$3/2_{8}$}&0        &   5.9\%            &65.5\%    &  5.4\%     & 2.4\%     & 11.5\%             \\
                     &                             &$-$50    &   6.6\%            &59.5\%    &  2.6\%     & 6.9\%     & 12.6\%             \\ \cline{2-8}
                     & \multirow{2}{*}{$1/2_{13}$} &0        &  21.8\%            &24.4\%    &  7.7\%     & 0.7\%     & 27.1\%             \\
                     &                             &$-$50    &  25.8\%            &20.1\%    &  2.6\%     & 2.1\%     & 30.1\%             \\ \hline
\multirow{4}{*}{PKO2}& \multirow{2}{*}{$3/2_{8}$}  &0        &   2.1\%            &43.8\%    & 25.3\%     &  4.0\%    & 2.4\%              \\
                     &                             &$-$50    &   6.5\%            &53.8\%    &  6.2\%     &  3.5\%    & {\bfseries 20.2\%} \\ \cline{2-8}
                     & \multirow{2}{*}{$1/2_{13}$} &0        &   2.9\%            &23.5\%    & 37.7\%     &  3.3\%    &  5.0\%             \\
                     &                             &$-$50    & {\bfseries 24.7\%} &  2.8\%   &  8.8\%     &  1.1\%    & {\bfseries 30.1\%} \\ \hline\hline
\end{tabular}
\end{table}

In order to ensure the completeness of the expansion, it is necessary to consider the negative energy states of the spherical DWS basis in expanding the single-particle wave functions, despite the expansion proportions being rather small. It is already illustrated in Figs. \ref{Fig:Convergence} and \ref{Fig:Lev-CUT} that the RHF calculations with PKO2 and PKO3 appear to be more sensitive to the negative energy cutoff $E_-^C$ than the RMF one. Because of the Fock terms, more two-body correlations are considered in the RHF models than the RMF one that contains only the Hartree terms. Upon the alteration of the $E_-^C$ value from $0$ to $-$50 MeV, the whole expansion of the single-particle orbits can be influenced, due to more negative energy states involved in the expansion. As demonstrated in Table \ref{tab:Ba144-ECUT}, the expansions on the spherical DWS basis described by PKO2 undergo more notable redistribution, in comparison to the result of DD-ME2. In particular, the expansion properties of the components $1h_{11/2}$ and $2d_{5/2}$ (emphasized in bold types), which are essential for the octupole deformation, are largely changed from $E_-^C=0$ to $-50$ MeV. Such changes are necessitated to promise the completeness, particularly for the exchange correlations introduced by the Fock terms, as demonstrated by Figs. \ref{Fig:Convergence}-\ref{Fig:Lev-CUT} and Table \ref{tab:Ba144-ECUT}. In comparison to the total energy $E$, the deformations ($\beta_2, \beta_3$) are more sensitive to the redistribution of the expansions when the $E_-^C$ changing from $0$ to $-50$ MeV, as deduced from Fig. \ref{Fig:Convergence}.

\subsection{Octupole deformation effects in $^{144}$Ba}

With given space truncations, namely $E_+^C = 350$ MeV, $E_-^C=-150$ MeV, $m_{\max}=15/2$ and $K_{m}=4$, we performed the OD-RHF calculations for the octupole nucleus $^{144}$Ba, using the RHF Lagrangians PKO$i$ ($i=1,2,3$) \cite{Long2006PLB640.150, Long2008EPL82.12001} and the RMF one DD-ME2 \cite{Lalazissis2005PRC71.024312}. Table \ref{Tab:Bulk} lists the total energy $E$ (MeV), the deformations $(\beta_2, \beta_3)$ and charge radius $r_c$ (fm) of $^{144}$Ba, for the ground state (GS), the quadruple local minimum (Quad) and the case with the spherical shape (Sph).

\begin{table}[htbp]
\caption{Total energy $E$ (MeV), quadruple and octupole deformations $(\beta_2, \beta_3)$, and charge radius $r_c$ (fm) of $^{144}$Ba, calculated by PKO$i$ ($i=1,2,3$) and DD-ME2 for the ground state (GS), the quadruple local minima (Quad) and the spherical shape (Sph). The experimental values read as $E=-1190.22$ MeV \cite{Wang2017CPC41.030003}, $(\beta_2, \beta_3)=(0.18, 0.17)$ \cite{Bucher2016PRL116} and $r_c = 4.9236(0.0112)$ fm \cite{Angeli2013ADNDT99}. The last column $\Delta E$ (MeV) is the energy differences between the GS and Quad cases, and the ones between the Quad and Sph cases.} \renewcommand{\arraystretch}{1.5}\setlength{\tabcolsep}{0.4em}\label{Tab:Bulk}
\begin{tabular}{c|c|ccc|c}  \hline\hline
                        &     &   $E$                &  $(\beta_2, \beta_3)$& $r_c$     & $\Delta E$ \\ \hline

\multirow{3}{*}{DD-ME2} &GS   &   $-$1187.95          &   (0.23, 0.09)        &  5.0340  & $-$1.51  \\
                        &Quad &   $-$1186.44          &   (0.20, 0.00)        &  5.0013  & $-$5.77  \\
                        &Sph  &   $-$1180.87          &   (0.00, 0.00)        &  4.9013  &          \\ \hline

\multirow{3}{*}{PKO2}   &GS   &   $-$1189.47          &   (0.22, 0.14)        &  5.0234  & $-$3.69  \\
                        &Quad &   $-$1185.78          &   (0.19, 0.00)        &  4.9971  & $-$3.28  \\
                        &Sph  &   $-$1182.50          &   (0.00, 0.00)        &  4.9828  &          \\ \hline

\multirow{3}{*}{PKO1}   &GS   &   $-$1192.54          &   (0.24, 0.14)        &  5.0525  & $-$3.15 \\
                        &Quad &   $-$1189.39          &   (0.21, 0.00)        &  5.0224  & $-$4.27 \\
                        &Sph  &   $-$1185.12          &   (0.00, 0.00)        &  5.0047  &         \\ \hline

\multirow{3}{*}{PKO3}   &GS   &   $-$1191.77          &   (0.23, 0.13)        &  5.0365  & $-$2.69 \\
                        &Quad &   $-$1189.08          &   (0.21, 0.00)        &  5.0066  & $-$4.61 \\
                        &Sph  &   $-$1184.47          &   (0.00, 0.00)        &  4.9878  &         \\ \hline \hline
\end{tabular}
\end{table}

As illustrated in Table \ref{Tab:Bulk}, $^{144}$Ba becomes deeper bound from the spherical shape to the case with axial quadruple deformation, and the total energy becomes more closely aligned with the experimental value \cite{Wang2017CPC41.030003}. Further considering the octupole deformation, all the selected Lagrangians yield the octupole GS for $^{144}$Ba, and PKO2 shows the best agreement with the experimental data, including the total energy $E$, the deformations $(\beta_2, \beta_3)$ and charge radius $r_c$. From the quadruple local minima to the octupole GS, it is commonly seen that the obtained quadruple deformations $\beta_2$ increase slightly. Thus, the $\Delta E$ values in the last column of Table \ref{Tab:Bulk} can be approximately regarded as the effects of the octupole and quadruple deformations, respectively. As deduced from the $\Delta E$ values between the octupole GS and Quad cases, the RHF Lagrangians PKO$i$ ($i=1,2,3$) present stronger octupole enhancement than the RMF one DD-ME2. This can be simply attributed to the effects of the Fock terms, namely the exchange correlations.

However, it is unexpected that PKO1 and PKO3, which contain the degree of freedom associated with the $\pi$-PV coupling, yield weaker octupole enhancements than PKO2. From PKO2 to PKO1 and further to PKO3, it seems that the $\pi$-coupling plays a role against the octupole deformation, regarding the fact that PKO2 does not contain the $\pi$-PV coupling and PKO3 carries stronger $\pi$-PV coupling than PKO1. In contrast, given the $\Delta E$ values from the Sph to Quad cases in Table \ref{Tab:Bulk}, stronger quadruple enhancement is obtained by PKO3 than PKO1, and PKO1 show stronger enhancement than PKO2. Qualitatively such systematics of the octupole and quadruple enhancements can be understood, as combined with the single-particle spectra of $^{144}$Ba.

\begin{figure}[ht]
\begin{center}
\includegraphics[height=0.27\textwidth]{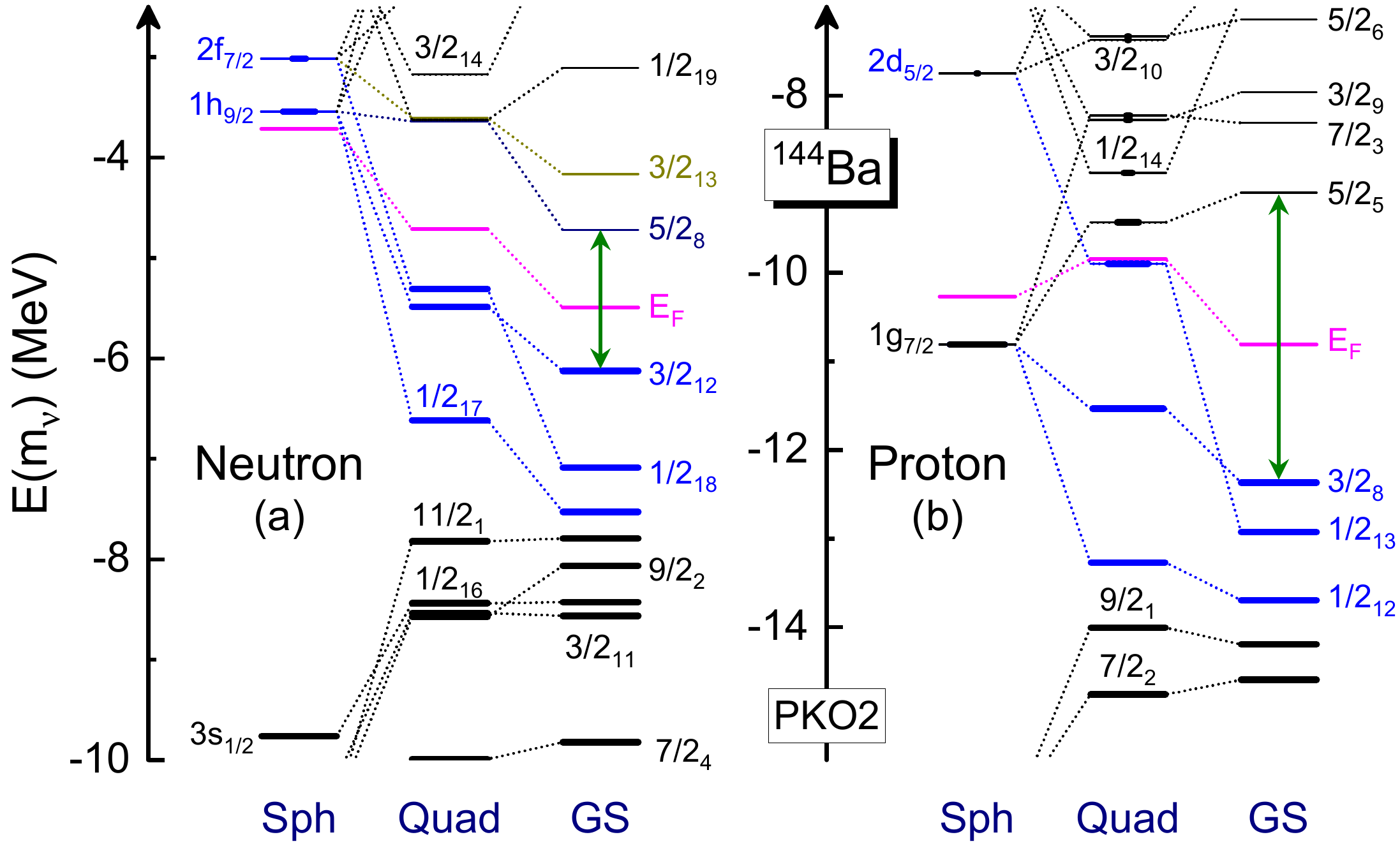}
\caption{Neutron (left) and proton (right) spectra of $^{144}$Ba at spherical shape (Sph) quadruple local minimum (Quad) and ground state (GS) given by PKO2. The ultrathick bars represent the occupation probabilities of the orbits $m_\nu$ and $E_F$ denotes the Fermi levels. }
\label{Fig:Lev-PKO2}
\end{center}
\end{figure}

Figures \ref{Fig:Lev-PKO2} (a) and (b) show the neutron and proton single-particle spectra of $^{144}$Ba given by PKO2, respectively. It shall be commented that the single-particle spectra given by PKO2 are not significantly different from the others, which are not shown here. As illustrated in Fig. \ref{Fig:Lev-PKO2}, from the quadruple minima to the octupole GS, both neutron and proton valence orbits (marked in blue color) become much deeper bound, and the induced large shell gaps (denoted by dashed arrows) stabilize the octupole deformation for the GS of $^{144}$Ba. In order to further understand the octupole effects, Table \ref{tab:Ba144-GS} shows the main expansion proportions of the neutron and proton valence orbits, in terms of the spherical DWS basis states. It is seen that the intrusions of both neutron $i1_{13/2}$ and proton $1h_{11/2}$ components play a key role in giving the octupole deformation, which is commonly supported by the selected models in this work.

Taking proton as an example, the couplings between the $2d$- and intruding $1h_{11/2}$-components, which fulfill the condition $\Delta l = 3$, are essential for given a stable octupole deformation. As illustrated in Fig. \ref{Fig:Lev-PKO2} (b) and the lower panel of Table \ref{tab:Ba144-GS}, the orbits $(3/2_8, 1/2_{12})$ and $1/2_{13}$ are dominated by the $1g_{7/2}$ and $2d_{5/2}$ components (in bold types) at the quadruple minima, respectively, and thus much more evident octupole deformation effects is observed for the $1/2_{13}$ orbit than the ones $(3/2_8, 1/2_{12})$. Similarly, combined Fig. \ref{Fig:Lev-PKO2} (a) and upper panel of Table \ref{tab:Ba144-GS}, it is also transparent for the systematics of neutron valence orbits, in which the octupole deformation effects can be well understood from the couplings between the $2f$- and intruding $1i1_{13/2}$-components.

Quantitatively, it is observed in Tables \ref{tab:Ba144-ECUT} and \ref{tab:Ba144-GS} that PKO2 presents larger $h_{11/2}$ proportion for proton orbit $3/2_8$ than DD-ME2, but similar $2d_{5/2}$ one. It is essential to enhance the effects of the octuple deformation. In fact, larger $i1_{13/2}$ proportions in neutron valence orbits are also obtained by the PKO2 calculations, in comparison to the DD-ME2 results which are not shown here. Thus, regarding the fact that the RHF Lagrangians PKO$i$ ($i=1,2,3$) present stronger octupole enhancement than the RMF one DD-ME2, as illustrated in Table \ref{Tab:Bulk}, the Fock terms are likely to enhance the couplings which are essential for the octuple deformation.

\begin{table}[htbp]
\caption{Proportions (in percentage) of the main expansion components of neutron orbits $3/2_{12}$, $1/2_{18}$ and $1/2_{17}$ (upper panel) and proton orbits $3/2_8$, $1/2_{13}$ and $1/2_{12}$(lower panel) of $^{144}$Ba. These results are calculated with the RHF Lagrangian PKO2. }   \label{tab:Ba144-GS}
\renewcommand{\arraystretch}{1.30}\setlength{\tabcolsep}{0.3em}
\begin{tabular}{c|c|rrrrrr}  \hline \hline
\multicolumn{2}{c|}{Neutron}         & $1i_{13/2}$        & $3p_{3/2}$ & $2f_{5/2}$ & $2f_{7/2}$ & $1h_{9/2}$ & $1h_{11/2}$  \\ \hline
\multirow{2}{*}{$3/2_{12}$} &Quad    &             0.0\%  &    2.3\%   &  16.5\%    &  11.7\%    &{\bfseries 62.4\%}  &       4.4\%  \\
                            &GS      & {\bfseries 24.5\%} &    2.9\%   &  12.1\%    &  22.5\%    &   11.5\%   &      14.1\%  \\ \hline
\multirow{2}{*}{$1/2_{18}$} &Quad    &             0.0\%  &   23.9\%   &   0.8\%    &{\bfseries 42.7\%}    &   18.8\%   &       8.3\%  \\
                            &GS      & {\bfseries 23.9\%} &    2.4\%   &  15.7\%    &  12.4\%    &   22.0\%   &       3.1\%  \\ \hline
\multirow{2}{*}{$1/2_{17}$} &Quad    &             0.0\%  &    6.5\%   &  27.4\%    &   4.0\%    & {\bfseries 48.2\%}   &       3.1\%  \\
                            &GS      & {\bfseries  3.5\%} &   15.6\%   &  15.7\%    &   5.6\%    &   11.3\%   &      14.6\%  \\ \hline \hline
\multicolumn{2}{c|}{Proton}          & $1h_{11/2}$        & $3s_{1/2}$ & $2d_{3/2}$ & $2d_{5/2}$ & $1g_{7/2}$ & $1g_{9/2}$   \\ \hline
\multirow{2}{*}{$3/2_{8}$}  &Quad    &             0.0\%  &   $-$~~    &   4.1\%    &   3.0\%    &{\bfseries 86.9\%}   &       3.8\%  \\
                            & GS     & {\bfseries 20.2\%} &   $-$~~    &   1.9\%    &   6.6\%    &   53.5\%   &       5.9\%  \\ \hline
\multirow{2}{*}{$1/2_{13}$} &Quad    &             0.0\%  &   14.1\%   &   4.4\%    &{\bfseries 55.7\%}    &   14.2\%   &       9.1\%  \\
                            & GS     & {\bfseries 29.8\%} &    7.1\%   &   5.7\%    &  25.3\%    &    2.3\%   &       8.6\%  \\ \hline
\multirow{2}{*}{$1/2_{12}$} &Quad    &             0.0\%  &    2.9\%   &  18.8\%    &   2.2\%    &{\bfseries 70.2\%}   &       2.6\%  \\
                            & GS     & {\bfseries  3.4\%} &    3.3\%   &  21.4\%    &   0.0\%    &   49.6\%   &       3.3\%  \\ \hline\hline
\end{tabular}
\end{table}

However, as previously stated, the $\pi$-PV coupling, which contributes only via the Fock terms, seems to present an effect against the octupole deformation, but enhance the effects of the quadruple deformation for $^{144}$Ba, which is demonstrated by the $\Delta E$ values in Table \ref{Tab:Bulk}. Combined Fig. \ref{Fig:Lev-PKO2} and Table \ref{tab:Ba144-GS}, it is not difficult to understand the role of the $\pi$-PV coupling, in which the tensor force components, namely the $\pi$-T coupling, give raise to the repulsive couplings between the $j_>$ ($j_<$) and $j_>$ ($j_<$) states, but attractive ones between the $j_<$ and $j_>$ states \cite{Otsuka2005PRL95.232502}. At the quadruple minima as illustrated in Fig. \ref{Fig:Lev-PKO2} and Table \ref{tab:Ba144-GS}, valence nucleons populate largely on the orbits which are dominated by the $j_< = l-1/2$ components of the spherical DWS basis, e.g., the neutron ($3/2_{12}, 1/2_{17}$) and proton $(3/2_8, 1/2_{12})$ orbits. In contrast, the core of $^{144}$Ba corresponds to $N=82$ and $Z=50$, both manifesting the feature of $j_> = l+1/2$ on the average. Thus, similar as revealed in Ref. \cite{Geng2020PRC101.064302}, PKO1 and PKO3, that contain the $\pi$-PV coupling, present a stronger enhancement from spherical to quadruple minima than PKO2, due to the attractive tensor force effects carried by the $\pi$-PV. Notice that the RMF Lagrangian DD-ME2, which do not contain either the Fock terms or the $\pi$-PV coupling, present a stronger quadruple enhancement than the RHF models PKO$i$ ($i=1,2,3$). This can be attributed to the fact that DD-ME2 gives too less bound spherical $^{144}$Ba than the RHF models.

In giving the stable octupole deformation of $^{144}$Ba, it is already demonstrated that the intrude neutron $1i_{13/2}$ and proton $1h_{11/2}$ components play an essential role, which manifest the feature of $j_>$. As deduced from the nature of tensor force, the $\pi$-T couplings between the core of $^{144}$Ba and these intrude components are repulsive. Meanwhile, from the quadruple minimum to the octupole GS as illustrated in Table \ref{tab:Ba144-GS}, the $j_<$ components in both neutron and proton valence orbits are largely reduced on average, which weakens the attractive $\pi$-T couplings between the core and valence nucleons. Consequently, it is not difficult to understand the fact that the $\pi$-PV coupling, mainly due to its tensor force components, manifests an effect against the octuple deformation for $^{144}$Ba, as deduced from the results in Table \ref{Tab:Bulk}.

\section{Summary}\label{sec:summary}
In this work, the axially octupole-quadruple deformed relativistic Hartree-Fock (OD-RHF) model is established, by utilizing the
spherical Dirac Woods-Saxon (DWS) basis to expand the single-particle wave functions. We present in details the formalism of the OD-RHF model in terms of the spherical DWS basis, and the space truncations are tested as well. Taking the octupole nucleus $^{144}$Ba as an example, the reliability of the newly developed OD-RHF model is illustrated, as well as the verification of the accuracy of the codes. Furthermore, we performed the calculations for $^{144}$Ba by using the RHF Lagrangians PKO$i$ ($i=1, 2, 3$) and the RMF one DD-ME2. It is
also demonstrated that the intrusion of neutron $1i_{13/2}$ and proton $1h_{11/2}$ components are essential in giving a stable octupole deformation for the ground state of $^{144}$Ba, as combined with the neutron and proton single-particle spectra.  Moreover, the intrude components can be enhanced by the Fock terms, which leads to stronger octupole enhancement than the RMF model that contains only the Hartree terms. However, due to the repulsive tensor couplings between the core of $^{144}$Ba and intrude components, the tensor force components carried by the $\pi$-PV coupling present an effect against the octupole deformation for $^{144}$Ba, which deserves special attentions in future study of octupole nuclei.

\section{Acknowledge}\label{sec:acknowledge}
This work is partly supported by the National Natural Science Foundation of China under Grant No. 12275111, the Strategic Priority Research Program of Chinese Academy of Sciences under Grant No. XDB34000000, the Fundamental Research Funds for the Central Universities lzujbky-2023-stlt01 and lzujbky-2022-sp02, and the Supercomputing Center of Lanzhou University.


\begin{thebibliography}{82}
\expandafter\ifx\csname natexlab\endcsname\relax\def\natexlab#1{#1}\fi
\expandafter\ifx\csname bibnamefont\endcsname\relax
  \def\bibnamefont#1{#1}\fi
\expandafter\ifx\csname bibfnamefont\endcsname\relax
  \def\bibfnamefont#1{#1}\fi
\expandafter\ifx\csname citenamefont\endcsname\relax
  \def\citenamefont#1{#1}\fi
\expandafter\ifx\csname url\endcsname\relax
  \def\url#1{\texttt{#1}}\fi
\expandafter\ifx\csname urlprefix\endcsname\relax\def\urlprefix{URL }\fi
\providecommand{\bibinfo}[2]{#2}
\providecommand{\eprint}[2][]{\url{#2}}

\bibitem[{\citenamefont{Sturm et~al.}(2010)\citenamefont{Sturm, Sharkov, and
  St$\ddot{\text{o}}$kerabcd}}]{Sturm2010NPA834.682c}
\bibinfo{author}{\bibfnamefont{C.}~\bibnamefont{Sturm}},
  \bibinfo{author}{\bibfnamefont{B.}~\bibnamefont{Sharkov}}, \bibnamefont{and}
  \bibinfo{author}{\bibfnamefont{H.}~\bibnamefont{St$\ddot{\text{o}}$kerabcd}},
  \bibinfo{journal}{Nucl. Phys. A} \textbf{\bibinfo{volume}{834}},
  \bibinfo{pages}{682c } (\bibinfo{year}{2010}).

\bibitem[{\citenamefont{Thoennessen}(2010)}]{Thoennessen2010NPA834.688c}
\bibinfo{author}{\bibfnamefont{M.}~\bibnamefont{Thoennessen}},
  \bibinfo{journal}{Nucl. Phys. A} \textbf{\bibinfo{volume}{834}},
  \bibinfo{pages}{688c } (\bibinfo{year}{2010}).

\bibitem[{\citenamefont{Zhan et~al.}(2010)\citenamefont{Zhan, Xu, Xiao, Xia,
  Zhao, Yuan, and Groupa}}]{Zhan2010NPA834.694c}
\bibinfo{author}{\bibfnamefont{W.~L.} \bibnamefont{Zhan}},
  \bibinfo{author}{\bibfnamefont{H.~S.} \bibnamefont{Xu}},
  \bibinfo{author}{\bibfnamefont{G.~Q.} \bibnamefont{Xiao}},
  \bibinfo{author}{\bibfnamefont{J.~W.} \bibnamefont{Xia}},
  \bibinfo{author}{\bibfnamefont{H.~W.} \bibnamefont{Zhao}},
  \bibinfo{author}{\bibfnamefont{Y.~J.} \bibnamefont{Yuan}}, \bibnamefont{and}
  \bibinfo{author}{\bibfnamefont{H.-C.} \bibnamefont{Groupa}},
  \bibinfo{journal}{Nucl. Phys. A} \textbf{\bibinfo{volume}{834}},
  \bibinfo{pages}{694c } (\bibinfo{year}{2010}).

\bibitem[{\citenamefont{Motobayashi}(2010)}]{Motobayashi2010NPA834.707c}
\bibinfo{author}{\bibfnamefont{T.}~\bibnamefont{Motobayashi}},
  \bibinfo{journal}{Nucl. Phys. A} \textbf{\bibinfo{volume}{834}},
  \bibinfo{pages}{707c } (\bibinfo{year}{2010}).

\bibitem[{\citenamefont{Gales}(2010)}]{Gales2010NPA834.717c}
\bibinfo{author}{\bibfnamefont{S.}~\bibnamefont{Gales}},
  \bibinfo{journal}{Nucl. Phys. A} \textbf{\bibinfo{volume}{834}},
  \bibinfo{pages}{717c } (\bibinfo{year}{2010}).

\bibitem[{\citenamefont{Jonson}(2004)}]{Jonson2004PhyRep389.1}
\bibinfo{author}{\bibfnamefont{B.}~\bibnamefont{Jonson}},
  \bibinfo{journal}{Phys. Rep.} \textbf{\bibinfo{volume}{389}},
  \bibinfo{pages}{1 } (\bibinfo{year}{2004}).

\bibitem[{\citenamefont{Tanihata}(1995)}]{Tanihata1995PPNP35.505}
\bibinfo{author}{\bibfnamefont{I.}~\bibnamefont{Tanihata}},
  \bibinfo{journal}{Prog. Part. Nucl. Phys.} \textbf{\bibinfo{volume}{35}},
  \bibinfo{pages}{505 } (\bibinfo{year}{1995}).

\bibitem[{\citenamefont{Jensen et~al.}(2004)\citenamefont{Jensen, Riisager,
  Fedorov, and Garrido}}]{Jensen2004RMP76.215}
\bibinfo{author}{\bibfnamefont{A.~S.} \bibnamefont{Jensen}},
  \bibinfo{author}{\bibfnamefont{K.}~\bibnamefont{Riisager}},
  \bibinfo{author}{\bibfnamefont{D.~V.} \bibnamefont{Fedorov}},
  \bibnamefont{and} \bibinfo{author}{\bibfnamefont{E.}~\bibnamefont{Garrido}},
  \bibinfo{journal}{Rev. Mod. Phys} \textbf{\bibinfo{volume}{76}},
  \bibinfo{pages}{215} (\bibinfo{year}{2004}).

\bibitem[{\citenamefont{Casten and Sherrill}(2000)}]{Casten2000PPNP45.S171}
\bibinfo{author}{\bibfnamefont{R.~F.} \bibnamefont{Casten}} \bibnamefont{and}
  \bibinfo{author}{\bibfnamefont{B.~M.} \bibnamefont{Sherrill}},
  \bibinfo{journal}{Prog. Part. Nucl. Phys.} \textbf{\bibinfo{volume}{45}},
  \bibinfo{pages}{S171 } (\bibinfo{year}{2000}).

\bibitem[{\citenamefont{Ershov et~al.}(2010)\citenamefont{Ershov, Grigorenko,
  Vaagen, and Zhukov}}]{Ershov2010JPG:NP37.064026}
\bibinfo{author}{\bibfnamefont{S.~N.} \bibnamefont{Ershov}},
  \bibinfo{author}{\bibfnamefont{L.~V.} \bibnamefont{Grigorenko}},
  \bibinfo{author}{\bibfnamefont{J.~S.} \bibnamefont{Vaagen}},
  \bibnamefont{and} \bibinfo{author}{\bibfnamefont{M.~V.}
  \bibnamefont{Zhukov}}, \bibinfo{journal}{J. Phys. G: Nucl. Phys.}
  \textbf{\bibinfo{volume}{37}}, \bibinfo{pages}{064026}
  (\bibinfo{year}{2010}).

\bibitem[{\citenamefont{Hoffman et~al.}(2008)\citenamefont{Hoffman, Baumann,
  Bazin, Brown, Christian, DeYoung, Finck, Frank, Hinnefeld, Howes
  et~al.}}]{Hoffman2008PRL100.152502}
\bibinfo{author}{\bibfnamefont{C.~R.} \bibnamefont{Hoffman}},
  \bibinfo{author}{\bibfnamefont{T.}~\bibnamefont{Baumann}},
  \bibinfo{author}{\bibfnamefont{D.}~\bibnamefont{Bazin}},
  \bibinfo{author}{\bibfnamefont{J.}~\bibnamefont{Brown}},
  \bibinfo{author}{\bibfnamefont{G.}~\bibnamefont{Christian}},
  \bibinfo{author}{\bibfnamefont{P.~A.} \bibnamefont{DeYoung}},
  \bibinfo{author}{\bibfnamefont{J.~E.} \bibnamefont{Finck}},
  \bibinfo{author}{\bibfnamefont{N.}~\bibnamefont{Frank}},
  \bibinfo{author}{\bibfnamefont{J.}~\bibnamefont{Hinnefeld}},
  \bibinfo{author}{\bibfnamefont{R.}~\bibnamefont{Howes}},
  \bibnamefont{et~al.}, \bibinfo{journal}{Phys. Rev. Lett.}
  \textbf{\bibinfo{volume}{100}}, \bibinfo{pages}{152502}
  (\bibinfo{year}{2008}).

\bibitem[{\citenamefont{Simon et~al.}(1999)\citenamefont{Simon, Aleksandrov,
  Aumann, Axelsson, Baumann, Borge, Chulkov, Collatz, Cub, Dostal
  et~al.}}]{Simon1999PRL83.496}
\bibinfo{author}{\bibfnamefont{H.}~\bibnamefont{Simon}},
  \bibinfo{author}{\bibfnamefont{D.}~\bibnamefont{Aleksandrov}},
  \bibinfo{author}{\bibfnamefont{T.}~\bibnamefont{Aumann}},
  \bibinfo{author}{\bibfnamefont{L.}~\bibnamefont{Axelsson}},
  \bibinfo{author}{\bibfnamefont{T.}~\bibnamefont{Baumann}},
  \bibinfo{author}{\bibfnamefont{M.~J.~G.} \bibnamefont{Borge}},
  \bibinfo{author}{\bibfnamefont{L.~V.} \bibnamefont{Chulkov}},
  \bibinfo{author}{\bibfnamefont{R.}~\bibnamefont{Collatz}},
  \bibinfo{author}{\bibfnamefont{J.}~\bibnamefont{Cub}},
  \bibinfo{author}{\bibfnamefont{W.}~\bibnamefont{Dostal}},
  \bibnamefont{et~al.}, \bibinfo{journal}{Phys. Rev. Lett.}
  \textbf{\bibinfo{volume}{83}}, \bibinfo{pages}{496} (\bibinfo{year}{1999}).

\bibitem[{\citenamefont{Motobayashi et~al.}(1995)\citenamefont{Motobayashi,
  Ikeda, Ando, Ieki, Inoue, Iwasa, Kikuchi, Kurokawa, Moriya, Ogawa
  et~al.}}]{Motobayashi1995PLB346.9}
\bibinfo{author}{\bibfnamefont{T.}~\bibnamefont{Motobayashi}},
  \bibinfo{author}{\bibfnamefont{Y.}~\bibnamefont{Ikeda}},
  \bibinfo{author}{\bibfnamefont{Y.}~\bibnamefont{Ando}},
  \bibinfo{author}{\bibfnamefont{K.}~\bibnamefont{Ieki}},
  \bibinfo{author}{\bibfnamefont{M.}~\bibnamefont{Inoue}},
  \bibinfo{author}{\bibfnamefont{N.}~\bibnamefont{Iwasa}},
  \bibinfo{author}{\bibfnamefont{T.}~\bibnamefont{Kikuchi}},
  \bibinfo{author}{\bibfnamefont{M.}~\bibnamefont{Kurokawa}},
  \bibinfo{author}{\bibfnamefont{S.}~\bibnamefont{Moriya}},
  \bibinfo{author}{\bibfnamefont{S.}~\bibnamefont{Ogawa}},
  \bibnamefont{et~al.}, \bibinfo{journal}{Phys. Lett. B}
  \textbf{\bibinfo{volume}{346}}, \bibinfo{pages}{9} (\bibinfo{year}{1995}).

\bibitem[{\citenamefont{Tshoo et~al.}(2012)\citenamefont{Tshoo, Satou, Bhang,
  Choi, Nakamura, Kondo, Deguchi, Kawada, Kobayashi, Nakayama
  et~al.}}]{TshooPRL109.022501}
\bibinfo{author}{\bibfnamefont{K.}~\bibnamefont{Tshoo}},
  \bibinfo{author}{\bibfnamefont{Y.}~\bibnamefont{Satou}},
  \bibinfo{author}{\bibfnamefont{H.}~\bibnamefont{Bhang}},
  \bibinfo{author}{\bibfnamefont{S.}~\bibnamefont{Choi}},
  \bibinfo{author}{\bibfnamefont{T.}~\bibnamefont{Nakamura}},
  \bibinfo{author}{\bibfnamefont{Y.}~\bibnamefont{Kondo}},
  \bibinfo{author}{\bibfnamefont{S.}~\bibnamefont{Deguchi}},
  \bibinfo{author}{\bibfnamefont{Y.}~\bibnamefont{Kawada}},
  \bibinfo{author}{\bibfnamefont{N.}~\bibnamefont{Kobayashi}},
  \bibinfo{author}{\bibfnamefont{Y.}~\bibnamefont{Nakayama}},
  \bibnamefont{et~al.}, \bibinfo{journal}{Phys. Rev. Lett.}
  \textbf{\bibinfo{volume}{109}}, \bibinfo{pages}{022501}
  (\bibinfo{year}{2012}).

\bibitem[{\citenamefont{Ozawa et~al.}(2000)\citenamefont{Ozawa, Kobayashi,
  Suzuki, and Tanihata}}]{Ozawa2000PRL84.5493}
\bibinfo{author}{\bibfnamefont{A.}~\bibnamefont{Ozawa}},
  \bibinfo{author}{\bibfnamefont{T.}~\bibnamefont{Kobayashi}},
  \bibinfo{author}{\bibfnamefont{T.}~\bibnamefont{Suzuki}}, \bibnamefont{and}
  \bibinfo{author}{\bibfnamefont{K.~Y.~I.} \bibnamefont{Tanihata}},
  \bibinfo{journal}{Phys. Rev. Lett.} \textbf{\bibinfo{volume}{84}},
  \bibinfo{pages}{5493} (\bibinfo{year}{2000}).

\bibitem[{\citenamefont{Kanungo et~al.}(2009)\citenamefont{Kanungo, Nociforo,
  Prochazka, Aumann, Boutin, Cortina-Gil, Davids, Diakaki, Farinon, Geissel
  et~al.}}]{Kanungo2009PRL102.152501}
\bibinfo{author}{\bibfnamefont{R.}~\bibnamefont{Kanungo}},
  \bibinfo{author}{\bibfnamefont{C.}~\bibnamefont{Nociforo}},
  \bibinfo{author}{\bibfnamefont{A.}~\bibnamefont{Prochazka}},
  \bibinfo{author}{\bibfnamefont{T.}~\bibnamefont{Aumann}},
  \bibinfo{author}{\bibfnamefont{D.}~\bibnamefont{Boutin}},
  \bibinfo{author}{\bibfnamefont{D.}~\bibnamefont{Cortina-Gil}},
  \bibinfo{author}{\bibfnamefont{B.}~\bibnamefont{Davids}},
  \bibinfo{author}{\bibfnamefont{M.}~\bibnamefont{Diakaki}},
  \bibinfo{author}{\bibfnamefont{F.}~\bibnamefont{Farinon}},
  \bibinfo{author}{\bibfnamefont{H.}~\bibnamefont{Geissel}},
  \bibnamefont{et~al.}, \bibinfo{journal}{Phys. Rev. Lett.}
  \textbf{\bibinfo{volume}{102}}, \bibinfo{pages}{152501}
  (\bibinfo{year}{2009}).

\bibitem[{\citenamefont{Gade et~al.}(2006)\citenamefont{Gade, Janssens, Bazin,
  Broda, Brown, Campbell, Carpenter, Cook, Deacon, Dinca
  et~al.}}]{Gade2006PRC74.021302}
\bibinfo{author}{\bibfnamefont{A.}~\bibnamefont{Gade}},
  \bibinfo{author}{\bibfnamefont{R.~V.~F.} \bibnamefont{Janssens}},
  \bibinfo{author}{\bibfnamefont{D.}~\bibnamefont{Bazin}},
  \bibinfo{author}{\bibfnamefont{R.}~\bibnamefont{Broda}},
  \bibinfo{author}{\bibfnamefont{B.~A.} \bibnamefont{Brown}},
  \bibinfo{author}{\bibfnamefont{C.~M.} \bibnamefont{Campbell}},
  \bibinfo{author}{\bibfnamefont{M.~P.} \bibnamefont{Carpenter}},
  \bibinfo{author}{\bibfnamefont{J.~M.} \bibnamefont{Cook}},
  \bibinfo{author}{\bibfnamefont{A.~N.} \bibnamefont{Deacon}},
  \bibinfo{author}{\bibfnamefont{D.-C.} \bibnamefont{Dinca}},
  \bibnamefont{et~al.}, \bibinfo{journal}{Phys. Rev. C}
  \textbf{\bibinfo{volume}{74}}, \bibinfo{pages}{021302}
  (\bibinfo{year}{2006}).

\bibitem[{\citenamefont{Steppenbeck et~al.}(2015)\citenamefont{Steppenbeck,
  Takeuchi, Aoi, Doornenbal, Matsushita, Wang, Utsuno, Baba, Go, Lee
  et~al.}}]{Steppenbeck2015PRL114.252501}
\bibinfo{author}{\bibfnamefont{D.}~\bibnamefont{Steppenbeck}},
  \bibinfo{author}{\bibfnamefont{S.}~\bibnamefont{Takeuchi}},
  \bibinfo{author}{\bibfnamefont{N.}~\bibnamefont{Aoi}},
  \bibinfo{author}{\bibfnamefont{P.}~\bibnamefont{Doornenbal}},
  \bibinfo{author}{\bibfnamefont{M.}~\bibnamefont{Matsushita}},
  \bibinfo{author}{\bibfnamefont{H.}~\bibnamefont{Wang}},
  \bibinfo{author}{\bibfnamefont{Y.}~\bibnamefont{Utsuno}},
  \bibinfo{author}{\bibfnamefont{H.}~\bibnamefont{Baba}},
  \bibinfo{author}{\bibfnamefont{S.}~\bibnamefont{Go}},
  \bibinfo{author}{\bibfnamefont{J.}~\bibnamefont{Lee}}, \bibnamefont{et~al.},
  \bibinfo{journal}{Phys. Rev. Lett.} \textbf{\bibinfo{volume}{114}},
  \bibinfo{pages}{252501} (\bibinfo{year}{2015}).

\bibitem[{\citenamefont{Steppenbeck et~al.}(2013)\citenamefont{Steppenbeck,
  Takeuchi, Aoi, Doornenbal, Matsushita, Wang, Baba, Fukuda, Go, Honma
  et~al.}}]{Steppenbeck2013Nature502.207}
\bibinfo{author}{\bibfnamefont{D.}~\bibnamefont{Steppenbeck}},
  \bibinfo{author}{\bibfnamefont{S.}~\bibnamefont{Takeuchi}},
  \bibinfo{author}{\bibfnamefont{N.}~\bibnamefont{Aoi}},
  \bibinfo{author}{\bibfnamefont{P.}~\bibnamefont{Doornenbal}},
  \bibinfo{author}{\bibfnamefont{M.}~\bibnamefont{Matsushita}},
  \bibinfo{author}{\bibfnamefont{H.}~\bibnamefont{Wang}},
  \bibinfo{author}{\bibfnamefont{H.}~\bibnamefont{Baba}},
  \bibinfo{author}{\bibfnamefont{N.}~\bibnamefont{Fukuda}},
  \bibinfo{author}{\bibfnamefont{S.}~\bibnamefont{Go}},
  \bibinfo{author}{\bibfnamefont{M.}~\bibnamefont{Honma}},
  \bibnamefont{et~al.}, \bibinfo{journal}{Nature}
  \textbf{\bibinfo{volume}{502}}, \bibinfo{pages}{207} (\bibinfo{year}{2013}).

\bibitem[{\citenamefont{Minamisono et~al.}(1992)\citenamefont{Minamisono,
  Ohtsubo, Minami, Fukuda, Kitagawa, Fukuda, Matsuta, Nojiri, Takeda, Sagawa
  et~al.}}]{Minamisono1992PRL69.2058}
\bibinfo{author}{\bibfnamefont{T.}~\bibnamefont{Minamisono}},
  \bibinfo{author}{\bibfnamefont{T.}~\bibnamefont{Ohtsubo}},
  \bibinfo{author}{\bibfnamefont{I.}~\bibnamefont{Minami}},
  \bibinfo{author}{\bibfnamefont{S.}~\bibnamefont{Fukuda}},
  \bibinfo{author}{\bibfnamefont{A.}~\bibnamefont{Kitagawa}},
  \bibinfo{author}{\bibfnamefont{M.}~\bibnamefont{Fukuda}},
  \bibinfo{author}{\bibfnamefont{K.}~\bibnamefont{Matsuta}},
  \bibinfo{author}{\bibfnamefont{Y.}~\bibnamefont{Nojiri}},
  \bibinfo{author}{\bibfnamefont{S.}~\bibnamefont{Takeda}},
  \bibinfo{author}{\bibfnamefont{H.}~\bibnamefont{Sagawa}},
  \bibnamefont{et~al.}, \bibinfo{journal}{Phys. Rev. Lett.}
  \textbf{\bibinfo{volume}{69}}, \bibinfo{pages}{2058} (\bibinfo{year}{1992}).

\bibitem[{\citenamefont{Tanihata et~al.}(1985)\citenamefont{Tanihata, Hamagaki,
  Hashimoto, Shida, Yoshikawa, Sugimoto, Yamakawa, Kobayashi, and
  Takahashi}}]{Tanihata1985PRL55.2676}
\bibinfo{author}{\bibfnamefont{I.}~\bibnamefont{Tanihata}},
  \bibinfo{author}{\bibfnamefont{H.}~\bibnamefont{Hamagaki}},
  \bibinfo{author}{\bibfnamefont{O.}~\bibnamefont{Hashimoto}},
  \bibinfo{author}{\bibfnamefont{Y.}~\bibnamefont{Shida}},
  \bibinfo{author}{\bibfnamefont{N.}~\bibnamefont{Yoshikawa}},
  \bibinfo{author}{\bibfnamefont{K.}~\bibnamefont{Sugimoto}},
  \bibinfo{author}{\bibfnamefont{O.}~\bibnamefont{Yamakawa}},
  \bibinfo{author}{\bibfnamefont{T.}~\bibnamefont{Kobayashi}},
  \bibnamefont{and}
  \bibinfo{author}{\bibfnamefont{N.}~\bibnamefont{Takahashi}},
  \bibinfo{journal}{Phys. Rev. Lett.} \textbf{\bibinfo{volume}{55}},
  \bibinfo{pages}{2676} (\bibinfo{year}{1985}).

\bibitem[{\citenamefont{Schwab et~al.}(1995)\citenamefont{Schwab, Geissel,
  Lenske, Behr, Br$\ddot{\text{u}}$nle, Burkard, Irnich, Kobayashi, Kraus,
  Magel et~al.}}]{Schwab1995ZPA350.283}
\bibinfo{author}{\bibfnamefont{W.}~\bibnamefont{Schwab}},
  \bibinfo{author}{\bibfnamefont{H.}~\bibnamefont{Geissel}},
  \bibinfo{author}{\bibfnamefont{H.}~\bibnamefont{Lenske}},
  \bibinfo{author}{\bibfnamefont{K.-H.} \bibnamefont{Behr}},
  \bibinfo{author}{\bibfnamefont{A.}~\bibnamefont{Br$\ddot{\text{u}}$nle}},
  \bibinfo{author}{\bibfnamefont{K.}~\bibnamefont{Burkard}},
  \bibinfo{author}{\bibfnamefont{H.}~\bibnamefont{Irnich}},
  \bibinfo{author}{\bibfnamefont{T.}~\bibnamefont{Kobayashi}},
  \bibinfo{author}{\bibfnamefont{G.}~\bibnamefont{Kraus}},
  \bibinfo{author}{\bibfnamefont{A.}~\bibnamefont{Magel}},
  \bibnamefont{et~al.}, \bibinfo{journal}{Z. Phys. A}
  \textbf{\bibinfo{volume}{350}}, \bibinfo{pages}{283} (\bibinfo{year}{1995}).

\bibitem[{\citenamefont{Asaro et~al.}(1953)\citenamefont{Asaro, Jr, and
  Perlman}}]{Asaro1953PR92.1495}
\bibinfo{author}{\bibfnamefont{F.}~\bibnamefont{Asaro}},
  \bibinfo{author}{\bibfnamefont{F.~S.} \bibnamefont{Jr}}, \bibnamefont{and}
  \bibinfo{author}{\bibfnamefont{I.}~\bibnamefont{Perlman}},
  \bibinfo{journal}{Phys. Rev.} \textbf{\bibinfo{volume}{92}},
  \bibinfo{pages}{1495} (\bibinfo{year}{1953}).

\bibitem[{\citenamefont{Jr et~al.}(1954)\citenamefont{Jr, Asaro, and
  Perlman}}]{Stephens1954PR96.1568}
\bibinfo{author}{\bibfnamefont{F.~S.} \bibnamefont{Jr}},
  \bibinfo{author}{\bibfnamefont{F.}~\bibnamefont{Asaro}}, \bibnamefont{and}
  \bibinfo{author}{\bibfnamefont{I.}~\bibnamefont{Perlman}},
  \bibinfo{journal}{Phys. Rev.} \textbf{\bibinfo{volume}{96}},
  \bibinfo{pages}{1568} (\bibinfo{year}{1954}).

\bibitem[{\citenamefont{Jr et~al.}(1955)\citenamefont{Jr, Asaro, and
  Perlman}}]{Stephens1955PR100.1543}
\bibinfo{author}{\bibfnamefont{F.~S.} \bibnamefont{Jr}},
  \bibinfo{author}{\bibfnamefont{F.}~\bibnamefont{Asaro}}, \bibnamefont{and}
  \bibinfo{author}{\bibfnamefont{I.}~\bibnamefont{Perlman}},
  \bibinfo{journal}{Phys. Rev.} \textbf{\bibinfo{volume}{100}},
  \bibinfo{pages}{1543} (\bibinfo{year}{1955}).

\bibitem[{\citenamefont{Lee and Inglis}(1957)}]{LeePR1957108.774}
\bibinfo{author}{\bibfnamefont{K.}~\bibnamefont{Lee}} \bibnamefont{and}
  \bibinfo{author}{\bibfnamefont{D.~R.} \bibnamefont{Inglis}},
  \bibinfo{journal}{Phys. Rev.} \textbf{\bibinfo{volume}{108}},
  \bibinfo{pages}{774} (\bibinfo{year}{1957}).

\bibitem[{\citenamefont{Butler and Nazarewicz}(1996)}]{Butler1996RMP68}
\bibinfo{author}{\bibfnamefont{P.~A.} \bibnamefont{Butler}} \bibnamefont{and}
  \bibinfo{author}{\bibfnamefont{W.}~\bibnamefont{Nazarewicz}},
  \bibinfo{journal}{Rev. Mod. Phys.} \textbf{\bibinfo{volume}{68}},
  \bibinfo{pages}{349} (\bibinfo{year}{1996}).

\bibitem[{\citenamefont{Gaffney et~al.}(2013)\citenamefont{Gaffney, Butler,
  Scheck, Hayes, Wenander, Albers, Bastin, Bauer, Blazhev, B\"onig
  et~al.}}]{Gaffney2013NATURE477}
\bibinfo{author}{\bibfnamefont{L.~P.} \bibnamefont{Gaffney}},
  \bibinfo{author}{\bibfnamefont{P.~A.} \bibnamefont{Butler}},
  \bibinfo{author}{\bibfnamefont{M.}~\bibnamefont{Scheck}},
  \bibinfo{author}{\bibfnamefont{A.~B.} \bibnamefont{Hayes}},
  \bibinfo{author}{\bibfnamefont{F.}~\bibnamefont{Wenander}},
  \bibinfo{author}{\bibfnamefont{M.}~\bibnamefont{Albers}},
  \bibinfo{author}{\bibfnamefont{B.}~\bibnamefont{Bastin}},
  \bibinfo{author}{\bibfnamefont{C.}~\bibnamefont{Bauer}},
  \bibinfo{author}{\bibfnamefont{A.}~\bibnamefont{Blazhev}},
  \bibinfo{author}{\bibfnamefont{S.}~\bibnamefont{B\"onig}},
  \bibnamefont{et~al.}, \bibinfo{journal}{Nature}
  \textbf{\bibinfo{volume}{497}}, \bibinfo{pages}{199} (\bibinfo{year}{2013}).

\bibitem[{\citenamefont{Bucher et~al.}(2016{\natexlab{a}})\citenamefont{Bucher,
  Zhu, Wu, Janssens, Cline, Hayes, Albers, Ayangeakaa, Butler, Campbell
  et~al.}}]{Bucher2016PRL116}
\bibinfo{author}{\bibfnamefont{B.}~\bibnamefont{Bucher}},
  \bibinfo{author}{\bibfnamefont{S.}~\bibnamefont{Zhu}},
  \bibinfo{author}{\bibfnamefont{C.~Y.} \bibnamefont{Wu}},
  \bibinfo{author}{\bibfnamefont{R.~V.~F.} \bibnamefont{Janssens}},
  \bibinfo{author}{\bibfnamefont{D.}~\bibnamefont{Cline}},
  \bibinfo{author}{\bibfnamefont{A.~B.} \bibnamefont{Hayes}},
  \bibinfo{author}{\bibfnamefont{M.}~\bibnamefont{Albers}},
  \bibinfo{author}{\bibfnamefont{A.~D.} \bibnamefont{Ayangeakaa}},
  \bibinfo{author}{\bibfnamefont{P.~A.} \bibnamefont{Butler}},
  \bibinfo{author}{\bibfnamefont{C.~M.} \bibnamefont{Campbell}},
  \bibnamefont{et~al.}, \bibinfo{journal}{Phys. Rev. Lett.}
  \textbf{\bibinfo{volume}{116}}, \bibinfo{pages}{112503}
  (\bibinfo{year}{2016}{\natexlab{a}}).

\bibitem[{\citenamefont{Bucher et~al.}(2016{\natexlab{b}})\citenamefont{Bucher,
  Zhu, Wu, Janssens, Bernard, Robledo, Rodr\'{\i}guez, Cline, Hayes, Ayangeakaa
  et~al.}}]{Bucher2016PRL118}
\bibinfo{author}{\bibfnamefont{B.}~\bibnamefont{Bucher}},
  \bibinfo{author}{\bibfnamefont{S.}~\bibnamefont{Zhu}},
  \bibinfo{author}{\bibfnamefont{C.~Y.} \bibnamefont{Wu}},
  \bibinfo{author}{\bibfnamefont{R.~V.~F.} \bibnamefont{Janssens}},
  \bibinfo{author}{\bibfnamefont{R.~N.} \bibnamefont{Bernard}},
  \bibinfo{author}{\bibfnamefont{L.~M.} \bibnamefont{Robledo}},
  \bibinfo{author}{\bibfnamefont{T.~R.} \bibnamefont{Rodr\'{\i}guez}},
  \bibinfo{author}{\bibfnamefont{D.}~\bibnamefont{Cline}},
  \bibinfo{author}{\bibfnamefont{A.~B.} \bibnamefont{Hayes}},
  \bibinfo{author}{\bibfnamefont{A.~D.} \bibnamefont{Ayangeakaa}},
  \bibnamefont{et~al.}, \bibinfo{journal}{Phys. Rev. Lett.}
  \textbf{\bibinfo{volume}{118}}, \bibinfo{pages}{152504}
  (\bibinfo{year}{2016}{\natexlab{b}}).

\bibitem[{\citenamefont{Chishti et~al.}(2020)\citenamefont{Chishti,
  $O'$Donnell, Battaglia, Bowry, Jaroszynski, Singh, Scheck, Spagnoletti, and
  Smith}}]{Chishti2020NARURE16}
\bibinfo{author}{\bibfnamefont{M.~M.~R.} \bibnamefont{Chishti}},
  \bibinfo{author}{\bibfnamefont{D.}~\bibnamefont{$O'$Donnell}},
  \bibinfo{author}{\bibfnamefont{G.}~\bibnamefont{Battaglia}},
  \bibinfo{author}{\bibfnamefont{M.}~\bibnamefont{Bowry}},
  \bibinfo{author}{\bibfnamefont{D.~A.} \bibnamefont{Jaroszynski}},
  \bibinfo{author}{\bibfnamefont{B.~S.~N.} \bibnamefont{Singh}},
  \bibinfo{author}{\bibfnamefont{M.}~\bibnamefont{Scheck}},
  \bibinfo{author}{\bibfnamefont{P.}~\bibnamefont{Spagnoletti}},
  \bibnamefont{and} \bibinfo{author}{\bibfnamefont{J.~F.} \bibnamefont{Smith}},
  \bibinfo{journal}{Nat. Phys.} \textbf{\bibinfo{volume}{16}},
  \bibinfo{pages}{853} (\bibinfo{year}{2020}).

\bibitem[{\citenamefont{Zhang and Jia}(2022)}]{Zhang2022PRL128}
\bibinfo{author}{\bibfnamefont{C.}~\bibnamefont{Zhang}} \bibnamefont{and}
  \bibinfo{author}{\bibfnamefont{J.}~\bibnamefont{Jia}},
  \bibinfo{journal}{Phys. Rev. Lett.} \textbf{\bibinfo{volume}{128}},
  \bibinfo{pages}{022301} (\bibinfo{year}{2022}).

\bibitem[{\citenamefont{Butler}(2016)}]{Butler2016JPG43.073002}
\bibinfo{author}{\bibfnamefont{P.~A.} \bibnamefont{Butler}},
  \bibinfo{journal}{J. Phys. G: Nucl. Part. Phys.}
  \textbf{\bibinfo{volume}{43}}, \bibinfo{pages}{073002}
  (\bibinfo{year}{2016}).

\bibitem[{\citenamefont{Bouyssy et~al.}(1987)\citenamefont{Bouyssy, Mathiot,
  Giai, and Marcos}}]{Bouyssy1987PRC36.380}
\bibinfo{author}{\bibfnamefont{A.}~\bibnamefont{Bouyssy}},
  \bibinfo{author}{\bibfnamefont{J.~F.} \bibnamefont{Mathiot}},
  \bibinfo{author}{\bibfnamefont{N.~V.} \bibnamefont{Giai}}, \bibnamefont{and}
  \bibinfo{author}{\bibfnamefont{S.}~\bibnamefont{Marcos}},
  \bibinfo{journal}{Phys. Rev.} \textbf{\bibinfo{volume}{C 36}},
  \bibinfo{pages}{380} (\bibinfo{year}{1987}).

\bibitem[{\citenamefont{Long et~al.}(2006{\natexlab{a}})\citenamefont{Long,
  Van~Giai, and Meng}}]{Long2006PLB640.150}
\bibinfo{author}{\bibfnamefont{W.~H.} \bibnamefont{Long}},
  \bibinfo{author}{\bibfnamefont{N.}~\bibnamefont{Van~Giai}}, \bibnamefont{and}
  \bibinfo{author}{\bibfnamefont{J.}~\bibnamefont{Meng}},
  \bibinfo{journal}{Phys. Lett. B} \textbf{\bibinfo{volume}{640}},
  \bibinfo{pages}{150} (\bibinfo{year}{2006}{\natexlab{a}}).

\bibitem[{\citenamefont{Long et~al.}(2007)\citenamefont{Long, Sagawa, Van~Giai,
  and Meng}}]{Long2007PRC76.034314}
\bibinfo{author}{\bibfnamefont{W.~H.} \bibnamefont{Long}},
  \bibinfo{author}{\bibfnamefont{H.}~\bibnamefont{Sagawa}},
  \bibinfo{author}{\bibfnamefont{N.}~\bibnamefont{Van~Giai}}, \bibnamefont{and}
  \bibinfo{author}{\bibfnamefont{J.}~\bibnamefont{Meng}},
  \bibinfo{journal}{Phys. Rev.} \textbf{\bibinfo{volume}{C 76}},
  \bibinfo{pages}{034314} (\bibinfo{year}{2007}).

\bibitem[{\citenamefont{Long et~al.}(2010{\natexlab{a}})\citenamefont{Long,
  Ring, Giai, and Meng}}]{Long2010PRC81.024308}
\bibinfo{author}{\bibfnamefont{W.~H.} \bibnamefont{Long}},
  \bibinfo{author}{\bibfnamefont{P.}~\bibnamefont{Ring}},
  \bibinfo{author}{\bibfnamefont{N.~V.} \bibnamefont{Giai}}, \bibnamefont{and}
  \bibinfo{author}{\bibfnamefont{J.}~\bibnamefont{Meng}},
  \bibinfo{journal}{Phys. Rev.} \textbf{\bibinfo{volume}{C 81}},
  \bibinfo{pages}{024308} (\bibinfo{year}{2010}{\natexlab{a}}).

\bibitem[{\citenamefont{Geng et~al.}(2020)\citenamefont{Geng, Xiang, Sun, and
  Long}}]{Geng2020PRC101.064302}
\bibinfo{author}{\bibfnamefont{J.}~\bibnamefont{Geng}},
  \bibinfo{author}{\bibfnamefont{J.}~\bibnamefont{Xiang}},
  \bibinfo{author}{\bibfnamefont{B.~Y.} \bibnamefont{Sun}}, \bibnamefont{and}
  \bibinfo{author}{\bibfnamefont{W.~H.} \bibnamefont{Long}},
  \bibinfo{journal}{Phys. Rev. C} \textbf{\bibinfo{volume}{101}},
  \bibinfo{pages}{064302} (\bibinfo{year}{2020}).

\bibitem[{\citenamefont{Geng and Long}(2022)}]{Geng2022PRC105.034329}
\bibinfo{author}{\bibfnamefont{J.}~\bibnamefont{Geng}} \bibnamefont{and}
  \bibinfo{author}{\bibfnamefont{W.~H.} \bibnamefont{Long}},
  \bibinfo{journal}{Phys. Rev. C} \textbf{\bibinfo{volume}{105}},
  \bibinfo{pages}{034329} (\bibinfo{year}{2022}).

\bibitem[{\citenamefont{Long et~al.}(2008)\citenamefont{Long, Sagawa, Meng, and
  Van~Giai}}]{Long2008EPL82.12001}
\bibinfo{author}{\bibfnamefont{W.~H.} \bibnamefont{Long}},
  \bibinfo{author}{\bibfnamefont{H.}~\bibnamefont{Sagawa}},
  \bibinfo{author}{\bibfnamefont{J.}~\bibnamefont{Meng}}, \bibnamefont{and}
  \bibinfo{author}{\bibfnamefont{N.}~\bibnamefont{Van~Giai}},
  \bibinfo{journal}{Europhys. Lett.} \textbf{\bibinfo{volume}{82}},
  \bibinfo{pages}{12001} (\bibinfo{year}{2008}).

\bibitem[{\citenamefont{Long et~al.}(2009)\citenamefont{Long, Nakatsukasa,
  Sagawa, Meng, Nakada, and Zhang}}]{Long2009PLB680.428}
\bibinfo{author}{\bibfnamefont{W.~H.} \bibnamefont{Long}},
  \bibinfo{author}{\bibfnamefont{T.}~\bibnamefont{Nakatsukasa}},
  \bibinfo{author}{\bibfnamefont{H.}~\bibnamefont{Sagawa}},
  \bibinfo{author}{\bibfnamefont{J.}~\bibnamefont{Meng}},
  \bibinfo{author}{\bibfnamefont{H.}~\bibnamefont{Nakada}}, \bibnamefont{and}
  \bibinfo{author}{\bibfnamefont{Y.}~\bibnamefont{Zhang}},
  \bibinfo{journal}{Phys. Lett.} \textbf{\bibinfo{volume}{B 680}},
  \bibinfo{pages}{428 } (\bibinfo{year}{2009}).

\bibitem[{\citenamefont{Li et~al.}(2016)\citenamefont{Li, Margueron, Long, and
  Giai}}]{Li2016PLB753.97}
\bibinfo{author}{\bibfnamefont{J.~J.} \bibnamefont{Li}},
  \bibinfo{author}{\bibfnamefont{J.}~\bibnamefont{Margueron}},
  \bibinfo{author}{\bibfnamefont{W.~H.} \bibnamefont{Long}}, \bibnamefont{and}
  \bibinfo{author}{\bibfnamefont{N.~V.} \bibnamefont{Giai}},
  \bibinfo{journal}{Phys. Lett. B} \textbf{\bibinfo{volume}{753}},
  \bibinfo{pages}{97} (\bibinfo{year}{2016}).

\bibitem[{\citenamefont{Wang et~al.}(2013)\citenamefont{Wang, Dong, and
  Long}}]{Wang2013PRC87.047301}
\bibinfo{author}{\bibfnamefont{L.~J.} \bibnamefont{Wang}},
  \bibinfo{author}{\bibfnamefont{J.~M.} \bibnamefont{Dong}}, \bibnamefont{and}
  \bibinfo{author}{\bibfnamefont{W.~H.} \bibnamefont{Long}},
  \bibinfo{journal}{Phys. Rev.} \textbf{\bibinfo{volume}{C 87}},
  \bibinfo{pages}{047301} (\bibinfo{year}{2013}).

\bibitem[{\citenamefont{Sun et~al.}(2008)\citenamefont{Sun, Long, Meng, and
  Lombardo}}]{Sun2008PRC78.065805}
\bibinfo{author}{\bibfnamefont{B.~Y.} \bibnamefont{Sun}},
  \bibinfo{author}{\bibfnamefont{W.~H.} \bibnamefont{Long}},
  \bibinfo{author}{\bibfnamefont{J.}~\bibnamefont{Meng}}, \bibnamefont{and}
  \bibinfo{author}{\bibfnamefont{U.}~\bibnamefont{Lombardo}},
  \bibinfo{journal}{Phys. Rev.} \textbf{\bibinfo{volume}{C 78}},
  \bibinfo{pages}{065805} (\bibinfo{year}{2008}).

\bibitem[{\citenamefont{Long et~al.}(2012)\citenamefont{Long, Sun, Hagino, and
  Sagawa}}]{Long2012PRC85.025806}
\bibinfo{author}{\bibfnamefont{W.~H.} \bibnamefont{Long}},
  \bibinfo{author}{\bibfnamefont{B.~Y.} \bibnamefont{Sun}},
  \bibinfo{author}{\bibfnamefont{K.}~\bibnamefont{Hagino}}, \bibnamefont{and}
  \bibinfo{author}{\bibfnamefont{H.}~\bibnamefont{Sagawa}},
  \bibinfo{journal}{Phys. Rev.} \textbf{\bibinfo{volume}{C 85}},
  \bibinfo{pages}{025806} (\bibinfo{year}{2012}).

\bibitem[{\citenamefont{Zhao et~al.}(2015)\citenamefont{Zhao, sun, and
  Long}}]{Zhao2015JPG42.095101}
\bibinfo{author}{\bibfnamefont{Q.}~\bibnamefont{Zhao}},
  \bibinfo{author}{\bibfnamefont{B.~Y.} \bibnamefont{sun}}, \bibnamefont{and}
  \bibinfo{author}{\bibfnamefont{W.~H.} \bibnamefont{Long}},
  \bibinfo{journal}{J. Phys. G: Nucl. Part. Phys.}
  \textbf{\bibinfo{volume}{42}}, \bibinfo{pages}{095101}
  (\bibinfo{year}{2015}).

\bibitem[{\citenamefont{Li et~al.}(2014)\citenamefont{Li, Long, Margueron, and
  Van~Giai}}]{Li2014PLB732.169}
\bibinfo{author}{\bibfnamefont{J.~J.} \bibnamefont{Li}},
  \bibinfo{author}{\bibfnamefont{W.~H.} \bibnamefont{Long}},
  \bibinfo{author}{\bibfnamefont{J.~M.} \bibnamefont{Margueron}},
  \bibnamefont{and} \bibinfo{author}{\bibfnamefont{N.}~\bibnamefont{Van~Giai}},
  \bibinfo{journal}{Phys. Lett.} \textbf{\bibinfo{volume}{B 732}},
  \bibinfo{pages}{169 } (\bibinfo{year}{2014}).

\bibitem[{\citenamefont{Li et~al.}(2019)\citenamefont{Li, Long, Margueron, and
  Giai}}]{Li2019PLB788.192}
\bibinfo{author}{\bibfnamefont{J.~J.} \bibnamefont{Li}},
  \bibinfo{author}{\bibfnamefont{W.~H.} \bibnamefont{Long}},
  \bibinfo{author}{\bibfnamefont{J.}~\bibnamefont{Margueron}},
  \bibnamefont{and} \bibinfo{author}{\bibfnamefont{N.~V.} \bibnamefont{Giai}},
  \bibinfo{journal}{Phys. Lett. B} \textbf{\bibinfo{volume}{788}},
  \bibinfo{pages}{192} (\bibinfo{year}{2019}).

\bibitem[{\citenamefont{Long et~al.}(2010{\natexlab{b}})\citenamefont{Long,
  Ring, Meng, Van~Giai, and Bertulani}}]{Long2010PRC81.031302}
\bibinfo{author}{\bibfnamefont{W.~H.} \bibnamefont{Long}},
  \bibinfo{author}{\bibfnamefont{P.}~\bibnamefont{Ring}},
  \bibinfo{author}{\bibfnamefont{J.}~\bibnamefont{Meng}},
  \bibinfo{author}{\bibfnamefont{N.}~\bibnamefont{Van~Giai}}, \bibnamefont{and}
  \bibinfo{author}{\bibfnamefont{C.~A.} \bibnamefont{Bertulani}},
  \bibinfo{journal}{Phys. Rev.} \textbf{\bibinfo{volume}{C 81}},
  \bibinfo{pages}{031302} (\bibinfo{year}{2010}{\natexlab{b}}).

\bibitem[{\citenamefont{Lu et~al.}(2013)\citenamefont{Lu, Sun, and
  Long}}]{Lu2013PRC87.034311}
\bibinfo{author}{\bibfnamefont{X.~L.} \bibnamefont{Lu}},
  \bibinfo{author}{\bibfnamefont{B.~Y.} \bibnamefont{Sun}}, \bibnamefont{and}
  \bibinfo{author}{\bibfnamefont{W.~H.} \bibnamefont{Long}},
  \bibinfo{journal}{Phys. Rev.} \textbf{\bibinfo{volume}{C 87}},
  \bibinfo{pages}{034311} (\bibinfo{year}{2013}).

\bibitem[{\citenamefont{Long et~al.}(2006{\natexlab{b}})\citenamefont{Long,
  Sagawa, Meng, and Van~Giai}}]{Long2006PLB639.242}
\bibinfo{author}{\bibfnamefont{W.~H.} \bibnamefont{Long}},
  \bibinfo{author}{\bibfnamefont{H.}~\bibnamefont{Sagawa}},
  \bibinfo{author}{\bibfnamefont{J.}~\bibnamefont{Meng}}, \bibnamefont{and}
  \bibinfo{author}{\bibfnamefont{N.}~\bibnamefont{Van~Giai}},
  \bibinfo{journal}{Phys. Lett. B} \textbf{\bibinfo{volume}{639}},
  \bibinfo{pages}{242} (\bibinfo{year}{2006}{\natexlab{b}}).

\bibitem[{\citenamefont{Liang et~al.}(2010)\citenamefont{Liang, Long, Meng, and
  Giai}}]{Liang2010EPJA44.119}
\bibinfo{author}{\bibfnamefont{H.}~\bibnamefont{Liang}},
  \bibinfo{author}{\bibfnamefont{W.~H.} \bibnamefont{Long}},
  \bibinfo{author}{\bibfnamefont{J.}~\bibnamefont{Meng}}, \bibnamefont{and}
  \bibinfo{author}{\bibfnamefont{N.~V.} \bibnamefont{Giai}},
  \bibinfo{journal}{Eur. Phys. J. A} \textbf{\bibinfo{volume}{44}},
  \bibinfo{pages}{119} (\bibinfo{year}{2010}).

\bibitem[{\citenamefont{Geng et~al.}(2019)\citenamefont{Geng, Li, Long, Niu,
  and Chang}}]{Geng2019PRC100.051301R}
\bibinfo{author}{\bibfnamefont{J.}~\bibnamefont{Geng}},
  \bibinfo{author}{\bibfnamefont{J.~J.} \bibnamefont{Li}},
  \bibinfo{author}{\bibfnamefont{W.~H.} \bibnamefont{Long}},
  \bibinfo{author}{\bibfnamefont{Y.~F.} \bibnamefont{Niu}}, \bibnamefont{and}
  \bibinfo{author}{\bibfnamefont{S.~Y.} \bibnamefont{Chang}},
  \bibinfo{journal}{Phys. Rev. C} \textbf{\bibinfo{volume}{100}},
  \bibinfo{pages}{051301(R)} (\bibinfo{year}{2019}).

\bibitem[{\citenamefont{Liang et~al.}(2008)\citenamefont{Liang, Van~Giai, and
  Meng}}]{Liang2008PRL101.122502}
\bibinfo{author}{\bibfnamefont{H.~Z.} \bibnamefont{Liang}},
  \bibinfo{author}{\bibfnamefont{N.}~\bibnamefont{Van~Giai}}, \bibnamefont{and}
  \bibinfo{author}{\bibfnamefont{J.}~\bibnamefont{Meng}},
  \bibinfo{journal}{Phys. Rev. Lett.} \textbf{\bibinfo{volume}{101}},
  \bibinfo{pages}{122502} (\bibinfo{year}{2008}).

\bibitem[{\citenamefont{Liang et~al.}(2009)\citenamefont{Liang, Van~Giai, and
  Meng}}]{Liang2009PRC79.064316}
\bibinfo{author}{\bibfnamefont{H.~Z.} \bibnamefont{Liang}},
  \bibinfo{author}{\bibfnamefont{N.}~\bibnamefont{Van~Giai}}, \bibnamefont{and}
  \bibinfo{author}{\bibfnamefont{J.}~\bibnamefont{Meng}},
  \bibinfo{journal}{Phys. Rev. C} \textbf{\bibinfo{volume}{79}},
  \bibinfo{pages}{064316} (\bibinfo{year}{2009}).

\bibitem[{\citenamefont{Liang et~al.}(2012)\citenamefont{Liang, Zhao, and
  Meng}}]{Liang2012PRC85.064302}
\bibinfo{author}{\bibfnamefont{H.~Z.} \bibnamefont{Liang}},
  \bibinfo{author}{\bibfnamefont{P.~W.} \bibnamefont{Zhao}}, \bibnamefont{and}
  \bibinfo{author}{\bibfnamefont{J.}~\bibnamefont{Meng}},
  \bibinfo{journal}{Phys. Rev. C} \textbf{\bibinfo{volume}{85}},
  \bibinfo{pages}{064302} (\bibinfo{year}{2012}).

\bibitem[{\citenamefont{Niu et~al.}(2013)\citenamefont{Niu, Niu, Liang, Long,
  Nik\v{s}i\'{c}, Vretenar, and Meng}}]{Niu2013PLB723.172}
\bibinfo{author}{\bibfnamefont{Z.}~\bibnamefont{Niu}},
  \bibinfo{author}{\bibfnamefont{Y.}~\bibnamefont{Niu}},
  \bibinfo{author}{\bibfnamefont{H.}~\bibnamefont{Liang}},
  \bibinfo{author}{\bibfnamefont{W.}~\bibnamefont{Long}},
  \bibinfo{author}{\bibfnamefont{T.}~\bibnamefont{Nik\v{s}i\'{c}}},
  \bibinfo{author}{\bibfnamefont{D.}~\bibnamefont{Vretenar}}, \bibnamefont{and}
  \bibinfo{author}{\bibfnamefont{J.}~\bibnamefont{Meng}},
  \bibinfo{journal}{Phys. Lett. B} \textbf{\bibinfo{volume}{723}},
  \bibinfo{pages}{172 } (\bibinfo{year}{2013}).

\bibitem[{\citenamefont{Niu et~al.}(2017)\citenamefont{Niu, Niu, Liang, Long,
  and Meng}}]{Niu2017PRC95.044301}
\bibinfo{author}{\bibfnamefont{Z.~M.} \bibnamefont{Niu}},
  \bibinfo{author}{\bibfnamefont{Y.~F.} \bibnamefont{Niu}},
  \bibinfo{author}{\bibfnamefont{H.~Z.} \bibnamefont{Liang}},
  \bibinfo{author}{\bibfnamefont{W.~H.} \bibnamefont{Long}}, \bibnamefont{and}
  \bibinfo{author}{\bibfnamefont{J.}~\bibnamefont{Meng}},
  \bibinfo{journal}{Phys. Rev. C} \textbf{\bibinfo{volume}{95}},
  \bibinfo{pages}{044031} (\bibinfo{year}{2017}).

\bibitem[{\citenamefont{Jiang et~al.}(2015{\natexlab{a}})\citenamefont{Jiang,
  Yang, Dong, and Long}}]{Jiang2015PRC91.025802}
\bibinfo{author}{\bibfnamefont{L.~J.} \bibnamefont{Jiang}},
  \bibinfo{author}{\bibfnamefont{S.}~\bibnamefont{Yang}},
  \bibinfo{author}{\bibfnamefont{J.~M.} \bibnamefont{Dong}}, \bibnamefont{and}
  \bibinfo{author}{\bibfnamefont{W.~H.} \bibnamefont{Long}},
  \bibinfo{journal}{Phys. Rev. C} \textbf{\bibinfo{volume}{91}},
  \bibinfo{pages}{025802} (\bibinfo{year}{2015}{\natexlab{a}}).

\bibitem[{\citenamefont{Jiang et~al.}(2015{\natexlab{b}})\citenamefont{Jiang,
  Yang, Sun, Long, and Gu}}]{Jiang2015PRC91.034326}
\bibinfo{author}{\bibfnamefont{L.~J.} \bibnamefont{Jiang}},
  \bibinfo{author}{\bibfnamefont{S.}~\bibnamefont{Yang}},
  \bibinfo{author}{\bibfnamefont{B.~Y.} \bibnamefont{Sun}},
  \bibinfo{author}{\bibfnamefont{W.~H.} \bibnamefont{Long}}, \bibnamefont{and}
  \bibinfo{author}{\bibfnamefont{H.~Q.} \bibnamefont{Gu}},
  \bibinfo{journal}{Phys. Rev. C} \textbf{\bibinfo{volume}{91}},
  \bibinfo{pages}{034326} (\bibinfo{year}{2015}{\natexlab{b}}).

\bibitem[{\citenamefont{Zong and Sun}(2018)}]{Zong2018CPC42.024101}
\bibinfo{author}{\bibfnamefont{Y.-Y.} \bibnamefont{Zong}} \bibnamefont{and}
  \bibinfo{author}{\bibfnamefont{B.-Y.} \bibnamefont{Sun}},
  \bibinfo{journal}{Chin. Phys. C} \textbf{\bibinfo{volume}{42}},
  \bibinfo{pages}{024101} (\bibinfo{year}{2018}).

\bibitem[{\citenamefont{Bai et~al.}(2009{\natexlab{a}})\citenamefont{Bai,
  Sagawa, Zhang, Zhang, Col\`o, and Xu}}]{Bai2009Phys.Lett.B675.28}
\bibinfo{author}{\bibfnamefont{C.~L.} \bibnamefont{Bai}},
  \bibinfo{author}{\bibfnamefont{H.}~\bibnamefont{Sagawa}},
  \bibinfo{author}{\bibfnamefont{H.~Q.} \bibnamefont{Zhang}},
  \bibinfo{author}{\bibfnamefont{X.~Z.} \bibnamefont{Zhang}},
  \bibinfo{author}{\bibfnamefont{G.}~\bibnamefont{Col\`o}}, \bibnamefont{and}
  \bibinfo{author}{\bibfnamefont{F.~R.} \bibnamefont{Xu}},
  \bibinfo{journal}{Phys. Lett. B} \textbf{\bibinfo{volume}{675}},
  \bibinfo{pages}{28} (\bibinfo{year}{2009}{\natexlab{a}}).

\bibitem[{\citenamefont{Bai et~al.}(2009{\natexlab{b}})\citenamefont{Bai,
  Zhang, Zhang, Xu, Sagawa, and Col\`o}}]{Bai2009Phys.Rev.C79.041301}
\bibinfo{author}{\bibfnamefont{C.~L.} \bibnamefont{Bai}},
  \bibinfo{author}{\bibfnamefont{H.~Q.} \bibnamefont{Zhang}},
  \bibinfo{author}{\bibfnamefont{X.~Z.} \bibnamefont{Zhang}},
  \bibinfo{author}{\bibfnamefont{F.~R.} \bibnamefont{Xu}},
  \bibinfo{author}{\bibfnamefont{H.}~\bibnamefont{Sagawa}}, \bibnamefont{and}
  \bibinfo{author}{\bibfnamefont{G.}~\bibnamefont{Col\`o}},
  \bibinfo{journal}{Phys. Rev. C} \textbf{\bibinfo{volume}{79}},
  \bibinfo{pages}{041301} (\bibinfo{year}{2009}{\natexlab{b}}).

\bibitem[{\citenamefont{Bai et~al.}(2010)\citenamefont{Bai, Zhang, Sagawa,
  Zhang, Col\`o, and Xu}}]{Bai2010Phys.Rev.Lett.105.072501}
\bibinfo{author}{\bibfnamefont{C.~L.} \bibnamefont{Bai}},
  \bibinfo{author}{\bibfnamefont{H.~Q.} \bibnamefont{Zhang}},
  \bibinfo{author}{\bibfnamefont{H.}~\bibnamefont{Sagawa}},
  \bibinfo{author}{\bibfnamefont{X.~Z.} \bibnamefont{Zhang}},
  \bibinfo{author}{\bibfnamefont{G.}~\bibnamefont{Col\`o}}, \bibnamefont{and}
  \bibinfo{author}{\bibfnamefont{F.~R.} \bibnamefont{Xu}},
  \bibinfo{journal}{Phys. Rev. Lett.} \textbf{\bibinfo{volume}{105}},
  \bibinfo{pages}{072501} (\bibinfo{year}{2010}).

\bibitem[{\citenamefont{Bai et~al.}(2011{\natexlab{a}})\citenamefont{Bai,
  Zhang, Sagawa, Zhang, Col\`o, and Xu}}]{Bai2011Phys.Rev.C83.054316}
\bibinfo{author}{\bibfnamefont{C.~L.} \bibnamefont{Bai}},
  \bibinfo{author}{\bibfnamefont{H.~Q.} \bibnamefont{Zhang}},
  \bibinfo{author}{\bibfnamefont{H.}~\bibnamefont{Sagawa}},
  \bibinfo{author}{\bibfnamefont{X.~Z.} \bibnamefont{Zhang}},
  \bibinfo{author}{\bibfnamefont{G.}~\bibnamefont{Col\`o}}, \bibnamefont{and}
  \bibinfo{author}{\bibfnamefont{F.~R.} \bibnamefont{Xu}},
  \bibinfo{journal}{Phys. Rev. C} \textbf{\bibinfo{volume}{83}},
  \bibinfo{pages}{054316} (\bibinfo{year}{2011}{\natexlab{a}}).

\bibitem[{\citenamefont{Bai et~al.}(2011{\natexlab{b}})\citenamefont{Bai,
  Sagawa, Col\`o, Zhang, and Zhang}}]{Bai2011Phys.Rev.C84.044329}
\bibinfo{author}{\bibfnamefont{C.~L.} \bibnamefont{Bai}},
  \bibinfo{author}{\bibfnamefont{H.}~\bibnamefont{Sagawa}},
  \bibinfo{author}{\bibfnamefont{G.}~\bibnamefont{Col\`o}},
  \bibinfo{author}{\bibfnamefont{H.~Q.} \bibnamefont{Zhang}}, \bibnamefont{and}
  \bibinfo{author}{\bibfnamefont{X.~Z.} \bibnamefont{Zhang}},
  \bibinfo{journal}{Phys. Rev. C} \textbf{\bibinfo{volume}{84}},
  \bibinfo{pages}{044329} (\bibinfo{year}{2011}{\natexlab{b}}).

\bibitem[{\citenamefont{Pannert et~al.}(1987)\citenamefont{Pannert, Ring, and
  Boguta}}]{Pannert1987PRL59.2420}
\bibinfo{author}{\bibfnamefont{W.}~\bibnamefont{Pannert}},
  \bibinfo{author}{\bibfnamefont{P.}~\bibnamefont{Ring}}, \bibnamefont{and}
  \bibinfo{author}{\bibfnamefont{J.}~\bibnamefont{Boguta}},
  \bibinfo{journal}{Phys. Rev. Lett.} \textbf{\bibinfo{volume}{59}},
  \bibinfo{pages}{2420} (\bibinfo{year}{1987}).

\bibitem[{\citenamefont{Price and Walker}(1987)}]{Price1987PRC36.354}
\bibinfo{author}{\bibfnamefont{C.~E.} \bibnamefont{Price}} \bibnamefont{and}
  \bibinfo{author}{\bibfnamefont{G.~E.} \bibnamefont{Walker}},
  \bibinfo{journal}{Phys. Rev. C} \textbf{\bibinfo{volume}{36}},
  \bibinfo{pages}{354} (\bibinfo{year}{1987}).

\bibitem[{\citenamefont{Gambhir et~al.}(1990)\citenamefont{Gambhir, Ring, and
  Thimet}}]{Gambhir1990AP198.132}
\bibinfo{author}{\bibfnamefont{Y.~K.} \bibnamefont{Gambhir}},
  \bibinfo{author}{\bibfnamefont{P.}~\bibnamefont{Ring}}, \bibnamefont{and}
  \bibinfo{author}{\bibfnamefont{A.}~\bibnamefont{Thimet}},
  \bibinfo{journal}{Ann. Phys.} \textbf{\bibinfo{volume}{198}},
  \bibinfo{pages}{132} (\bibinfo{year}{1990}).

\bibitem[{\citenamefont{Zhou et~al.}(2003)\citenamefont{Zhou, Meng, and
  Ring}}]{Zhou2003PRC68.034323}
\bibinfo{author}{\bibfnamefont{S.-G.} \bibnamefont{Zhou}},
  \bibinfo{author}{\bibfnamefont{J.}~\bibnamefont{Meng}}, \bibnamefont{and}
  \bibinfo{author}{\bibfnamefont{P.}~\bibnamefont{Ring}},
  \bibinfo{journal}{Phys. Rev. C} \textbf{\bibinfo{volume}{68}},
  \bibinfo{pages}{034323} (\bibinfo{year}{2003}).

\bibitem[{\citenamefont{Li et~al.}(2012)\citenamefont{Li, Meng, Ring, Zhao, and
  Zhou}}]{Li2012PRC85.024312}
\bibinfo{author}{\bibfnamefont{L.~L.} \bibnamefont{Li}},
  \bibinfo{author}{\bibfnamefont{J.}~\bibnamefont{Meng}},
  \bibinfo{author}{\bibfnamefont{P.}~\bibnamefont{Ring}},
  \bibinfo{author}{\bibfnamefont{E.-G.} \bibnamefont{Zhao}}, \bibnamefont{and}
  \bibinfo{author}{\bibfnamefont{S.-G.} \bibnamefont{Zhou}},
  \bibinfo{journal}{Phys. Rev. C} \textbf{\bibinfo{volume}{85}},
  \bibinfo{pages}{024312} (\bibinfo{year}{2012}).

\bibitem[{\citenamefont{Chen et~al.}(2012)\citenamefont{Chen, Li, Liang, and
  Meng}}]{Chen2012PRC85.067301}
\bibinfo{author}{\bibfnamefont{Y.}~\bibnamefont{Chen}},
  \bibinfo{author}{\bibfnamefont{L.~L.} \bibnamefont{Li}},
  \bibinfo{author}{\bibfnamefont{H.~Z.} \bibnamefont{Liang}}, \bibnamefont{and}
  \bibinfo{author}{\bibfnamefont{J.}~\bibnamefont{Meng}},
  \bibinfo{journal}{Phys. Rev. C} \textbf{\bibinfo{volume}{85}},
  \bibinfo{pages}{067301} (\bibinfo{year}{2012}).

\bibitem[{\citenamefont{Ebran et~al.}(2011)\citenamefont{Ebran, Khan, Pe\~na
  Arteaga, and Vretenar}}]{Ebran2011PRC83.064323}
\bibinfo{author}{\bibfnamefont{J.-P.} \bibnamefont{Ebran}},
  \bibinfo{author}{\bibfnamefont{E.}~\bibnamefont{Khan}},
  \bibinfo{author}{\bibfnamefont{D.}~\bibnamefont{Pe\~na Arteaga}},
  \bibnamefont{and} \bibinfo{author}{\bibfnamefont{D.}~\bibnamefont{Vretenar}},
  \bibinfo{journal}{Phys. Rev.} \textbf{\bibinfo{volume}{C 83}},
  \bibinfo{pages}{064323} (\bibinfo{year}{2011}).

\bibitem[{\citenamefont{Geng et~al.}(2023)\citenamefont{Geng, Niu, and
  Long}}]{Geng2023CPC47.044102}
\bibinfo{author}{\bibfnamefont{J.}~\bibnamefont{Geng}},
  \bibinfo{author}{\bibfnamefont{Y.~F.} \bibnamefont{Niu}}, \bibnamefont{and}
  \bibinfo{author}{\bibfnamefont{W.~H.} \bibnamefont{Long}},
  \bibinfo{journal}{Chin. Phys. C} \textbf{\bibinfo{volume}{47}},
  \bibinfo{pages}{044102} (\bibinfo{year}{2023}).

\bibitem[{\citenamefont{Berger et~al.}(1984)\citenamefont{Berger, Girod, and
  Gogny}}]{Berger1984NPA428.23}
\bibinfo{author}{\bibfnamefont{J.~F.} \bibnamefont{Berger}},
  \bibinfo{author}{\bibfnamefont{M.}~\bibnamefont{Girod}}, \bibnamefont{and}
  \bibinfo{author}{\bibfnamefont{D.}~\bibnamefont{Gogny}},
  \bibinfo{journal}{Nucl. Phys. A} \textbf{\bibinfo{volume}{428}},
  \bibinfo{pages}{23} (\bibinfo{year}{1984}).

\bibitem[{\citenamefont{Typel and Wolter}(1999)}]{Typel1999NPA656.331}
\bibinfo{author}{\bibfnamefont{S.}~\bibnamefont{Typel}} \bibnamefont{and}
  \bibinfo{author}{\bibfnamefont{H.~H.} \bibnamefont{Wolter}},
  \bibinfo{journal}{Nucl. Phys. A} \textbf{\bibinfo{volume}{656}},
  \bibinfo{pages}{331} (\bibinfo{year}{1999}).

\bibitem[{\citenamefont{Varshalovich et~al.}(1988)\citenamefont{Varshalovich,
  Moskalev, and Khersonskii}}]{Varshalovich1988.165}
\bibinfo{author}{\bibfnamefont{D.~A.} \bibnamefont{Varshalovich}},
  \bibinfo{author}{\bibfnamefont{A.~N.} \bibnamefont{Moskalev}},
  \bibnamefont{and} \bibinfo{author}{\bibfnamefont{V.~K.}
  \bibnamefont{Khersonskii}}, \emph{\bibinfo{title}{{Q}uantum theory of angular
  momentum}} (\bibinfo{publisher}{World Scientific},
  \bibinfo{address}{Singapore}, \bibinfo{year}{1988}).

\bibitem[{\citenamefont{Lalazissis et~al.}(2005)\citenamefont{Lalazissis,
  Nik$\check{\rm s}$i$\acute{\rm c}$, Vretenar, and
  Ring}}]{Lalazissis2005PRC71.024312}
\bibinfo{author}{\bibfnamefont{G.~A.} \bibnamefont{Lalazissis}},
  \bibinfo{author}{\bibfnamefont{T.}~\bibnamefont{Nik$\check{\rm
  s}$i$\acute{\rm c}$}},
  \bibinfo{author}{\bibfnamefont{D.}~\bibnamefont{Vretenar}}, \bibnamefont{and}
  \bibinfo{author}{\bibfnamefont{P.}~\bibnamefont{Ring}},
  \bibinfo{journal}{Phys. Rev. C} \textbf{\bibinfo{volume}{71}},
  \bibinfo{pages}{024312} (\bibinfo{year}{2005}).

\bibitem[{\citenamefont{Wang et~al.}(2017)\citenamefont{Wang, Audi, Kondev,
  Huang, Naimi, and Xu}}]{Wang2017CPC41.030003}
\bibinfo{author}{\bibfnamefont{M.}~\bibnamefont{Wang}},
  \bibinfo{author}{\bibfnamefont{G.}~\bibnamefont{Audi}},
  \bibinfo{author}{\bibfnamefont{F.~G.} \bibnamefont{Kondev}},
  \bibinfo{author}{\bibfnamefont{W.~J.} \bibnamefont{Huang}},
  \bibinfo{author}{\bibfnamefont{S.}~\bibnamefont{Naimi}}, \bibnamefont{and}
  \bibinfo{author}{\bibfnamefont{X.}~\bibnamefont{Xu}},
  \bibinfo{journal}{Chinese Physics C} \textbf{\bibinfo{volume}{41}}
  (\bibinfo{year}{2017}).

\bibitem[{\citenamefont{Angeli and Marinova}(2013)}]{Angeli2013ADNDT99}
\bibinfo{author}{\bibfnamefont{I.}~\bibnamefont{Angeli}} \bibnamefont{and}
  \bibinfo{author}{\bibfnamefont{K.}~\bibnamefont{Marinova}},
  \bibinfo{journal}{Atomic Data and Nuclear Data Tables}
  \textbf{\bibinfo{volume}{99}}, \bibinfo{pages}{69} (\bibinfo{year}{2013}).

\bibitem[{\citenamefont{Otsuka et~al.}(2005)\citenamefont{Otsuka, Suzuki,
  Fujimoto, Grawe, and Akaishi}}]{Otsuka2005PRL95.232502}
\bibinfo{author}{\bibfnamefont{T.}~\bibnamefont{Otsuka}},
  \bibinfo{author}{\bibfnamefont{T.}~\bibnamefont{Suzuki}},
  \bibinfo{author}{\bibfnamefont{R.}~\bibnamefont{Fujimoto}},
  \bibinfo{author}{\bibfnamefont{H.}~\bibnamefont{Grawe}}, \bibnamefont{and}
  \bibinfo{author}{\bibfnamefont{Y.}~\bibnamefont{Akaishi}},
  \bibinfo{journal}{Phys. Rev. Lett.} \textbf{\bibinfo{volume}{95}},
  \bibinfo{pages}{232502} (\bibinfo{year}{2005}).

\bibitem[{\citenamefont{Koepf and Ring}(1991)}]{Koepf1991ZPA339.81}
\bibinfo{author}{\bibfnamefont{W.}~\bibnamefont{Koepf}} \bibnamefont{and}
  \bibinfo{author}{\bibfnamefont{P.}~\bibnamefont{Ring}}, \bibinfo{journal}{Z.
  Phys. A} \textbf{\bibinfo{volume}{339}}, \bibinfo{pages}{81}
  (\bibinfo{year}{1991}).

\end{thebibliography}

\appendix

\begin{widetext}
\section{Energy functionals and self-energies in various coupling channels}

\subsection{Compounded symbols}

In this work, since the expansion of the wave functions $\psi_{\nu m}$ is carried on the spherical Dirac Woods-Saxon (DWS) base $\psi_{n\kappa m}$, the integrations in the energy functional is performed with respect to the angle variables $\svec\Omega=(\vartheta,\varphi)$. Such integration contains the Harmonic functions given by the expansions of the propagators (\ref{eq:expansionD1}), the coupling strengths (\ref{eq:ExpansionP}), and the couplings between the spherical spinors,
\begin{align} \label{eq:inter onega}
\sqrt{2\pi}\int d\svec \Omega Y_{\lambda_d\mu_d} (\svec \Omega) Y_{\lambda_y-\mu_y} (\svec \Omega)  Y_{\lambda_p0} (\svec \Omega) =& \frac{1}{\sqrt{2}} \hat{\lambda}_d \hat{\lambda}_y \hat{\lambda}_p^{-1} C_{\lambda_d0 \lambda_y0}^{\lambda_p0} C_{\lambda_d\mu \lambda_y -\mu}^{\lambda_p 0},
\end{align}
where $\mu = \mu_d = \mu_y$, and $(\lambda_d, \mu_d)$, $(\lambda_y, \mu_y)$ and $\lambda_p$ denote the terms due to the couplings between the spinors, the expansions of the propagators and the coupling strengths, respectively. As an abbreviation, the symbol $\Theta$ is introduced to denote the above integration,
\begin{align}
  \Theta_{\lambda_d\lambda_p}^{\lambda_y\mu} \equiv & (-1)^\mu\frac{1}{\sqrt{2}} \hat{\lambda}_d \hat{\lambda}_y \hat{\lambda}_p^{-1} C_{\lambda_d0 \lambda_y0}^{\lambda_p0} C_{\lambda_d\mu \lambda_y -\mu}^{\lambda_p 0}. \label{eq:Theta}
\end{align}
For the couplings between spherical spinors, we introduce the following symbols $\scr D$ ($\bar{\scr D}$) and $\cals Q$ ($\bar{\cals Q}$) as,
\begin{align}
  \scr D_{\kappa_1m_1;\kappa_2m_2}^{\lambda\mu} =&\frac{1}{\sqrt 2} \hat j_1\hat j_2 \hat \lambda^{-1}C_{j_1\ff2 j_2-\ff2}^{L0} C_{j_1-m_1j_2m_2}^{\lambda\mu}, \label{eq:Dsymbol}\\ \bar{\scr D}_{\kappa_1m_1;\kappa_2m_2}^{\lambda \bar\mu} = & (-1)^{\kappa_1 + \pi_1 } \scr D_{\kappa_1-m_1;\kappa_2m_2}^{\lambda \bar\mu}, \\
  \cals Q_{\kappa_1m_1;\kappa_2m_2}^{\lambda\mu\sigma} \equiv & (-1)^{j_1+l_1-\ff2}\sqrt{3}\hat j_1\hat j_2\hat l_1\hat l_2\sum_{J} C_{l_10l_20}^{\lambda0}C_{\lambda\mu1\sigma}^{JM}  \begin{Bmatrix} j_1 & j_2 & J \\ l_1 & l_2 & \lambda \\ \ff2 & \ff2 & 1 \end{Bmatrix} C_{j_1-m_1j_2m_2}^{JM},\\
  \bar{\cals Q}_{\kappa_1m_1;\kappa_2m_2}^{\lambda\bar\mu\sigma} \equiv &(-1)^{\kappa_1+ \pi_1} \cals Q_{\kappa_1-m_1;\kappa_2m_2}^{\lambda\bar\mu\sigma}.
\end{align}
Where $\pi_1=0$ for $\kappa_1$ corresponding to even parity, and $\pi_1=1$ for $\kappa_1$ corresponding to odd parity. In the symbols $\scr D$ and $\bar{\scr D}$, $\mu = m_2-m_1$ and $\bar\mu = m_2+m_1$, and for $\cals Q$ and $\bar{\cals Q}$, one can find $\mu+\sigma = m_2-m_1$ and $\bar\mu+\sigma = m_2+m_1$. With the help of these symbols, the couplings of the spherical spinors can be simply expressed as,
\begin{align}
  \Omega_{j_1m_1}^{l_{1}\dag}\Omega_{j_2m_2}^{l_{2}} = & \frac{(-1)^{m_1+\ff2}}{\sqrt{2\pi}}\sum_{\lambda_d} \scr D_{\kappa_1 m_1;\kappa_2m_2}^{\lambda_d\mu_d} Y_{\lambda_d\mu_d},&
  \Omega_{j_1m_1}^{l_1}\svec\sigma \Omega_{j_2m_2}^{l_2} = & \frac{(-1)^{m_1-\ff2}}{\sqrt{2\pi}}\sum_{\lambda\mu\sigma} \cals Q_{\kappa_1m_1;\kappa_2m_2}^{\lambda_d\mu_d\sigma} Y_{\lambda_d\mu_d} \svec e_\sigma,
\end{align}
with $\svec e_\sigma$ being the covariant spherical base vector with $\sigma = -1, 0, +1$.

\subsection{Energy functionals and self-energies from the Hartree terms} \label{app:Hartree}
For the $\sigma$-S coupling and the time component of $\omega$-V one, the self-energies from the Hartree terms can be expressed as,
\begin{subequations}\label{eq:sv-self}
\begin{align}
  \Sigma_{S,\sigma}^{\lambda_d}(r) = & -2\pi \sum_{\lambda_p} g_\sigma^{\lambda_p}(r) \int r'^2 dr'\sum_{\lambda_y} \Theta_{\lambda_d\lambda_p}^{\lambda_y 0}R_{\lambda_y\lambda_y}^{\sigma}(r,r')  \sum_{\lambda_p'\lambda_d'} \Theta_{\lambda_d'\lambda_p'}^{\lambda_y 0} g_\sigma^{\lambda_p'}(r')\rho_s^{\lambda_d'}(r'),\\
  \Sigma_{0,\omega}^{\lambda_d}(r) = & +2\pi \sum_{\lambda_p} g_\omega^{\lambda_p}(r) \int r'^2 dr'\sum_{\lambda_y} \Theta_{\lambda_d\lambda_p}^{\lambda_y 0}R_{\lambda_y\lambda_y}^{\omega}(r,r')  \sum_{\lambda_p'\lambda_d'} \Theta_{\lambda_d'\lambda_p'}^{\lambda_y 0} g_\omega^{\lambda_p'}(r')\rho_b^{\lambda_d'}(r').
\end{align}
\end{subequations}
Thus, the relevant energy functionals can be derived as,
\begin{subequations}\label{eq:HartreeE}
\begin{align}
  E_\sigma^D = & + \frac{2\pi}{2} \int r^2 dr \sum_{\lambda_d} \rho_s^{\lambda_d}(r) \Sigma_{S,\sigma}^{\lambda_d}(r),&
  E_\omega^D = & + \frac{2\pi}{2} \int r^2 dr \sum_{\lambda_d} \rho_b^{\lambda_d}(r) \Sigma_{0,\omega}^{\lambda_d}(r).
\end{align}
\end{subequations}
For the time component of $\rho$-V coupling, such expressions can be obtained by replacing $g_\omega^{\lambda_p}$ ($g_\omega^{\lambda_p'}$) with $g_\rho^{\lambda_p}$ ($g_\rho^{\lambda_p'}$) and $\rho_b^{\lambda_d}$ ($\rho_b^{\lambda_d'}$) with $\rho_{b,3}^{\lambda_d}$ ($\rho_{b,3}^{\lambda_d'}$), where $\rho_{b,3}^{\lambda_d} = \rho_{b,n}^{\lambda_d} - \rho_{b,p}^{\lambda_d}$. For the Coulomb field ($A$-V coupling), these expressions can be deduced by setting $\lambda_p = \lambda_p' =0$ and replacing the nucleon density $\rho_b$ as the proton one $\rho_p$. For the Hartree terms of the space components of the vector couplings, as well as the $\pi$-PV couplings, the contributions are derived as zero.

Since the coupling strengths $g_\sigma$, $g_\omega$ and $g_\rho$ are density-dependent, the variations of the energy functionals (\ref{eq:HartreeE}) may lead to the additional rearrangement terms as,
\begin{align}
  \Sigma_{R,\sigma}^{D,\lambda_d}(r) = &-2\pi \sum_{\lambda_y} \sum_{\lambda_p\lambda_d} \Theta^{\lambda_y0}_{\lambda_d\lambda_p}\Big[ \frac{\partial g_\sigma^{\lambda_p}}{\partial\rho_b^{\lambda_d}} \rho_s^{\lambda_d}\Big]_{r}\int r'^2dr'  R_{\lambda_y\lambda_y}^\sigma (r, r')\sum_{\lambda_p'\lambda_d'} \Theta^{\lambda_y0}_{\lambda_d'\lambda_p'} \lrs{g_{\sigma}^{\lambda_p'}\rho_{s}^{\lambda_d'}}_{r'},\\
  \Sigma_{R,\omega}^{D,\lambda_d}(r) = &+2\pi\sum_{\lambda_y} \sum_{\lambda_p\lambda_d} \Theta^{\lambda_y0}_{\lambda_d\lambda_p}\Big[ \frac{\partial g_\omega^{\lambda_p}}{\partial\rho_b^{\lambda_d}} \rho_b^{\lambda_d}\Big]_{r} \int r'^2dr' R_{\lambda_y\lambda_y}^\omega (r, r')\sum_{\lambda_p'\lambda_d'} \Theta^{\lambda_y0}_{\lambda_d'\lambda_p'} \lrs{g_{\omega}^{\lambda_p'}\rho_{b}^{\lambda_d'}}_{r'}.
\end{align}

\subsection{Energy functionals and self-energies from the Fock terms}\label{app:Fock}
In this work, we set the angular momentum projection $m$ to be positive for convenience in deriving both the Hartree and Fock terms. Thus, we need to revise the expansion of the spinor $\psi_{\nu m}$. In principle, the time conjugation partner can be set as $\psi_{\nu -m}\equiv \widehat{P}_t \psi_{\nu m}$, where $\widehat{P}_t$ is the time reversal operator. The expansion of $\psi_{\nu  -m}$ is as following,
\begin{align}
\psi_{\nu -m} = &\sum_\kappa (-1)^{j+l_u - m} \psi_{\nu\kappa -m},
\end{align}
where, by definition, $\psi_{\nu\kappa m}$ and $\psi_{\nu\kappa -m}$ share the radial components $\cals G_{i\kappa}$ and $\cals F_{i\kappa}$. For the Hartree terms, it does not bring additional complexity. However, in the context of the Fock terms, it is necessary to exercise caution when dealing with the couplings between the orbits $m$ and $-m'$ ($m,m'>0$). It is of interest to note that the partner $(-m,-m')$ contributes identically to the one $(m, m')$, as well as the partners $(m,-m')$ and $(-m,m')$.

To express the contributions of the Fock terms in compact form, we introduce the symbol $\widehat{\cals D}$ for $\sigma$-S coupling and the time component of the vector ones as,
\begin{align}
  \widehat{\cals D}_{\kappa_1\kappa_2m;\kappa_1'\kappa_2'm'}^{\lambda_p\lambda_p',\lambda_y;++} = & \frac{1}{2}\sum_{\lambda_d\lambda_d'}\Big[\scr D_{\kappa_1 m;\kappa_1'm'}^{\lambda_d\mu } \scr D_{\kappa_2' m';\kappa_2m}^{\lambda_d'\mu }\Theta_{\lambda_d\lambda_p}^{\lambda_y \mu}\Theta_{\lambda_d'\lambda_p'}^{\lambda_y \mu} + \bar{\scr D}_{\kappa_1 m;\kappa_1'm'}^{\lambda_d\bar\mu }\bar{ \scr D}_{\kappa_2' m';\kappa_2m}^{\lambda_d'\bar\mu }\Theta_{\lambda_d\lambda_p}^{\lambda_y \bar\mu}\Theta_{\lambda_d'\lambda_p'}^{\lambda_y\bar \mu}\Big],\\
  \widehat{\cals D}_{\kappa_1\kappa_2m;\kappa_1'\kappa_2'm'}^{\lambda_p\lambda_p',\lambda_y;+-} = & \frac{1}{2}\sum_{\lambda_d\lambda_d'}\Big[\scr D_{\kappa_1 m;\kappa_1'm'}^{\lambda_d\mu } \scr D_{-\kappa_2' m';-\kappa_2m}^{\lambda_d'\mu }\Theta_{\lambda_d\lambda_p}^{\lambda_y \mu}\Theta_{\lambda_d'\lambda_p'}^{\lambda_y \mu} + \bar{\scr D}_{\kappa_1 m;\kappa_1'm'}^{\lambda_d\bar\mu }\bar{ \scr D}_{-\kappa_2' m';-\kappa_2m}^{\lambda_d'\bar\mu }\Theta_{\lambda_d\lambda_p}^{\lambda_y \bar\mu}\Theta_{\lambda_d'\lambda_p'}^{\lambda_y\bar \mu}\Big],\\
  \widehat{\cals D}_{\kappa_1\kappa_2m;\kappa_1'\kappa_2'm'}^{\lambda_p\lambda_p',\lambda_y;-+} = & \frac{1}{2}\sum_{\lambda_d\lambda_d'}\Big[\scr D_{-\kappa_1 m;-\kappa_1'm'}^{\lambda_d\mu } \scr D_{\kappa_2' m';\kappa_2m}^{\lambda_d'\mu }\Theta_{\lambda_d\lambda_p}^{\lambda_y \mu}\Theta_{\lambda_d'\lambda_p'}^{\lambda_y \mu} + \bar{\scr D}_{-\kappa_1 m;-\kappa_1'm'}^{\lambda_d\bar\mu }\bar{ \scr D}_{\kappa_2' m';\kappa_2m}^{\lambda_d'\bar\mu }\Theta_{\lambda_d\lambda_p}^{\lambda_y \bar\mu}\Theta_{\lambda_d'\lambda_p'}^{\lambda_y\bar \mu}\Big],\\
  \widehat{\cals D}_{\kappa_1\kappa_2m;\kappa_1'\kappa_2'm'}^{\lambda_p\lambda_p',\lambda_y;--} = & \frac{1}{2}\sum_{\lambda_d\lambda_d'}\Big[\scr D_{-\kappa_1 m;-\kappa_1'm'}^{\lambda_d\mu } \scr D_{-\kappa_2' m';-\kappa_2m}^{\lambda_d'\mu }\Theta_{\lambda_d\lambda_p}^{\lambda_y \mu}\Theta_{\lambda_d'\lambda_p'}^{\lambda_y \mu} + \bar{\scr D}_{-\kappa_1 m;-\kappa_1'm'}^{\lambda_d\bar\mu }\bar{ \scr D}_{-\kappa_2' m';-\kappa_2m}^{\lambda_d'\bar\mu }\Theta_{\lambda_d\lambda_p}^{\lambda_y \bar\mu}\Theta_{\lambda_d'\lambda_p'}^{\lambda_y\bar \mu}\Big].
\end{align}
It should be noted that the terms containing the symbols $\scr D$ correspond to the contributions from the partners $(m,m')$ and $(-m,-m')$, and those containing the symbols $\bar{\scr D}$ correspond to the contributions from the partners $(m,-m')$ and $(-m,m')$. Consequently, the non-local self-energies of the $\sigma$-S coupling can be expressed as follows,
\begin{subequations}\label{eq:XY_sigma}
\begin{align}
  Y_{G, m}^{\kappa_1,\kappa_2; \sigma} = & +\frac{1}{2\pi} \sum_{m'} \delta_{\tau\tau'} \sum_{\kappa_1'\kappa_2'} \cals R_{\kappa_1'\kappa_2',m'}^{++}(r,r') \sum_{\lambda_p\lambda_p'} g_\sigma^{\lambda_p}(r) g_\sigma^{\lambda_p'}(r')\sum_{\lambda_y} R_{\lambda_y\lambda_y}^\sigma(r,r')\widehat{\cals D}_{\kappa_1'\kappa_2'm';\kappa_1\kappa_2m}^{\lambda_p\lambda_p',\lambda_y;++},\\
  Y_{F, m}^{\kappa_1,\kappa_2; \sigma} = & -\frac{1}{2\pi} \sum_{m'} \delta_{\tau\tau'} \sum_{\kappa_1'\kappa_2'} \cals R_{\kappa_1'\kappa_2',m'}^{+-}(r,r') \sum_{\lambda_p\lambda_p'} g_\sigma^{\lambda_p}(r) g_\sigma^{\lambda_p'}(r')\sum_{\lambda_y} R_{\lambda_y\lambda_y}^\sigma(r,r')\widehat{\cals D}_{\kappa_1'\kappa_2'm';\kappa_1\kappa_2m}^{\lambda_p\lambda_p',\lambda_y;+-},\\
  X_{G, m}^{\kappa_1,\kappa_2; \sigma} = & -\frac{1}{2\pi} \sum_{m'} \delta_{\tau\tau'} \sum_{\kappa_1'\kappa_2'} \cals R_{\kappa_1'\kappa_2',m'}^{-+}(r,r') \sum_{\lambda_p\lambda_p'} g_\sigma^{\lambda_p}(r) g_\sigma^{\lambda_p'}(r')\sum_{\lambda_y} R_{\lambda_y\lambda_y}^\sigma(r,r')\widehat{\cals D}_{\kappa_1'\kappa_2'm';\kappa_1\kappa_2m}^{\lambda_p\lambda_p',\lambda_y;-+},\\
  X_{F, m}^{\kappa_1,\kappa_2; \sigma} = & +\frac{1}{2\pi} \sum_{m'} \delta_{\tau\tau'} \sum_{\kappa_1'\kappa_2'} \cals R_{\kappa_1'\kappa_2',m'}^{--}(r,r') \sum_{\lambda_p\lambda_p'} g_\sigma^{\lambda_p}(r) g_\sigma^{\lambda_p'}(r')\sum_{\lambda_y} R_{\lambda_y\lambda_y}^\sigma(r,r')\widehat{\cals D}_{\kappa_1'\kappa_2'm';\kappa_1\kappa_2m}^{\lambda_p\lambda_p',\lambda_y;--},
\end{align}
\end{subequations}
where the sum over $\nu'$ has been absorbed into the non-local densities $\cals R$, and the factor $\delta_{\tau\tau'}$ indicates that the orbits $ m$ and $ m'$ should be both neutron or proton ones. In terms of the non-local self-energies, the energy functional of the $\sigma$-S coupling reads as,
\begin{align}
  E_{\sigma}^E =& \ff2\int drdr' \sum_{i} v_i^2\sum_{\kappa_1\kappa_2}\begin{pmatrix} \cals G_{i\kappa_1} & \cals F_{i\kappa_1} \end{pmatrix}_r\begin{pmatrix} Y_{G, m}^{\kappa_1,\kappa_2; \sigma} & Y_{F, m}^{\kappa_1,\kappa_2; \sigma}  \\[0.5em] X_{G, m}^{\kappa_1,\kappa_2; \sigma} & X_{F, m}^{\kappa_1,\kappa_2; \sigma} \end{pmatrix}_{r,r'}\begin{pmatrix} \cals G_{i\kappa_2} \\[0.5em] \cals F_{i\kappa_2} \end{pmatrix}_{r'},
\end{align}
where $v_i^2$ ($\in[0,2]$) is the occupation number of the orbit $(\nu m)$. For the rearrangement terms, the contribution to the self-energy $\Sigma_0^{\lambda_d}$ can be similarly expressed as,
\begin{align}
  \Sigma_{R,\sigma}^{E,\lambda_d} = & \int dr'\sum_i v_i^2\sum_{\kappa_1\kappa_2}\begin{pmatrix} \cals G_{i\kappa_1} & \cals F_{i\kappa_1} \end{pmatrix}_r\begin{pmatrix} P_{G, m,\lambda_d}^{\kappa_1,\kappa_2; \sigma} & P_{F, m,\lambda_d}^{\kappa_1,\kappa_2; \sigma}  \\[0.5em] Q_{G, m,\lambda_d}^{\kappa_1,\kappa_2; \sigma} & Q_{F, m,\lambda_d}^{\kappa_1,\kappa_2; \sigma} \end{pmatrix}_{r,r'}\begin{pmatrix} \cals G_{i\kappa_2} \\[0.5em] \cals F_{i\kappa_2} \end{pmatrix}_{r'},
\end{align}
where the terms $P$ and $Q$ read as,
\begin{subequations}\label{eq:PQ_sigma}
\begin{align}
  P_{G, m,\lambda_d}^{\kappa_1,\kappa_2; \sigma} = & +\frac{1}{2\pi} \sum_{\pi'm'} \delta_{\tau\tau'} \sum_{\kappa_1'\kappa_2'} \cals R_{\kappa_1'\kappa_2',m'}^{++}(r,r') \sum_{\lambda_p\lambda_p'} \frac{\partial g_\sigma^{\lambda_p}(r) }{\partial\rho_b^{\lambda_d}(r)} g_\sigma^{\lambda_p'}(r')\sum_{\lambda_y} R_{\lambda_y\lambda_y}^\sigma(r,r')\widehat{\cals D}_{\kappa_1'\kappa_2'm';\kappa_1\kappa_2m}^{\lambda_p\lambda_p',\lambda_y;++},\\
  P_{F, m,\lambda_d}^{\kappa_1,\kappa_2; \sigma} = & -\frac{1}{2\pi} \sum_{\pi'm'} \delta_{\tau\tau'} \sum_{\kappa_1'\kappa_2'} \cals R_{\kappa_1'\kappa_2',m'}^{+-}(r,r') \sum_{\lambda_p\lambda_p'} \frac{\partial g_\sigma^{\lambda_p}(r) }{\partial\rho_b^{\lambda_d}(r)} g_\sigma^{\lambda_p'}(r')\sum_{\lambda_y} R_{\lambda_y\lambda_y}^\sigma(r,r')\widehat{\cals D}_{\kappa_1'\kappa_2'm';\kappa_1\kappa_2m}^{\lambda_p\lambda_p',\lambda_y;+-},\\
  Q_{G, m,\lambda_d}^{\kappa_1,\kappa_2; \sigma} = & -\frac{1}{2\pi} \sum_{\pi'm'} \delta_{\tau\tau'} \sum_{\kappa_1'\kappa_2'} \cals R_{\kappa_1'\kappa_2',m'}^{-+}(r,r') \sum_{\lambda_p\lambda_p'} \frac{\partial g_\sigma^{\lambda_p}(r) }{\partial\rho_b^{\lambda_d}(r)} g_\sigma^{\lambda_p'}(r')\sum_{\lambda_y} R_{\lambda_y\lambda_y}^\sigma(r,r')\widehat{\cals D}_{\kappa_1'\kappa_2'm';\kappa_1\kappa_2m}^{\lambda_p\lambda_p',\lambda_y;-+},\\
  Q_{F, m,\lambda_d}^{\kappa_1,\kappa_2; \sigma} = & +\frac{1}{2\pi} \sum_{\pi'm'} \delta_{\tau\tau'} \sum_{\kappa_1'\kappa_2'} \cals R_{\kappa_1'\kappa_2',m'}^{--}(r,r') \sum_{\lambda_p\lambda_p'} \frac{\partial g_\sigma^{\lambda_p}(r) }{\partial\rho_b^{\lambda_d}(r)} g_\sigma^{\lambda_p'}(r')\sum_{\lambda_y} R_{\lambda_y\lambda_y}^\sigma(r,r')\widehat{\cals D}_{\kappa_1'\kappa_2'm';\kappa_1\kappa_2m}^{\lambda_p\lambda_p',\lambda_y;--}.
\end{align}
\end{subequations}
For the time components of the vector couplings, analogous expressions can be derived by replacing the expansion terms of the propagator $R_{\lambda_y\lambda_y}^\sigma$ and the coupling strengths $g_\sigma^{\lambda_p}$ in Eqs. (\ref{eq:XY_sigma}, \ref{eq:PQ_sigma}) with the corresponding ones, and additionally the plus signs in the $Y_G$, $X_F$, $P_G$ and $Q_F$ terms should be changed as the minus one. It is also important to note that for the isovector $\rho$-V coupling, the factor $\delta_{\tau\tau'}$ should be replaced by $(2-\delta_{\tau\tau'})$. For the Fock terms of the Coulomb interaction, there is no arrangement term and for the non-local self-energies [see Eqs. (\ref{eq:XY_sigma})] only $\lambda_p = \lambda_p'=0$ terms remain.

With regard to the space component of the vector couplings, we take the $\omega$-V coupling as an example, and the others can be similarly deduced. In order to write the expressions in a compact form, we introduce the symbols $\widehat{\cals B}$ as,
\begin{align}
  \widehat{\cals B}_{\kappa_1'\kappa_2'm';\kappa_1\kappa_2m}^{\lambda_p\lambda_p',\lambda_y;++} \equiv & \frac{1}{2}\sum_{\lambda_d\lambda_d'\sigma} \Big[ \cals Q_{-\kappa_1'm', \kappa_1m}^{\lambda_d \mu\sigma}\cals Q_{-\kappa_2'm',\kappa_2m}^{\lambda_d'\mu\sigma} \Theta_{\lambda_d\lambda_p}^{\lambda_y\mu} \Theta_{\lambda_d'\lambda_p'}^{\lambda_y\mu} + \bar{\cals Q}_{-\kappa_1'm', \kappa_1m}^{\lambda_d\bar \mu\sigma}\bar{\cals Q}_{-\kappa_2'm',\kappa_2m}^{\lambda_d'\bar \mu\sigma} \Theta_{\lambda_d\lambda_p}^{\lambda_y\bar \mu} \Theta_{\lambda_d'\lambda_p'}^{\lambda_y\bar \mu}\Big],\\
  \widehat{\cals B}_{\kappa_1'\kappa_2'm';\kappa_1\kappa_2m}^{\lambda_p\lambda_p',\lambda_y;+-} \equiv & \frac{1}{2}\sum_{\lambda_d\lambda_d'\sigma} \Big[ \cals Q_{-\kappa_1'm', \kappa_1m}^{\lambda_d \mu\sigma}\cals Q_{\kappa_2'm',-\kappa_2m}^{\lambda_d'\mu\sigma} \Theta_{\lambda_d\lambda_p}^{\lambda_y\mu} \Theta_{\lambda_d'\lambda_p'}^{\lambda_y\mu} + \bar{\cals Q}_{-\kappa_1'm', \kappa_1m}^{\lambda_d\bar \mu\sigma}\bar{\cals Q}_{\kappa_2'm',-\kappa_2m}^{\lambda_d'\bar \mu\sigma} \Theta_{\lambda_d\lambda_p}^{\lambda_y\bar \mu} \Theta_{\lambda_d'\lambda_p'}^{\lambda_y\bar \mu}\Big],\\
  \widehat{\cals B}_{\kappa_1'\kappa_2'm';\kappa_1\kappa_2m}^{\lambda_p\lambda_p',\lambda_y;-+} \equiv & \frac{1}{2}\sum_{\lambda_d\lambda_d'\sigma} \Big[ \cals Q_{\kappa_1'm', -\kappa_1m}^{\lambda_d \mu\sigma}\cals Q_{-\kappa_2'm',\kappa_2m}^{\lambda_d'\mu\sigma} \Theta_{\lambda_d\lambda_p}^{\lambda_y\mu} \Theta_{\lambda_d'\lambda_p'}^{\lambda_y\mu} + \bar{\cals Q}_{\kappa_1'm', -\kappa_1m}^{\lambda_d\bar \mu\sigma}\bar{\cals Q}_{-\kappa_2'm',\kappa_2m}^{\lambda_d'\bar \mu\sigma} \Theta_{\lambda_d\lambda_p}^{\lambda_y\bar \mu} \Theta_{\lambda_d'\lambda_p'}^{\lambda_y\bar \mu}\Big],\\
  \widehat{\cals B}_{\kappa_1'\kappa_2'm';\kappa_1\kappa_2m}^{\lambda_p\lambda_p',\lambda_y;--} \equiv & \frac{1}{2}\sum_{\lambda_d\lambda_d'\sigma} \Big[ \cals Q_{\kappa_1'm', -\kappa_1m}^{\lambda_d \mu\sigma}\cals Q_{\kappa_2'm',-\kappa_2m}^{\lambda_d'\mu\sigma} \Theta_{\lambda_d\lambda_p}^{\lambda_y\mu} \Theta_{\lambda_d'\lambda_p'}^{\lambda_y\mu} + \bar{\cals Q}_{\kappa_1'm', -\kappa_1m}^{\lambda_d\bar \mu\sigma}\bar{\cals Q}_{\kappa_2'm',-\kappa_2m}^{\lambda_d'\bar \mu\sigma} \Theta_{\lambda_d\lambda_p}^{\lambda_y\bar \mu} \Theta_{\lambda_d'\lambda_p'}^{\lambda_y\bar \mu}\Big].
\end{align}
Thus, the non-local self-energies can be expressed as,
\begin{align}
  Y_{G, m}^{\kappa_1,\kappa_2; \svec\omega} = & +\frac{1}{2\pi} \sum_{m'} \delta_{\tau\tau'} \sum_{\kappa_1'\kappa_2'} \cals R_{\kappa_1'\kappa_2',m'}^{--}(r,r') \sum_{\lambda_p\lambda_p'} g_\omega^{\lambda_p}(r) g_\omega^{\lambda_p'}(r')\sum_{\lambda_y} R_{\lambda_y\lambda_y}^\omega(r,r')\widehat{\cals B}_{\kappa_1'\kappa_2'm';\kappa_1\kappa_2m}^{\lambda_p\lambda_p',\lambda_y;++},\\
  Y_{F, m}^{\kappa_1,\kappa_2; \svec\omega} = & -\frac{1}{2\pi} \sum_{m'} \delta_{\tau\tau'} \sum_{\kappa_1'\kappa_2'} \cals R_{\kappa_1'\kappa_2',m'}^{-+}(r,r') \sum_{\lambda_p\lambda_p'} g_\omega^{\lambda_p}(r) g_\omega^{\lambda_p'}(r')\sum_{\lambda_y} R_{\lambda_y\lambda_y}^\omega(r,r')\widehat{\cals B}_{\kappa_1'\kappa_2'm';\kappa_1\kappa_2m}^{\lambda_p\lambda_p',\lambda_y;+-},\\
  X_{G, m}^{\kappa_1,\kappa_2; \svec\omega} = & -\frac{1}{2\pi} \sum_{m'} \delta_{\tau\tau'} \sum_{\kappa_1'\kappa_2'} \cals R_{\kappa_1'\kappa_2',m'}^{+-}(r,r') \sum_{\lambda_p\lambda_p'} g_\omega^{\lambda_p}(r) g_\omega^{\lambda_p'}(r')\sum_{\lambda_y} R_{\lambda_y\lambda_y}^\omega(r,r')\widehat{\cals B}_{\kappa_1'\kappa_2'm';\kappa_1\kappa_2m}^{\lambda_p\lambda_p',\lambda_y;-+},\\
  X_{F, m}^{\kappa_1,\kappa_2; \svec\omega} = & +\frac{1}{2\pi} \sum_{m'} \delta_{\tau\tau'} \sum_{\kappa_1'\kappa_2'} \cals R_{\kappa_1'\kappa_2',m'}^{++}(r,r') \sum_{\lambda_p\lambda_p'} g_\omega^{\lambda_p}(r) g_\omega^{\lambda_p'}(r')\sum_{\lambda_y} R_{\lambda_y\lambda_y}^\omega(r,r')\widehat{\cals B}_{\kappa_1'\kappa_2'm';\kappa_1\kappa_2m}^{\lambda_p\lambda_p',\lambda_y;--}.
\end{align}
Here we use the bold type $\svec\omega$ to represent the space component. With regard to the aforementioned non-local self-energies, the energy functional associated with the space component of the $\omega$-V coupling may be expressed as follows,
\begin{align}
  E_{\svec\omega}^E =& \ff2\int drdr' \sum_i v_i^2\sum_{\kappa_1\kappa_2}\begin{pmatrix} \cals G_{i\kappa_1} & \cals F_{i\kappa_1} \end{pmatrix}_r\begin{pmatrix} Y_{G, m}^{\kappa_1,\kappa_2; \svec\omega} & Y_{F, m}^{\kappa_1,\kappa_2; \svec\omega}  \\[0.5em] X_{G, m}^{\kappa_1,\kappa_2; \svec\omega} & X_{F, m}^{\kappa_1,\kappa_2; \svec\omega} \end{pmatrix}_{r,r'}\begin{pmatrix} \cals G_{i\kappa_2} \\[0.5em] \cals F_{i\kappa_2} \end{pmatrix}_{r'}.
\end{align}
For the rearrangement terms, the contribution to the self-energy can be similarly expressed as,
\begin{align}
  \Sigma_{R,\svec\omega}^{E,\lambda_d} = & \int dr'\sum_i v_i^2\sum_{\kappa_1\kappa_2}\begin{pmatrix} \cals G_{i\kappa_1} & \cals F_{i\kappa_1} \end{pmatrix}_r\begin{pmatrix} P_{G, m,\lambda_d}^{\kappa_1,\kappa_2; \svec\omega} & P_{F, m,\lambda_d}^{\kappa_1,\kappa_2; \svec\omega}  \\[0.5em] Q_{G, m,\lambda_d}^{\kappa_1,\kappa_2; \svec\omega} & Q_{F, m,\lambda_d}^{\kappa_1,\kappa_2; \svec\omega} \end{pmatrix}_{r,r'}\begin{pmatrix} \cals G_{i\kappa_2} \\[0.5em] \cals F_{i\kappa_2} \end{pmatrix}_{r'},
\end{align}
where the $P$ and $Q$ terms read as,
\begin{align}
  P_{G, m, \lambda_d}^{\kappa_1,\kappa_2; \svec\omega} = & +\frac{1}{2\pi} \sum_{m'} \delta_{\tau\tau'} \sum_{\kappa_1'\kappa_2'} \cals R_{\kappa_1'\kappa_2',m'}^{--}(r,r') \sum_{\lambda_p\lambda_p'} \frac{\partial g_\omega^{\lambda_p}(r)}{\partial \rho_b^{\lambda_d}(r)} g_\omega^{\lambda_p'}(r')\sum_{\lambda_y} R_{\lambda_y\lambda_y}^\omega(r,r')\widehat{\cals B}_{\kappa_1'\kappa_2'm';\kappa_1\kappa_2m}^{\lambda_p\lambda_p',\lambda_y;++},\\
  P_{F, m, \lambda_d}^{\kappa_1,\kappa_2; \svec\omega} = & -\frac{1}{2\pi} \sum_{m'} \delta_{\tau\tau'} \sum_{\kappa_1'\kappa_2'} \cals R_{\kappa_1'\kappa_2',m'}^{-+}(r,r') \sum_{\lambda_p\lambda_p'} \frac{\partial g_\omega^{\lambda_p}(r)}{\partial \rho_b^{\lambda_d}(r)} g_\omega^{\lambda_p'}(r')\sum_{\lambda_y} R_{\lambda_y\lambda_y}^\omega(r,r')\widehat{\cals B}_{\kappa_1'\kappa_2'm';\kappa_1\kappa_2m}^{\lambda_p\lambda_p',\lambda_y;+-},\\
  Q_{G, m, \lambda_d}^{\kappa_1,\kappa_2; \svec\omega} = & -\frac{1}{2\pi} \sum_{m'} \delta_{\tau\tau'} \sum_{\kappa_1'\kappa_2'} \cals R_{\kappa_1'\kappa_2',m'}^{+-}(r,r') \sum_{\lambda_p\lambda_p'} \frac{\partial g_\omega^{\lambda_p}(r)}{\partial \rho_b^{\lambda_d}(r)} g_\omega^{\lambda_p'}(r')\sum_{\lambda_y} R_{\lambda_y\lambda_y}^\omega(r,r')\widehat{\cals B}_{\kappa_1'\kappa_2'm';\kappa_1\kappa_2m}^{\lambda_p\lambda_p',\lambda_y;-+},\\
  Q_{F, m, \lambda_d}^{\kappa_1,\kappa_2; \svec\omega} = & +\frac{1}{2\pi} \sum_{m'} \delta_{\tau\tau'} \sum_{\kappa_1'\kappa_2'} \cals R_{\kappa_1'\kappa_2',m'}^{++}(r,r') \sum_{\lambda_p\lambda_p'} \frac{\partial g_\omega^{\lambda_p}(r)}{\partial \rho_b^{\lambda_d}(r)} g_\omega^{\lambda_p'}(r')\sum_{\lambda_y} R_{\lambda_y\lambda_y}^\omega(r,r')\widehat{\cals B}_{\kappa_1'\kappa_2'm';\kappa_1\kappa_2m}^{\lambda_p\lambda_p',\lambda_y;--}.
\end{align}
The expressions for the other vector couplings can be obtained by replacing the expansion terms of the propagator and coupling strengths with the relevant ones. In addition, for the isovector $\rho$-V coupling, one needs to replace the isospin factor $\delta_{\tau\tau'}$ by $(2-\delta_{\tau\tau'})$, and for the Coulomb interaction only the $\lambda_p=\lambda_p'=0$ terms remain.

For the $\pi$-PV coupling, the situation becomes even more complicated. Firstly there exist gradient operations over the propagator, which can be expressed as,
\begin{align}
  \svec\nabla_{\svec r} \svec\nabla_{\svec r'} D_\pi(\svec r, \svec r')  = & m_\pi^2 \sum_{L=0}^\infty \sum_{\lambda_y\lambda_y'}^{L\pm 1} C_{L010}^{\lambda_y0} C_{L010}^{\lambda_y'0} \cals V_{L}^{\lambda_y\lambda_y'}(m_\pi, r, r') \nonumber\\
  &\hspace{2em}\times\sum_M (-1)^M\sum_{\mu_y\sigma_y} C_{\lambda_y\mu_y 1\sigma_y}^{LM} Y_{\lambda_y\mu_y}(\vartheta,\varphi)\svec e_{\sigma_y}\sum_{\mu'_y\sigma'_y} C_{\lambda_y'\mu'_y 1\sigma'_y}^{LM} Y_{\lambda_y'-\mu'_y}(\vartheta',\varphi')\svec e_{-\sigma'_y},
\end{align}
where the radial terms read as,
\begin{align}\label{eq:pi-prop}
\cals V_L^{\lambda_y \lambda_y'}(m_\pi;r,r') \equiv -R_{\lambda_y \lambda_y'}(m_\pi; r,r') + \frac{1}{m_\pi^2 r^2}\delta(r-r').
\end{align}
To have compact expressions, we introduce the symbols $\cals P$ and $\bar{\cals P}$ as,
\begin{align}
\sum_{\lambda_d} C_{L010}^{\lambda_y0} C_{\lambda_y\mu1\sigma}^{L\cals M} \cals Q_{\kappa_1'm',\kappa_1m}^{\lambda_d\mu\sigma} \Theta_{\lambda_d\lambda_p}^{\lambda_y \mu} \equiv& \cals P_{\kappa_1'm',\kappa_1 m}^{\lambda_p, L\lambda_y;\mu \sigma}, &
\sum_{\lambda_d} C_{L010}^{\lambda_y0} C_{\lambda_y\mu1\sigma}^{L\cals M} \bar{\cals Q}_{\kappa_1'm',\kappa_1m}^{\lambda_d\mu\sigma} \Theta_{\lambda_d\lambda_p}^{\lambda_y \mu} \equiv& \bar{\cals P}_{\kappa_1'm',\kappa_1 m}^{\lambda_p, L\lambda_y; \bar\mu\sigma},
\end{align}
and further the symbols $\widehat{\cals A}$ for the combinations of $\cals P$ and $\bar{\cals P}$ as,
\begin{subequations}\label{eq:pipvsymbs}
\begin{align}
  \widehat{\cals A}_{\kappa_1'\kappa_2'm';\kappa_1\kappa_2m}^{\lambda_p\lambda_p',L\lambda_y\lambda_y'; ++} \equiv & \frac{1}{2}\Bigg\{\sum_{\sigma\sigma'}\cals P_{ \kappa_1'm', \kappa_1m}^{\lambda_p,L\lambda_y;\mu\sigma} \cals P_{ \kappa_2'm', \kappa_2m}^{\lambda_p',L\lambda_y';\mu'\sigma'} + \sum_{\sigma\sigma'}\bar{\cals P}_{ \kappa_1'm', \kappa_1m}^{\lambda_p,L\lambda_y;\bar\mu\sigma} \bar{\cals P}_{ \kappa_2'm', \kappa_2m}^{\lambda_p',L\lambda_y';\bar\mu'\sigma'}\Bigg\},\\
  \widehat{\cals A}_{\kappa_1'\kappa_2'm';\kappa_1\kappa_2m}^{\lambda_p\lambda_p',L\lambda_y\lambda_y'; +-} \equiv & \frac{1}{2}\Bigg\{\sum_{\sigma\sigma'}\cals P_{ \kappa_1'm', \kappa_1m}^{\lambda_p,L\lambda_y;\mu\sigma} \cals P_{-\kappa_2'm',-\kappa_2m}^{\lambda_p',L\lambda_y';\mu'\sigma'} + \sum_{\sigma\sigma'}\bar{\cals P}_{ \kappa_1'm', \kappa_1m}^{\lambda_p,L\lambda_y;\bar\mu\sigma} \bar{\cals P}_{-\kappa_2'm',-\kappa_2m}^{\lambda_p',L\lambda_y';\bar\mu'\sigma'}\Bigg\},\\
  \widehat{\cals A}_{\kappa_1'\kappa_2'm';\kappa_1\kappa_2m}^{\lambda_p\lambda_p',L\lambda_y\lambda_y'; -+} \equiv & \frac{1}{2}\Bigg\{\sum_{\sigma\sigma'}\cals P_{-\kappa_1'm',-\kappa_1m}^{\lambda_p,L\lambda_y;\mu\sigma} \cals P_{ \kappa_2'm', \kappa_2m}^{\lambda_p',L\lambda_y';\mu'\sigma'} + \sum_{\sigma\sigma'}\bar{\cals P}_{-\kappa_1'm',-\kappa_1m}^{\lambda_p,L\lambda_y;\bar\mu\sigma} \bar{\cals P}_{ \kappa_2'm', \kappa_2m}^{\lambda_p',L\lambda_y';\bar\mu'\sigma'}\Bigg\},\\
  \widehat{\cals A}_{\kappa_1'\kappa_2'm';\kappa_1\kappa_2m}^{\lambda_p\lambda_p',L\lambda_y\lambda_y'; --} \equiv & \frac{1}{2}\Bigg\{\sum_{\sigma\sigma'}\cals P_{-\kappa_1'm',-\kappa_1m}^{\lambda_p,L\lambda_y;\mu\sigma} \cals P_{-\kappa_2'm',-\kappa_2m}^{\lambda_p',L\lambda_y';\mu'\sigma'} + \sum_{\sigma\sigma'}\bar{\cals P}_{-\kappa_1'm',-\kappa_1m}^{\lambda_p,L\lambda_y;\bar\mu\sigma} \bar{\cals P}_{-\kappa_2'm',-\kappa_2m}^{\lambda_p',L\lambda_y';\bar\mu'\sigma'}\Bigg\},
\end{align}
\end{subequations}
where $\mu +\sigma = \mu'+\sigma' = m-m'$ and $\bar\mu +\sigma = \bar\mu'+\sigma' = m+m'$. With the defined symbols, the self-energies from the $\pi$-PV coupling can be expressed as,
\begin{align}
Y_{G, m}^{\kappa_1\kappa_2,\pi} = &\frac{1}{2\pi} \sum_{m'} (2-\delta_{\tau\tau'}) \sum_{\kappa_1'\kappa_2'} \cals R_{\kappa_1'\kappa_2',m'}^{++}(r,r')\sum_{\lambda_p\lambda_p'}f_\pi^{\lambda_p}(r)f_\pi^{\lambda_p'}(r')\sum_L \sum_{\lambda_y \lambda_y'}^{L\pm1} \cals V_{L}^{\lambda_y\lambda_y'} (r,r')\widehat{\cals A}_{\kappa_1'\kappa_2'm';\kappa_1\kappa_2m}^{\lambda_p\lambda_p',L\lambda_y\lambda_y'; ++} ;\\
Y_{F, m}^{\kappa_1\kappa_2,\pi} = &\frac{1}{2\pi} \sum_{m'} (2-\delta_{\tau\tau'}) \sum_{\kappa_1'\kappa_2'} \cals R_{\kappa_1'\kappa_2',m'}^{+-}(r,r')\sum_{\lambda_p\lambda_p'}f_\pi^{\lambda_p}(r)f_\pi^{\lambda_p'}(r')\sum_L \sum_{\lambda_y \lambda_y'}^{L\pm1} \cals V_{L}^{\lambda_y\lambda_y'} (r,r')\widehat{\cals A}_{\kappa_1'\kappa_2'm';\kappa_1\kappa_2m}^{\lambda_p\lambda_p',L\lambda_y\lambda_y'; +-} ;\\
X_{G, m}^{\kappa_1\kappa_2,\pi} = &\frac{1}{2\pi} \sum_{m'} (2-\delta_{\tau\tau'}) \sum_{\kappa_1'\kappa_2'} \cals R_{\kappa_1'\kappa_2',m'}^{-+}(r,r')\sum_{\lambda_p\lambda_p'}f_\pi^{\lambda_p}(r)f_\pi^{\lambda_p'}(r')\sum_L \sum_{\lambda_y \lambda_y'}^{L\pm1} \cals V_{L}^{\lambda_y\lambda_y'} (r,r')\widehat{\cals A}_{\kappa_1'\kappa_2'm';\kappa_1\kappa_2m}^{\lambda_p\lambda_p',L\lambda_y\lambda_y'; -+} ; \\
X_{F, m}^{\kappa_1\kappa_2,\pi} = &\frac{1}{2\pi} \sum_{m'} (2-\delta_{\tau\tau'}) \sum_{\kappa_1'\kappa_2'} \cals R_{\kappa_1'\kappa_2',m'}^{--}(r,r')\sum_{\lambda_p\lambda_p'}f_\pi^{\lambda_p}(r)f_\pi^{\lambda_p'}(r')\sum_L \sum_{\lambda_y \lambda_y'}^{L\pm1} \cals V_{L}^{\lambda_y\lambda_y'} (r,r')\widehat{\cals A}_{\kappa_1'\kappa_2'm';\kappa_1\kappa_2m}^{\lambda_p\lambda_p',L\lambda_y\lambda_y'; --} .
\end{align}
In terms of the non-local self-energies, the energy functional from the $\pi$-PV coupling read as,
\begin{align}
E_{\pi-\text{PV}}^{E} =& \frac{1}{2} \int drdr' \sum_{i}v_i^{2} \sum_{\kappa_1\kappa_2} \left(\begin{array}{cc} \cals G_{i\kappa_1}(r) & \cals F_{i\kappa_1}(r) \end{array}\right) \left(\begin{array}{cc} Y_{G, m}^{\kappa_1\kappa_2,\pi} & Y_{F, m}^{\kappa_1\kappa_2,\pi}\\[0.5em] X_{G, m}^{\kappa_1\kappa_2,\pi} & X_{F, m}^{\kappa_1\kappa_2,\pi} \end{array}\right)_{r,r'} \left(\begin{array}{c} \cals G_{i\kappa_2}(r') \\[0.5em] \cals F_{i\kappa_2}(r')\end{array}\right).
\end{align}
Similarly the rearrangement term $\Sigma_{R,\pi}^{E,\lambda_d}$ can be expressed as,
\begin{align}
\Sigma_{R,\pi}^{E,\lambda_d} = &\int dr' \sum_{i} v_i^2 \sum_{\kappa_1\kappa_2} \left(\begin{array}{cc}\cals G_{i\kappa_1}(r) & \cals F_{i\kappa_1}(r) \end{array}\right) \left(\begin{array}{cc} P_{G, m,\lambda_d}^{\kappa_a\kappa_b, \pi } & P_{F, m,\lambda_d}^{\kappa_a\kappa_b, \pi} \\ [0.5em] Q_{G, m,\lambda_d}^{\kappa_a\kappa_b, \pi } & Q_{F, m,\lambda_d}^{\kappa_a\kappa_b, \pi }\end{array}\right)_{r,r'} \left(\begin{array}{c} \cals G_{i\kappa_2}(r')\\ [0.5em] \cals F_{i\kappa_2}(r')\end{array}\right) ,
\end{align}
where the terms $P$ and $Q$ read as,
\begin{align}
  P_{G, m,\lambda_d}^{\kappa_1\kappa_2,\pi}=& \frac{1}{2\pi}\sum_{m'}(2-\delta_{\tau\tau'})  \sum_{\kappa_1'\kappa_2'} \cals R_{\kappa_1'\kappa_2',m'}^{++}(r,r')\sum_{\lambda_p\lambda_p'}\frac{\partial f_\pi^{\lambda_p}(r)}{\partial \rho_b^{\lambda_d}(r)} f_\pi^{\lambda_p'}(r')\sum_L \sum_{\lambda_y \lambda_y'}^{L\pm1} \cals V_{L}^{\lambda_y\lambda_y'}(r,r') \widehat{\cals A}_{\kappa_1'\kappa_2'm';\kappa_1\kappa_2m}^{\lambda_p\lambda_p',L\lambda_y\lambda_y'; ++} ;\\
  P_{F, m,\lambda_d}^{\kappa_1\kappa_2,\pi}=& \frac{1}{2\pi}\sum_{m'}(2-\delta_{\tau\tau'})  \sum_{\kappa_1'\kappa_2'} \cals R_{\kappa_1'\kappa_2',m'}^{+-}(r,r')\sum_{\lambda_p\lambda_p'}\frac{\partial f_\pi^{\lambda_p}(r)}{\partial \rho_b^{\lambda_d}(r)} f_\pi^{\lambda_p'}(r')\sum_L \sum_{\lambda_y \lambda_y'}^{L\pm1} \cals V_{L}^{\lambda_y\lambda_y'}(r,r') \widehat{\cals A}_{\kappa_1'\kappa_2'm';\kappa_1\kappa_2m}^{\lambda_p\lambda_p',L\lambda_y\lambda_y'; +-} ;\\
  Q_{G, m,\lambda_d}^{\kappa_1\kappa_2,\pi}=& \frac{1}{2\pi}\sum_{m'}(2-\delta_{\tau\tau'})  \sum_{\kappa_1'\kappa_2'} \cals R_{\kappa_1'\kappa_2',m'}^{-+}(r,r')\sum_{\lambda_p\lambda_p'}\frac{\partial f_\pi^{\lambda_p}(r)}{\partial \rho_b^{\lambda_d}(r)} f_\pi^{\lambda_p'}(r')\sum_L \sum_{\lambda_y \lambda_y'}^{L\pm1} \cals V_{L}^{\lambda_y\lambda_y'}(r,r') \widehat{\cals A}_{\kappa_1'\kappa_2'm';\kappa_1\kappa_2m}^{\lambda_p\lambda_p',L\lambda_y\lambda_y'; -+} ;\\
  Q_{F, m,\lambda_d}^{\kappa_1\kappa_2,\pi}=& \frac{1}{2\pi}\sum_{m'}(2-\delta_{\tau\tau'})  \sum_{\kappa_1'\kappa_2'} \cals R_{\kappa_1'\kappa_2',m'}^{--}(r,r')\sum_{\lambda_p\lambda_p'}\frac{\partial f_\pi^{\lambda_p}(r)}{\partial \rho_b^{\lambda_d}(r)} f_\pi^{\lambda_p'}(r')\sum_L \sum_{\lambda_y \lambda_y'}^{L\pm1} \cals V_{L}^{\lambda_y\lambda_y'}(r,r') \widehat{\cals A}_{\kappa_1'\kappa_2'm';\kappa_1\kappa_2m}^{\lambda_p\lambda_p',L\lambda_y\lambda_y'; --} .
\end{align}
Besides, the contact term is introduced to compensate the zero-range term in $\cals V_L^{\lambda_y\lambda_y'}$ [see Eq. (\ref{eq:pi-prop})]. The Hartree contributions from the contact term are derived as zero, and the Fock contributions read as,
\begin{align}
E_\pi^\delta = -\ff2 \times\frac{1}{3} \sum_{ii'}(2-\delta_{\tau_i \tau_{i'}}) \int d\svec r d\svec r' \Big[\frac{f_\pi}{m_\pi} \bar{\psi}_{\nu  m} \gamma_5 \svec \gamma \psi_{\nu'm'}\Big]_{\svec r} \cdot \Big[\frac{f_\pi}{m_\pi} \bar{\psi}_{\nu'  m'} \gamma_5 \svec \gamma \psi_{\nu m}\Big]_{\svec r'} \delta(\svec r - \svec r'),
\end{align}
where the $\delta$ function can be decomposed as,
\begin{align}
\delta (\svec r - \svec r') = \frac{\delta (r-r')}{r^2} \sum_{L=0}^{\infty}\sum_M (-1)^{M} Y_{LM}(\vartheta,\varphi) Y_{L-M}(\vartheta',\varphi').
\end{align}
To express the contact term compactly, we introduce the symbols $\scr B$ as,
\begin{align}
  \widehat{\scr B}_{\kappa_1'\kappa_2'm';\kappa_1\kappa_2m}^{\lambda_p\lambda_p';++} \equiv & \ff2\sum_{\lambda_d\lambda_d' L\sigma} \Big[ \cals Q_{\kappa_1'm', \kappa_1m}^{\lambda_d \mu\sigma}\cals Q_{\kappa_2'm',\kappa_2m}^{\lambda_d'\mu\sigma} \Theta_{\lambda_d\lambda_p}^{L\mu} \Theta_{\lambda_d'\lambda_p'}^{L\mu} + \bar{\cals Q}_{\kappa_1'm', \kappa_1m}^{\lambda_d\bar \mu\sigma}\bar{\cals Q}_{\kappa_2'm',\kappa_2m}^{\lambda_d'\bar \mu\sigma} \Theta_{\lambda_d\lambda_p}^{L\bar \mu} \Theta_{\lambda_d'\lambda_p'}^{L\bar \mu}\Big],\\
  \widehat{\scr B}_{\kappa_1'\kappa_2'm';\kappa_1\kappa_2m}^{\lambda_p\lambda_p';+-} \equiv & \ff2\sum_{\lambda_d\lambda_d' L\sigma} \Big[ \cals Q_{\kappa_1'm', \kappa_1m}^{\lambda_d \mu\sigma}\cals Q_{-\kappa_2'm',-\kappa_2m}^{\lambda_d'\mu\sigma} \Theta_{\lambda_d\lambda_p}^{L\mu} \Theta_{\lambda_d'\lambda_p'}^{L\mu} + \bar{\cals Q}_{\kappa_1'm', \kappa_1m}^{\lambda_d\bar \mu\sigma}\bar{\cals Q}_{-\kappa_2'm',-\kappa_2m}^{\lambda_d'\bar \mu\sigma} \Theta_{\lambda_d\lambda_p}^{L\bar \mu} \Theta_{\lambda_d'\lambda_p'}^{L\bar \mu}\Big],\\
  \widehat{\scr B}_{\kappa_1'\kappa_2'm';\kappa_1\kappa_2m}^{\lambda_p\lambda_p';-+} \equiv & \ff2\sum_{\lambda_d\lambda_d' L\sigma} \Big[ \cals Q_{-\kappa_1'm', -\kappa_1m}^{\lambda_d \mu\sigma}\cals Q_{\kappa_2'm',\kappa_2m}^{\lambda_d'\mu\sigma} \Theta_{\lambda_d\lambda_p}^{L\mu} \Theta_{\lambda_d'\lambda_p'}^{L\mu} + \bar{\cals Q}_{-\kappa_1'm', -\kappa_1m}^{\lambda_d\bar \mu\sigma}\bar{\cals Q}_{\kappa_2'm',\kappa_2m}^{\lambda_d'\bar \mu\sigma} \Theta_{\lambda_d\lambda_p}^{L\bar \mu} \Theta_{\lambda_d'\lambda_p'}^{L\bar \mu}\Big],\\
  \widehat{\scr B}_{\kappa_1'\kappa_2'm';\kappa_1\kappa_2m}^{\lambda_p\lambda_p';--} \equiv & \ff2\sum_{\lambda_d\lambda_d' L\sigma} \Big[ \cals Q_{-\kappa_1'm',-\kappa_1m}^{\lambda_d \mu\sigma}\cals Q_{-\kappa_2'm',-\kappa_2m}^{\lambda_d'\mu\sigma} \Theta_{\lambda_d\lambda_p}^{L\mu} \Theta_{\lambda_d'\lambda_p'}^{L\mu} + \bar{\cals Q}_{-\kappa_1'm',-\kappa_1m}^{\lambda_d\bar \mu\sigma}\bar{\cals Q}_{-\kappa_2'm',-\kappa_2m}^{\lambda_d'\bar \mu\sigma} \Theta_{\lambda_d\lambda_p}^{L\bar \mu} \Theta_{\lambda_d'\lambda_p'}^{L\bar \mu}\Big].
\end{align}
Thus, the energy functional of the contact term can be expressed as,
\begin{align}
E_{\pi }^{\delta} =& \frac{1}{2} \int dr \sum_{i}v_i^{2} \sum_{\kappa_1\kappa_2} \left(\begin{array}{cc} \cals G_{i\kappa_1}(r) & \cals F_{i\kappa_1}(r) \end{array}\right) \left(\begin{array}{cc} Y_{G, m}^{\kappa_1\kappa_2,\pi\delta} & Y_{F, m}^{\kappa_1\kappa_2,\pi\delta}\\[0.5em] X_{G, m}^{\kappa_1\kappa_2,\pi\delta} & X_{F, m}^{\kappa_1\kappa_2,\pi\delta} \end{array}\right)_{r,r} \left(\begin{array}{c} \cals G_{i\kappa_2}(r) \\[0.5em] \cals F_{i\kappa_2}(r)\end{array}\right),
\end{align}
where the self-energies read as,
\begin{align}
Y_{G, m}^{\kappa_1\kappa_2,\pi\delta} = &-\frac{1}{6\pi m_\pi^2 r^2}\sum_{m'} (2-\delta_{\tau\tau'}) \sum_{\kappa_1'\kappa_2'} \cals R_{\kappa_1'\kappa_2',m'}^{++}(r,r)  \sum_{\lambda_p \lambda_p'}f_\pi^{\lambda_p}(r) f_\pi^{\lambda_{p'}}(r) \widehat{\scr B}_{\kappa_1'\kappa_2'm';\kappa_1\kappa_2m}^{\lambda_p\lambda_p';++}, \\
Y_{F, m}^{\kappa_1\kappa_2,\pi\delta} = &-\frac{1}{6\pi m_\pi^2 r^2}\sum_{m'} (2-\delta_{\tau\tau'}) \sum_{\kappa_1'\kappa_2'} \cals R_{\kappa_1'\kappa_2',m'}^{+-}(r,r)  \sum_{\lambda_p \lambda_p'}f_\pi^{\lambda_p}(r) f_\pi^{\lambda_{p'}}(r) \widehat{\scr B}_{\kappa_1'\kappa_2'm';\kappa_1\kappa_2m}^{\lambda_p\lambda_p';+-}, \\
X_{G, m}^{\kappa_1\kappa_2,\pi\delta} = &-\frac{1}{6\pi m_\pi^2 r^2}\sum_{m'} (2-\delta_{\tau\tau'}) \sum_{\kappa_1'\kappa_2'} \cals R_{\kappa_1'\kappa_2',m'}^{-+}(r,r)  \sum_{\lambda_p \lambda_p'}f_\pi^{\lambda_p}(r) f_\pi^{\lambda_{p'}}(r) \widehat{\scr B}_{\kappa_1'\kappa_2'm';\kappa_1\kappa_2m}^{\lambda_p\lambda_p';-+}, \\
X_{F, m}^{\kappa_1\kappa_2,\pi\delta} = &-\frac{1}{6\pi m_\pi^2 r^2}\sum_{m'} (2-\delta_{\tau\tau'}) \sum_{\kappa_1'\kappa_2'} \cals R_{\kappa_1'\kappa_2',m'}^{--}(r,r)  \sum_{\lambda_p \lambda_p'}f_\pi^{\lambda_p}(r) f_\pi^{\lambda_{p'}}(r) \widehat{\scr B}_{\kappa_1'\kappa_2'm';\kappa_1\kappa_2m}^{\lambda_p\lambda_p';--}.
\end{align}
The rearrangement term in the self-energy can be derived as,
\begin{align}
\Sigma_{R,\pi}^{\delta,\lambda_d} = & \sum_{i} v_i^2 \sum_{\kappa_1\kappa_2} \left(\begin{array}{cc}\cals G_{i\kappa_1}(r) & \cals F_{i\kappa_1}(r) \end{array}\right) \left(\begin{array}{cc} P_{G, m,\lambda_d}^{\kappa_a\kappa_b, \pi \delta} & P_{F, m,\lambda_d}^{\kappa_a\kappa_b, \pi \delta} \\ [0.5em] Q_{G, m,\lambda_d}^{\kappa_a\kappa_b, \pi \delta} & Q_{F, m,\lambda_d}^{\kappa_a\kappa_b, \pi \delta}\end{array}\right)_{r,r} \left(\begin{array}{c} \cals G_{i\kappa_2}(r)\\ [0.5em] \cals F_{i\kappa_2}(r)\end{array}\right),
\end{align}
with the $P$ and $Q$ terms as,
\begin{align}
P_{G, m,\lambda_d}^{\kappa_1\kappa_2,\pi\delta} = &-\frac{1}{6\pi m_\pi^2 r^2} \sum_{m'}(2-\delta_{\tau\tau'}) \sum_{\kappa_1'\kappa_2'} \cals R_{\kappa_1'\kappa_2',m'}^{++}(r,r) \sum_{\lambda_p \lambda_p'}\frac{\partial f_\pi^{\lambda_p}(r)}{\partial \rho_b^{\lambda_d}(r)} f_\pi^{\lambda_{p'}}(r)\widehat{\scr B}_{\kappa_1'\kappa_2'm';\kappa_1\kappa_2m}^{\lambda_p\lambda_p';++},\\
P_{F, m,\lambda_d}^{\kappa_1\kappa_2,\pi\delta} = &-\frac{1}{6\pi m_\pi^2 r^2} \sum_{m'}(2-\delta_{\tau\tau'}) \sum_{\kappa_1'\kappa_2'} \cals R_{\kappa_1'\kappa_2',m'}^{+-}(r,r) \sum_{\lambda_p \lambda_p'}\frac{\partial f_\pi^{\lambda_p}(r)}{\partial \rho_b^{\lambda_d}(r)} f_\pi^{\lambda_{p'}}(r)\widehat{\scr B}_{\kappa_1'\kappa_2'm';\kappa_1\kappa_2m}^{\lambda_p\lambda_p';+-},\\
Q_{G, m,\lambda_d}^{\kappa_1\kappa_2,\pi\delta} = &-\frac{1}{6\pi m_\pi^2 r^2} \sum_{m'}(2-\delta_{\tau\tau'}) \sum_{\kappa_1'\kappa_2'} \cals R_{\kappa_1'\kappa_2',m'}^{-+}(r,r) \sum_{\lambda_p \lambda_p'}\frac{\partial f_\pi^{\lambda_p}(r)}{\partial \rho_b^{\lambda_d}(r)} f_\pi^{\lambda_{p'}}(r)\widehat{\scr B}_{\kappa_1'\kappa_2'm';\kappa_1\kappa_2m}^{\lambda_p\lambda_p';-+},\\
Q_{F, m,\lambda_d}^{\kappa_1\kappa_2,\pi\delta} = &-\frac{1}{6\pi m_\pi^2 r^2} \sum_{m'}(2-\delta_{\tau\tau'}) \sum_{\kappa_1'\kappa_2'} \cals R_{\kappa_1'\kappa_2',m'}^{--}(r,r) \sum_{\lambda_p \lambda_p'}\frac{\partial f_\pi^{\lambda_p}(r)}{\partial \rho_b^{\lambda_d}(r)} f_\pi^{\lambda_{p'}}(r)\widehat{\scr B}_{\kappa_1'\kappa_2'm';\kappa_1\kappa_2m}^{\lambda_p\lambda_p';--}.
\end{align}

\subsection{Process of the rearrangement term} \label{app:DDMNC}

For the octupole deformed nuclei, the coupling constant $g_\phi$ in equation ($\ref{eq:density-dependence}$) can be expressed in terms of Legendre polynomial as,
\begin{align}\label{eq:ExpansionP2}
  g_\phi(\rho_b) = & \sum_{\lambda_p} g_\phi^{\lambda_p}(\rho_b) P_{\lambda_p}(\cos\vartheta),
\end{align}
where the cutoff of $\lambda_p$ is decided by the convergence requirement. The expansion term $g_\phi^{\lambda_p}$ is calculated via the
inverse solution,
\begin{align}
  g_\phi^{\lambda_p} (\rho_b) = & \int_{-1}^1 d(\cos\vartheta) P_{\lambda_p}(\cos\vartheta) g_\phi(\rho_b).
\end{align}
Because of the density dependence in $g_\phi$, the rearrangement term $\Sigma_R$ should be taken into account to preserve the energy-momentum conservation \cite{Typel1999NPA656.331}. The variation of the coupling constants $g_\phi$ can be expressed as,
\begin{align}
  \delta g_\phi(\rho_b) = & \sum_{a,i} \sum_{\lambda_p} P_{\lambda_p}(\cos\vartheta) \sum_{\lambda_d} \frac{\partial g_\phi^{\lambda_p}}{\partial\rho_b^{\lambda_d}}\frac{\partial\rho_b^{\lambda_d}}{\partial C_{a,i}}\delta C_{a,i},
\end{align}
where the fraction term ${\partial\rho_b^{\lambda_d}}/{\partial C_{a,i}}$ can be deduced easily from the expression of $\rho_b^{\lambda_d}$ (\ref{eq:rhob-l}). The term ${\partial g_\phi^{\lambda_p}}/{\partial\rho_b^{\lambda_d}}$ is derived as,
\begin{align}
  \frac{\partial g_\phi^{\lambda_p}}{\partial\rho_b^{\lambda_d}} = & \sum_L\frac{\hat \lambda_p\hat\lambda_d}{\sqrt2\hat L} \big(C_{\lambda_p0\lambda_d0}^{L0}\big)^2 \int_{-1}^{1}dx P_L(x)\frac{\partial g_\phi}{\partial\rho_b},
\end{align}
where ${\partial g_\phi}/{\partial\rho_b}$ is determined by the density-dependent form (\ref{eq:density-dependence}), and $\hat L = \sqrt{2L+1}$. Due to the density dependencies in the coupling strengths, the rearrangement terms can be simply expressed as,
\begin{align}
  \Sigma_R^{\lambda_d} = & \sum_\phi \Big(\Sigma_{R,\phi}^{D,\lambda_d} + \Sigma_{R,\phi}^{E,\lambda_d}\Big).
\end{align}
For more detailed contributions from the Hartree and Fock terms, one can refer to Appendix \ref{app:Hartree} and \ref{app:Fock}, respectively.

\subsection{The non-local parts in pairing correlations with Gogny force}\label{app:pair}

In this work, the finite-range Gogny force D1S is adopted as the pairing force, and the full contributions from all
the $J$-components are considered. The non-local parts in interaction matrix elements read as
\begin{align}
    \bar{Y}_{\kappa \kappa'}^G(r,r') = & \sum_\lambda \lrs{A_\lambda(r,r') \widehat{\cals X}_{m\kappa\tilde\kappa; m' \kappa'\tilde\kappa'}^\lambda - D_\lambda(r,r') \widehat{\cals S}_{m\kappa\tilde\kappa; m' \kappa'\tilde\kappa'}^\lambda }, \\
    \bar{Y}_{\kappa \kappa'}^F(r,r') = & \sum_\lambda \lrs{A_\lambda(r,r') \widehat{\cals X}_{m\kappa-\tilde\kappa; m' \kappa'-\tilde\kappa'}^\lambda - D_\lambda(r,r') \widehat{\cals S}_{m\kappa-\tilde\kappa; m' \kappa'-\tilde\kappa'}^\lambda }, \\
    \bar{Y}_{ \kappa'\kappa}^G(r,r') = & \sum_\lambda \lrs{A_\lambda(r,r') \widehat{\cals X}_{m-\kappa\tilde\kappa; m' -\kappa'\tilde\kappa'}^\lambda - D_\lambda(r,r') \widehat{\cals S}_{m-\kappa\tilde\kappa; m' -\kappa'\tilde\kappa'}^\lambda }, \\
    \bar{Y}_{ \kappa'\kappa}^F(r,r') = & \sum_\lambda \lrs{A_\lambda(r,r') \widehat{\cals X}_{m-\kappa-\tilde\kappa; m'- \kappa'-\tilde\kappa'}^\lambda - D_\lambda(r,r') \widehat{\cals S}_{m-\kappa-\tilde\kappa; m' -\kappa'-\tilde\kappa'}^\lambda }.
\end{align}
In the above expressions, the symbols $\widehat{\cals X}$ and $\widehat{\cals S}$ read as,
\begin{align}
  \widehat{\cals X}_{m\kappa\tilde\kappa; m' \kappa'\tilde\kappa'}^{\lambda} \equiv&  \sum_{JLS}  f \cals S_{ m\kappa\tilde\kappa}^{JLS} \cals X_{\kappa\tilde\kappa;\kappa'\tilde\kappa'}^{L,\lambda} \cals S_{ m'\kappa'\tilde\kappa'}^{JLS}, &
  \widehat{\cals S}_{m\kappa\tilde\kappa; m'\kappa'\tilde\kappa'}^{\lambda } \equiv & \sum_{JLS} f (-1)^{S} \cals S_{m\kappa\tilde\kappa }^{JLS} \cals X_{\kappa\tilde\kappa;\kappa'\tilde\kappa'}^{L,\lambda} \cals S_{m'\kappa'\tilde\kappa'}^{JLS},\label{eq:scrXS}
\end{align}
with the symbols $\cals X$ and $\cals S$ as,
\begin{align}
    \cals X_{\kappa \tilde \kappa, \kappa'\tilde \kappa'}^{L,\lambda} = & C_{\tilde l_u' 0 \tilde l_u 0} ^{\lambda0} C_{l_u'0 l_u0}^{\lambda0}\Lrb{\begin{matrix} l_u' & l_u & \lambda\\ \tilde l_u & \tilde l_u' & L  \end{matrix}}, & \cals S_{m\kappa \tilde \kappa}^{JLS} = & \hat j\hat{\tilde j}\hat l_u\hat{\tilde l}_u\hat L\hat S \begin{Bmatrix} j &\tilde j & J \\ l_u & \tilde l_u & L \\ \ff2 & \ff2 & S \end{Bmatrix} C_{\tilde j-m jm }^{J0},
\end{align}
and the symbol factor $f = \ff4 (-1)^{\tilde{j}^{'}+ l_u^{'}-m^{'}  + \tilde{j}+l_u -m + \tilde{l}_u - L }[(-1)^{\tilde{l}_u^{'}}+(-1)^{l_u^{'}+J}]$.
In the $6j$ and $9j$ symbols, $l_u$ corresponds to the orbital angular momentum of upper component of the spherical Dirac spinor, and the one of the lower component reads as $l_d$. That is to say, for the terms with $-\kappa$, $l_u$ shall be replaced by $l_d$.

\subsection{Dirac Woods-Saxon potentials and deformation parameters} \label{app:DWS-DP}

In current work, the calculation is performed using the self-consistent iteration method which starts in general from an initial potential. The local Dirac Woods-Saxon potentials \cite{Koepf1991ZPA339.81} which are also used to determine the spherical DWS base are taken as the initial ones.
In order to chose an appropriate starting deformation point $(\beta_2,\beta_3)$, starting from which the energy functional will converge to a local
minimum nearby $(\beta_2,\beta_3)$, the effects of deformation need to be considered.

For convenience, the local self-energies, i.e., $\Sigma_0+\gamma^0\Sigma_S$, in Dirac equation are also referred as $\Sigma_{\pm} = \Sigma_0 \pm \Sigma_S$. it is worth noting that the local self-energy $\Sigma_T$ and the non-local ones in Dirac equation (\ref{eq:Dirac-x}) are set to be zero.
The Woods-Saxon type local $\Sigma_\pm$, the starting local self-energies, can be expressed as,
\begin{align}\label{eq:DWS}
  \Sigma_{+}^{\tau_3}(r ) = &+ V_0 \frac{1-a_0(N-Z)\tau_3/A}{1 + \exp\big[ (r-R_+^{\tau_3})/a_+^{\tau_3}\big]}+V_c^{\tau_3}, &
    \Sigma_-^{\tau_3}(r ) = & - V_0\frac{a_v^{\tau_3}\lrs{1-a_0(N-Z)\tau_3/A}}{1 + \exp\big[ (r-R_-^{\tau_3})/a_-^{\tau_3}\big]}+V_c^{\tau_3},
\end{align}
where the isospin projection operator $\tau_3$ is defined as  $\tau_3\lrlc{n} = \lrlc{n}$ and $\tau_3\lrlc{p} = -\lrlc{p}$, and $R_\pm^{\tau_3} = r_{0,\pm}^{\tau_3} A^{1/3}$ are the empirical radius of neutron or proton for the nucleus with mass number $A$, and $V_c^{\tau_3}$ represents the Coulomb potential between protons. The parameters in the Dirac Woods-Saxon potential, being used in this work to provide the spherical DWS base and initial potential, are given in Table \ref{tab:DWS}.

\begin{table}[h]
  \caption{Parameters of the Dirac Woods-Saxon potential, and $V_0=-71.28$ MeV and $a_0 = 0.4616$ \cite{Koepf1991ZPA339.81}. Except the dimensionless $a_v$, the other parameters are in fm. }\label{tab:DWS}\setlength{\tabcolsep}{0.6em}
  \begin{tabular}{c|rcccc}\hline\hline
     & $a_v$~~~~& $r_{0,+}$ & $r_{0,-}$ & $a_+$ & $a_-$ \\ \hline
    Neutron &11.1175 &1.2334&1.1443& 0.615&0.6476 \\
    Proton & 8.9698&1.2496& 1.1400 & 0.6124& 0.6469\\ \hline\hline
  \end{tabular}
\end{table}

For the Coulomb potential in the Dirac Woods-Saxon potentials (\ref{eq:DWS}), it is evaluated by assuming the charge distributed uniformly,
\begin{subequations}\label{eq:Coulomb}
\begin{align}
  V_c = & \alpha Z \Big(\frac{3}{2R_c} - \frac{r^2}{2R_c^3}\Big), & \text{when }r<R_c,\\
  V_c = & \alpha Z\frac{1}{r}, & \text{when } r\geqslant R_c,
\end{align}
\end{subequations}
where $R_c = R_+^{\tau_3=-1}$, and $\alpha$ is the fine-structure constant that represents the coupling strength of Coulomb interaction.

For the initial DWS potentials (\ref{eq:DWS}), one has to take the deformation effects into account. For the octupole deformed nuclei with pear-like shape, considering axial symmetry and reflect asymmetry, the surface with equal density can be described as,
\begin{align}
  R(\vartheta,\beta_2,\beta_3) = & R_0\Big[1 + \sqrt{\frac{5}{16\pi}}\beta_2 \big(3\cos^2\vartheta - 1\big)\Big.
                       + \sqrt{\frac{7}{16\pi}}\beta_3 \big(5\cos^3\vartheta - \cos\vartheta\big)\Big],
\end{align}
where $(\beta_2,\beta_3)$ describe the deformation of a nucleus with pear-like shape. Considering the deformed effect, the empirical radii $R_{\pm}^{\tau_3}$ shall be replaced by $R_{\pm}^{\tau_3}(\vartheta,\beta_2,\beta_3)$, which leads to $\Sigma_\pm^{\tau_3} = \Sigma_\pm^{\tau_3}(r,\vartheta;\beta_2,\beta_3)$,
\begin{align}
  R_\pm^{\tau_3}(\vartheta,\beta_2,\beta_3) = & R_{0,\pm}^{\tau_3} \bigg[1+\sqrt{\frac{5}{16\pi}}\beta_2 \big(3\cos^2\vartheta - 1\big)\bigg. + \sqrt{\frac{7}{16\pi}}\beta_3 \big(5\cos^3\vartheta - \cos\vartheta\big)\Big],\\
  R_{0,\pm}^{\tau_3} =& r_{0,\pm}^{\tau_3}\left[\frac{4\pi}{3}A \left(\frac{4\pi}{3}+\beta_2^2+\frac{1}{21}\sqrt{\frac{5}{\pi}}\beta_2^3 +\frac{10\beta_3^2}{3}+\frac{6\beta_2\beta_3^2}{\sqrt{5\pi}}\right)^{-1}\right]^{\frac{1}{3}},
\end{align}
where the values of $r_{0,\pm}^{\tau_3}$ are given in Table \ref{tab:DWS}.

For the initial DWS potentials $\Sigma_\pm^{\tau_3}(r,\vartheta,\beta_2,\beta_3)$, similar expansion terms as the local self-energies in the Hartree energy functional (\ref{eq:ED}) can be determined by,
\begin{align}
  \Sigma_{\pm}^{\tau_3,\lambda_d}(r) = & \int_0^\pi \Sigma_{\pm}^{\tau_3} (r,\vartheta) P_{\lambda}(\cos\vartheta) \sin\vartheta d\vartheta.
\end{align}
The initial matrix in the eigenvalue equation (\ref{eq:eigen}) is then obtained by replacing the expansion terms $\Sigma_0^{\lambda_d}\pm \Sigma_S^{\lambda_d}$  in $H^D_{aa'}$ (\ref{eq:HDaa'}) with $\Sigma_\pm^{\tau_3,\lambda_d}$, since the kinetic part $H_{aa'}^{\text{kin}}$ only depends on the DWS base and the non-local part $H_{aa'}^{E}$ is set as zero.

With the expansion of densities (\ref{eq:densities}), it is rather straightforward to deduce the intrinsic quadruple and octupole mass moment $Q_{2;3} $ for a deformed nucleus as,
\begin{align}
  Q_{2} = & 2\pi\sqrt{\frac{8}{5}} \int dr r^4 \rho_{b}^{\lambda_d=2}(r), &
  Q_{3} = & 2\pi\sqrt{\frac{8}{7}} \int dr r^5 \rho_{b}^{\lambda_d=3}(r).
\end{align}
With the mass moment $Q_{2,3}$, the quadruple and octupole deformation $\beta_2$ and $\beta_3$ can be evaluated approximately by
\begin{align}
  \beta_2 = & \frac{\sqrt{5\pi}Q_2}{3R_0^2A}, &
  \beta_3 = & \frac{\sqrt{7\pi}Q_3}{3R_0^3A}
\end{align}
where $R_0 = 1.2 A^{1/3}$.
\end{widetext}

\end{document}